\documentclass[preprint,12pt]{elsarticle}

\usepackage{lineno, hyperref, amsfonts, amsmath, textgreek, booktabs, amsthm}
\usepackage{makecell, chngcntr, graphicx, caption, subcaption, soul, color, placeins, amssymb}
\usepackage[title]{appendix}

\usepackage[integrals]{wasysym}
\SetSymbolFont{wasy}{bold}{U}{wasy}{m}{n}
\usepackage{setspace}

\usepackage{geometry}
\newtheorem*{remark}{Remark}
\modulolinenumbers[1]
\counterwithout{figure}{section}
\usepackage{framed}
\usepackage{multicol}
\usepackage{nomencl}
\makenomenclature
\renewcommand*\nompreamble{\begin{multicols}{2}}
\renewcommand*\nompostamble{\end{multicols}}
\usepackage{etoolbox}
\usepackage{xstring}

\renewcommand\nomgroup[1]{%
  \item[\bfseries
  \ifstrequal{#1}{A}{Abbreviations}{%
  \ifstrequal{#1}{S}{Symbols}{}}%
]}
\usepackage{multirow}
\usepackage[ruled,vlined,linesnumbered]{algorithm2e}
\usepackage{siunitx}
\newcommand{\bd}[1]{\boldsymbol{#1}}
\newcommand{\bm}[1]{\boldsymbol{\mathcal{#1}}}

\newcommand{\tripledots}{\mathbin{\vdots}}
\newcommand{\Transparency}{\mathcal{T}}
\newcommand{\sigdemand}{\hat{\sigma}_d}
\newcommand{\sigfroz}{\Sigma_{\mathrm{froz}}}

\begin{document}
\setstretch{1.5}

\begin{frontmatter}
\title{A Computational Model for Flexoelectricity-Driven Contact Electrification}

\author[label1]{Han Hu}
\author[label1,label2]{Xiaoying Zhuang\corref{mycorrespondingauthor}}
\cortext[mycorrespondingauthor]{Corresponding author}
\ead{zhuang@iop.uni-hannover.de}
\author[label3]{Timon Rabczuk}

\address[label1]{Institute of Photonics, Faculty of Mathematics and Physics, Leibniz University Hannover, Hannover, 30167, Germany}
\address[label2]{College of Civil Engineering, Department of Geotechnical Engineering, Tongji University, Shanghai, 200092, China}
\address[label3]{Institute of Structural Mechanics, Bauhaus University Weimar, 99423 Weimar, Germany}

\begin{abstract}
Recent theoretical studies show that nanoscale contact on dielectric substrates can induce flexoelectric polarization large enough to drive electron transfer. This has been supported by experimental evidence, indicating that contact electrification involves a significant electromechanical contribution. 
In this work, we develop a computational model for flexoelectricity-driven contact electrification that integrates finite-deformation flexoelectricity with contact mechanics. 
A tunneling transparency function inspired by the WKB approximation is introduced to regulate the interfacial channel, capturing the irreversible charge trapping during unloading.
Charge transfer rules are formulated for three contact scenarios: unbiased metal-dielectric contact, biased metal-dielectric contact, and dielectric-dielectric contact. 
The model is compared with atomic force microscopy measurements on PMMA and PDAP substrates under both biased and unbiased conditions.
For contact between identical dielectric materials, we show that geometric asymmetry in surface curvature is sufficient to induce charge separation, with polarity reversal occurring at a critical surface wavenumber.
Three-dimensional simulations on random rough surfaces reproduce the mosaic charge distributions observed experimentally without introducing any material inhomogeneity.
\end{abstract}

\end{frontmatter}

\section{Introduction}
\subsection{Background and motivation}
Contact electrification (CE), also known as triboelectrification when accompanied by friction, occurs when surfaces become net charged after contact and separation~\cite{Wang2019,Pan2019}. 
Although this phenomenon was first documented over 2,600 years ago by Thales of Miletus, who observed the electrostatic attraction of amber, its mechanism remains incompletely understood~\cite{Lowell1980,Harper1967}.
CE spans a vast range of length scales, from electrostatically induced segregation in pharmaceutical powders~\cite{Wong2015} and electrophotography~\cite{Schein2013} to high-energy macroscale phenomena such as volcanic lightning and dust storms~\cite{Saunders2008,Lacks2011}. 
Although CE was historically considered an industrial hazard (e.g., electrostatic discharge that damages electronics or dust explosions in coal plants~\cite{Greason1987,Glor1985,Matsusaka2010}), it has recently attracted renewed interest.

Over the past decade, the invention of the triboelectric nanogenerator (TENG) has transformed CE from a nuisance into a practical energy harvesting technology~\cite{Fan2012,Wang2020,Zi2015}. 
The operating mechanism of TENGs relies on coupling contact electrification with electrostatic induction to convert mechanical energy such as random vibrations, wind, ocean waves, or human motion, into electricity~\cite{chen2018scavenging,wang2016sustainably,wang2015progress,cheng2024influence}. 
Owing to their wide choice of materials, lightweight structure, and low cost, TENGs have been applied to power the Internet of Things, self-powered sensors, and wearable electronics~\cite{liu2021promoting,yi2019recent,dharmasena2017triboelectric}. 
Recent work has also extended TENG applications to chemical sensing, blue energy harvesting, and high-voltage power sources~\cite{tian2021self,xu2018coupled,zi2016effective,cheng2023triboelectric}.

However, the optimization of these devices is limited by a lack of consensus regarding the fundamental charge transfer mechanisms~\cite{cheng2023triboelectric,cao2024progress}. Although triboelectric series have been developed based on empirical observations~\cite{zou2019quantifying,chen2024quantifying}, quantitative prediction from intrinsic material properties remains challenging. 
Recent theoretical works propose that large strain gradients generated at contacting asperities induce sufficiently high levels of flexoelectric polarization to bend interfacial energy bands and drive electron tunneling~\cite{Mizzi2019does,persson2020role,Olson2022band}.
This mechanism has been experimentally supported by Lin et al.~\cite{Lin2022flexo} and Qiao et al.~\cite{Qiao2021mixed}, who demonstrated that altering the contact force and consequently the local strain gradient can manipulate the magnitude and even reverse the polarity of charge transfer. 
These findings suggest that CE involves a significant electromechanical contribution. Therefore, a rigorous computational approach is needed to quantify the thermodynamic drive provided by the interface's mechanical state~\cite{Olson2025complex}.

\subsection{Literature Review}
\subsubsection{Classical Mechanisms of Contact Electrification}
The physical origin of charge transfer in CE has been debated for over a century. Hypotheses can be historically categorized into three primary mechanisms: electron transfer, ion transfer, and material transfer~\cite{Wang2019,Xia2020material}. 
The electron transfer model suggests that thermodynamic equilibrium drives electrons
from a material with a higher Fermi level to one with a lower Fermi level until electrochemical potentials equalize \cite{Lowell1980}. This model is well-accepted for metal-metal and metal-semiconductor contacts \cite{Lowell1979,Harper1960}. For insulators, the surface state theory extends this framework by postulating that localized electronic states exist within the forbidden bandgap at the material surface, arising from lattice termination or surface defects \cite{Lowell1980,Harper1967}. These states define an effective Fermi level, enabling electron tunneling between contacting surfaces in a manner analogous to metals \cite{Vick1953,Cowley1965}. 
More recently, Wang et al. proposed the “overlapped electron-cloud model”, which provides an atomistic perspective: mechanical compression brings atoms into the repulsive force regime where interatomic distances fall below the normal bonding length, causing electron clouds to overlap and lowering the potential barrier for electron transition \cite{Lin2020overlap,Xu2018electron}.

In contrast, the ion transfer model suggests that mobile ions on the surface or within aqueous adlayers are the primary charge carriers \cite{Diaz1993,Liu2008,ma2025quantifying}. This mechanism is particularly relevant for polymers containing mobile ionic groups, or for contacts in high-humidity environments, where capillary water bridges facilitate ion migration \cite{Pence1994}. For example, researchers found that the partitioning of hydroxide ions (OH$^-$) and protons (H$^+$) on surfaces drives charging in insulating polymers \cite{McCarty2008}. However, this mechanism becomes negligible under dry conditions or when materials lack mobile ions \cite{Wang2019}.
The third mechanism is material transfer, where nanoscale fragments of material are mechanically torn and transferred between contacting bodies, carrying net charge with them \cite{Salaneck1976,Baytekin2012}. Experimental evidence correlates the amount of transferred material with charge density \cite{Pandey2018}, although material transfer can occur independently of charging \cite{Olson2025quantitative}. 

While no single model captures the full complexity of CE, researchers widely recognize that electron transfer dominates in metal-dielectric and dielectric-dielectric contacts under controlled conditions, whereas ion and material transfer contribute under specific environmental or tribological conditions \cite{Wang2019,Xia2020material,Tan2021electron}.

\subsubsection{The Role of Flexoelectricity in CE}
The classical mechanisms discussed above struggle to explain several anomalous observations, such as charge transfer between chemically identical materials and the reversal of charge polarity under varying contact pressure \cite{Xu2019curvature,Sow2012reversal,Lacks2011}.
A common feature of these anomalies is their dependence on the mechanical state of the contact, suggesting that deformation itself plays a fundamental role in the charging process. Mechanical contact at nanoscale asperities generates highly non-uniform strain fields, with strain gradients scaling inversely with the radius of curvature and reaching magnitudes of $10^{7}$--$10^{9}$~m$^{-1}$ for typical asperity dimensions of 10--100~nm \cite{Mizzi2019does,persson2020role}. At such magnitudes, the flexoelectric effect becomes significant. Flexoelectricity is the linear electromechanical coupling between strain gradients and electric polarization \cite{zubko2013flexoelectric,Wang2019progress}. Unlike piezoelectricity, which requires non-centrosymmetric crystal structures, flexoelectricity arises from the local breaking of inversion symmetry by non-uniform deformation and is therefore present in all dielectric materials regardless of their crystal class \cite{ma2001large,shu2011symmetry,nguyen2013nanoscale}.

The flexoelectric polarization induced at contact interfaces generates a localized flexoelectric potential difference, which can reach magnitudes of 1--10 V \cite{Mizzi2019does,Olson2022band}.
This potential difference bends the electronic energy bands relative to the Fermi level, and when the band bending exceeds the intrinsic work function difference, it can reverse the thermodynamic drive for electron transfer, providing a natural explanation for pressure-dependent polarity reversal and size-dependent charging in granular systems \cite{Forward2009,Bilici2014}. 
The theoretical framework of Mizzi et al.~\cite{Mizzi2019does} first quantified this mechanism, showing that the flexoelectric field at nanoscale asperity contacts is sufficient to drive charge separation. This mechanism was subsequently supported by atomic force microscopy experiments. For instance, Qiao et al.~\cite{Qiao2021mixed} and Lin et al.~\cite{Lin2022flexo} independently demonstrated that varying the contact force can induce energy band bending and reverse the polarity of charge transfer between a tip and various substrates. Building upon these experimental insights, Olson and Marks developed quantitative models that incorporate flexoelectric band bending and electron tunneling kinetics \cite{Olson2022band,Olson2025complex,Marks2025flexo}. Collectively, these findings point to flexoelectricity as an important driving force for contact electrification alongside the intrinsic work function difference.

\subsubsection{Computational Modeling of Flexoelectricity}
The recognition of the electromechanical contribution to CE necessitates computational frameworks capable of resolving the significant strain gradients at nanoscale contacts and their associated flexoelectric responses. Modeling flexoelectricity presents unique mathematical challenges because the presence of strain gradient terms yields fourth-order partial differential equations (PDEs) for the displacement field \cite{mao2014insights,Yudin2013fundamentals}. In weak form, these equations contain second-order derivatives, requiring basis functions with $C^1$-continuity across element boundaries \cite{Zhuang2020computational}.

To address this continuity requirement, three principal numerical strategies have emerged \cite{Zhuang2020computational}. Mixed finite element method (MFEM) treats displacement gradients as independent degrees of freedom, enforcing the kinematic constraint through Lagrange multipliers or penalty terms, thereby reducing the problem to one solvable with standard $C^0$ elements \cite{Mao2016,deng2017mixed}. While compatible with existing finite element architectures, this approach significantly increases the degrees of freedom. Meshfree methods construct inherently smooth shape functions that naturally accommodate higher-order derivatives, excelling in handling large deformations though at greater computational expense \cite{Abdollahi2014,Abdollahi2015}. Isogeometric analysis (IGA), which uses Non-Uniform Rational B-Splines (NURBS) as basis functions, is now widely adopted for flexoelectric modeling \cite{Hughes2005,nguyen2015isogeometric}. By construction, NURBS provide $C^{p-1}$ continuity for degree-$p$ splines, naturally satisfying the $C^1$ requirement without auxiliary variables, while maintaining exact geometric representation \cite{Zhuang2020computational}. These numerical tools have enabled investigations spanning topology optimization, soft dielectrics at finite deformations, semiconductor devices, and architected metamaterials \cite{nanthakumar2017topology,Ghasemi2017,Yvonnet2017,Nguyen2019,Codony2021mathematical,zhuang2025variationally,hu2025computational}.

While computational flexoelectricity is well established, its application to contact electrification modeling is relatively recent. Persson~\cite{persson2020role} applied the flexoelectric framework to randomly rough contacts and related the resulting fluctuating surface potential to the charge mosaics measured after separation. Silavnieks et al. \cite{Silavnieks2025flexo} developed a flexoelectricity-driven contact-separation model using asymptotic analytical solutions for Hertzian sphere-on-flat contacts, demonstrating that accounting for unloading processes significantly affects residual charge predictions as charge backflow occurs while interfacial gaps remain below a critical transfer distance. Similarly, Olson and Marks \cite{Olson2025quantitative} proposed a quantitative model successfully incorporating semiconductor depletion layers and flexoelectric polarization for metal-semiconductor systems. These studies represent the first steps toward integrating flexoelectric theory with contact mechanics for CE prediction.

Building upon these foundations, several opportunities for further development motivate the present work. 
To begin with, moving from analytical solutions to a fully coupled numerical framework with nanoscale adhesion makes it possible to resolve flexoelectric coupling and contact behavior under finite deformation.
Furthermore, the transition from equilibrium polarization-induced charge demand during loading to frozen residual charge after separation requires phenomenological rules informed by quantum tunneling physics to be incorporated into continuum frameworks. 
Additionally, since experimental measurements invariably capture only the final residual charge of the sample surfaces, the computational framework must resolve the charge redistribution that occurs as the contact area shrinks and tunneling channels close.
Finally, numerical frameworks naturally accommodate complex geometries, such as the surface roughness ubiquitous in realistic contact scenarios, allowing for the investigation of geometry-induced charge patterns.

\subsection{Contributions of the present work}
The present work addresses these gaps by developing a computational model for flexo\-electric\-ity-driven contact electrification. At the core of this model, a tunneling transparency function inspired by the WKB approximation is proposed to govern charge transfer across the nanoscale interface gap. Based on this framework, specific charge transfer hypotheses are formulated for three contact configurations. In unbiased metal-dielectric contact, the surface polarization charge induced by flexoelectricity drives electron transfer from the conducting tip; in biased contact, the externally applied field acts as a selective reservoir, restricting charge transfer to carriers of a single sign; and in dielectric-dielectric contact, surface states with finite capacity serve as the charge reservoir, allowing charge separation even between identical materials.
To evaluate the model's physical fidelity, the framework is qualitatively compared with atomic force microscopy measurements. It is subsequently applied to rough-surface contacts, where it reproduces several experimentally observed phenomena. These include curvature-dependent charge separation and spatially heterogeneous mosaic charge patterns, offering mechanistic explanations grounded in contact-induced local strain gradient variations. The model further predicts a polarity reversal controlled by surface waviness.

\subsection{Organization of the paper}
The remainder of this paper is organized as follows: 
Section~\ref{sec:theory} establishes the theoretical framework, covering finite-deformation flexoelectricity, contact mechanics with adhesion, and the specific charge transfer rules for the three contact scenarios. 
Building upon this formulation, Section~\ref{sec:validation} compares the model predictions with AFM experiments on metal-dielectric contacts under both unbiased and biased conditions, supplemented by parametric studies on tip radius and tunneling length. 
The framework is subsequently applied in Section~\ref{sec:application} to investigate dielectric-dielectric contact between identical materials, exploring geometry-induced charge separation on two-dimensional wavy surfaces and three-dimensional random rough profiles. 
Finally, Section~\ref{sec:discussion} critically discusses the model's limitations and relates our approach to existing literature, and Section~\ref{sec:conclusions} summarizes the conclusions and future directions.

\section{Theoretical Framework}
\label{sec:theory}

This section presents the theoretical foundations of the computational framework for flexoelectricity-driven contact electrification. We begin with finite-deformation flexoelectricity, then describe the contact mechanics constraints, and finally detail the charge-transfer mechanisms for different contact scenarios.

\subsection{Finite deformation flexoelectricity}
\label{sec:flexo_theory}

Consider a dielectric body occupying the domain $\mathcal{B}_0 \subset \mathbb{R}^3$ in the reference configuration, with boundary $\partial\mathcal{B}_0$. The deformation is described by the mapping $\bd{\chi}: \mathcal{B}_0 \to \mathcal{B}_t$, where $\mathcal{B}_t$ denotes the current configuration. The displacement field is $\mathbf{u} = \bd{\chi} - \mathbf{X}$, where $\mathbf{X}$ represents the material coordinates. The deformation gradient is defined as $\mathbf{F} = \partial\bd{\chi}/\partial\mathbf{X}$, with the Jacobian determinant $J = \det(\mathbf{F})$. The right Cauchy--Green deformation tensor is $\mathbf{C} = \mathbf{F}^\mathrm{T}\mathbf{F}$, and the Green--Lagrange strain tensor is
\begin{equation}
	\bm{E} = \frac{1}{2}(\mathbf{C} - \mathbf{I}),
\end{equation}
where $\mathbf{I}$ is the identity tensor. The strain gradient tensor is defined as $\bm{G} = \partial\bm{E}/\partial\mathbf{X}$.

The Lagrangian electric field is $\bd{E} = -\partial\varPhi/\partial\mathbf{X}$, where $\varPhi$ denotes the electric potential. We formulate the electromechanical coupled response within a variational framework characterized by the electric Gibbs free energy density per unit reference volume:
\begin{equation}
	\label{Eq:Gibbs}
	\bar{\Psi}^\mathrm{Gib}(\bm{E}, \bm{G}, \bd{E}) =
	\bar{\Psi}^\mathrm{Mech}(\bm{E}, \bm{G})
	+ \bar{\Psi}^\mathrm{Diele}(\bm{E}, \bd{E})
	+ \bar{\Psi}^\mathrm{Flexo}(\bm{E}, \bm{G}, \bd{E})
	- \bar{\Psi}^\mathrm{Elec}(\bm{E}, \bd{E}),
\end{equation}
with the individual contributions defined as \cite{codony2021modeling}:
\begin{align}
	\bar{\Psi}^\mathrm{Mech}(\bm{E}, \bm{G}) &= \bar{\Psi}^\mathrm{Elast}(\bm{E})
	+ \frac{1}{2}\bm{G} \tripledots \bar{\bd{h}} \tripledots \bm{G}, \label{eq:psi_mech}\\
	\bar{\Psi}^\mathrm{Diele}(\bm{E}, \bd{E}) &=
	-\frac{J}{2}(\epsilon - \epsilon_0)\bd{E} \cdot \mathbf{C}^{-1} \cdot \bd{E}, \label{eq:psi_diele}\\
	\bar{\Psi}^\mathrm{Flexo}(\bm{E}, \bm{G}, \bd{E}) &=
	-J\bd{E} \cdot \mathbf{C}^{-1} \cdot \bd{\mu} \tripledots \bm{G}, \label{eq:psi_flexo}\\
	\bar{\Psi}^\mathrm{Elec}(\bm{E}, \bd{E}) &=
	\frac{J\epsilon_0}{2}\bd{E} \cdot \mathbf{C}^{-1} \cdot \bd{E}, \label{eq:psi_elec}
\end{align}
where $\bar{\Psi}^\mathrm{Elast}$ is the hyperelastic strain energy density (we adopt the neo-Hookean model), $\bar{\bd{h}}$ is a sixth-order strain gradient elasticity tensor, $\bd{\mu}$ is the fourth-order flexoelectric coefficient tensor, $\epsilon_0$ is the vacuum permittivity, and $\epsilon$ is the material permittivity.

The constitutive relations are derived from the Gibbs free energy density. The second Piola--Kirchhoff stress tensor is
\begin{equation}
	\mathbf{S} := \frac{\partial\bar{\Psi}^\mathrm{Mech}}{\partial\bm{E}},
\end{equation}
and the second Piola--Kirchhoff double stress tensor is
\begin{equation}
	\bd{\Sigma} := \frac{\partial\bar{\Psi}^\mathrm{Gib}}{\partial\bm{G}} = \bar{\bd{h}} \tripledots \bm{G} - J\bd{E} \cdot \mathbf{C}^{-1} \cdot \bd{\mu}.
\end{equation}
The Lagrangian electric displacement is
\begin{equation}
	\label{eq:elec_disp}
	\bd{D} = -\frac{\partial\bar{\Psi}^\mathrm{Gib}}{\partial\bd{E}} = J\mathbf{C}^{-1} \cdot \left(\epsilon\bd{E} + \bd{\mu} \tripledots \bm{G}\right).
\end{equation}
In Eqs.~\eqref{eq:psi_diele}--\eqref{eq:elec_disp}, the $J\mathbf{C}^{-1}$ mapping terms account for the transformation of strictly spatial electrostatic definitions into the reference configuration. The second term in Eq.~\eqref{eq:elec_disp} represents the flexoelectric contribution to the electric displacement, which couples the strain gradient to the electric response.
The Maxwell stress tensor, which accounts for electrostatic body forces in the reference configuration, is given by the second Piola--Maxwell stress form \cite{zhuang2025variationally}
\begin{equation}
	\mathbf{S}^\mathrm{MW} = (\mathbf{C}^{-1}\bd{E}) \otimes \widetilde{\bd{D}} - \frac{1}{2}\left(\bd{E} \cdot \widetilde{\bd{D}}\right)\mathbf{C}^{-1},
\end{equation}
where $\widetilde{\bd{D}} = J\mathbf{C}^{-1} \cdot \left[\epsilon\bd{E} + 2\bd{\mu} \tripledots \bm{G}\right]$ is the auxiliary electric displacement~\cite{zhuang2025variationally}.

The governing equations in the reference configuration are:
\begin{equation}
	\label{eq:gov_mech}
	\nabla \cdot \left[\mathbf{F}\left(\mathbf{S} - \nabla \cdot \bd{\Sigma} + \mathbf{S}^\mathrm{MW}\right)\right] + \mathbf{b}_0 = \mathbf{0} \quad \text{in } \mathcal{B}_0,
\end{equation}
\begin{equation}
	\label{eq:gov_elec}
	\nabla \cdot \bd{D} = Q_0 \quad \text{in } \mathcal{B}_0,
\end{equation}
where $\mathbf{b}_0$ is the body force per unit reference volume and $Q_0$ is the body charge density per unit reference volume.

The boundary conditions for the mechanical field consist of prescribed displacements $\bar{\mathbf{u}}$ on the Dirichlet boundary $\partial\mathcal{B}_0^u$, prescribed normal displacement gradients $\bar{\mathbf{s}}$ on the higher-order Dirichlet boundary $\partial\mathcal{B}_0^s$, prescribed tractions $\bar{\mathbf{t}}$ on the Neumann boundary $\partial\mathcal{B}_0^t$, and prescribed higher-order tractions $\bar{\mathbf{r}}$ on the higher-order Neumann boundary $\partial\mathcal{B}_0^r$:
\begin{align}
	\mathbf{u} &= \bar{\mathbf{u}} \quad \text{on } \partial\mathcal{B}_0^u, \\
	\frac{\partial \mathbf{u}}{\partial \mathbf{n}} &= \bar{\mathbf{s}} \quad \text{on } \partial\mathcal{B}_0^s, \\
	\mathbf{F}(\mathbf{S} - \nabla \cdot \bd{\Sigma} + \mathbf{S}^\mathrm{MW}) \cdot \mathbf{N} + \widetilde{\mathbf{T}}^{t} &= \bar{\mathbf{t}} \quad \text{on } \partial\mathcal{B}_0^t, \\
	\widetilde{\mathbf{T}}^{n} &= \bar{\mathbf{r}} \quad \text{on } \partial\mathcal{B}_0^r,
\end{align}
where $\mathbf{N}$ is the outward unit normal in the reference configuration. The terms $\widetilde{\mathbf{T}}^{t}$ and $\widetilde{\mathbf{T}}^{n}$ emerge from the surface divergence theorem by splitting the variation of the deformation gradient into surface-tangential and surface-normal components. The detailed index notations for $\widetilde{\mathbf{T}}^{t}$ and $\widetilde{\mathbf{T}}^{n}$ are provided in \ref{Apdx1}. For the electrical field, the boundary conditions comprise prescribed potential $\bar{\varPhi}$ on the Dirichlet boundary $\partial\mathcal{B}_0^\varPhi$ and prescribed surface charge density $\bar{W}$ on the Neumann boundary $\partial\mathcal{B}_0^W$:
\begin{align}
	\varPhi &= \bar{\varPhi} \quad \text{on } \partial\mathcal{B}_0^\varPhi, \\
	\bd{D} \cdot \mathbf{N} &= \bar{W} \quad \text{on } \partial\mathcal{B}_0^W.
\end{align}

\subsection{Contact mechanics formulation}
\label{sec:contact_mech}

We consider a contact problem between two bodies: a ``slave'' body $\mathcal{B}_s$ and a ``master'' body $\mathcal{B}_m$. The contact surface is denoted by $\partial\mathcal{B}^C$. For a point $\bd{\chi}^s$ on the slave surface, its projection onto the master surface is $\bar{\bd{\chi}}^m$, determined by minimizing the distance function. The normal gap is defined as
\begin{equation}
	\label{eq:gap}
	g_N = (\bd{\chi}^s - \bar{\bd{\chi}}^m) \cdot \mathbf{n},
\end{equation}
where $\mathbf{n}$ is the outward unit normal of the master surface in the current configuration at the projection point.

For frictionless contact, the normal non-penetration constraints are formulated via the standard Karush--Kuhn--Tucker (KKT) conditions:
\begin{equation}
	\label{eq:KKT}
	g_N \geq 0, \quad t_N \leq 0, \quad g_N t_N = 0 \quad \text{on } \partial\mathcal{B}^C_t,
\end{equation}
where $t_N$ is the normal contact traction, representing a strictly repulsive pressure in this notation.

To capture the adhesive forces that influence nanoscale contact behavior, we incorporate an exponential cohesive zone model following the formulation of Xu and Needleman \cite{xu1993void} as modified by van den Bosch et al. \cite{van2006improved}. This model expresses the normal traction in terms of the normal separation while naturally satisfying the requirement of vanishing traction at complete separation. For the frictionless contact considered in this work, the normal cohesive traction takes the form \cite{dimitri2015coupled}:
\begin{equation}
	\label{eq:adhesion}
	t_N^\mathrm{adh} = \frac{\phi_N}{(g_N^\mathrm{max})^2}\,g_N \exp\left(-\frac{g_N}{g_N^\mathrm{max}}\right) \quad \text{for } g_N > 0,
\end{equation}
where a positive $t_N^\mathrm{adh}$ denotes an attractive pull between the separated surfaces. $\phi_N$ is the fracture energy and $g_N^\mathrm{max}$ is the characteristic length at which the cohesive traction reaches its maximum value $p_N^\mathrm{max}$. These parameters are related through $\phi_N = \mathrm{e}\,p_N^\mathrm{max} g_N^\mathrm{max}$, where $\mathrm{e}$ is Euler's number. This formulation ensures thermodynamic consistency and provides a smooth transition from adhesive attraction to complete separation~\cite{dimitri2015coupled}.

\subsection{Charge transfer mechanisms}
\label{sec:charge_transfer}

The framework postulates that interfacial charge transfer is driven by the thermodynamic force from flexoelectric surface bound charges~\cite{Mizzi2019does,Marks2025flexo,Silavnieks2025flexo} and is limited by tunneling kinetics~\cite{Olson2022band,Willatzen2019}. While the driving force and tunneling physics are universal, their manifestation differs across contact configurations depending on the nature of the contacting surfaces and the availability of charge carriers. This section develops a unified framework encompassing three contact scenarios: unbiased metal-dielectric contact, biased metal-dielectric contact, and dielectric-dielectric contact.

Before presenting the specific models, we outline the physical scenario common to all cases. Consider a contact experiment in which a probe tip is pressed against a dielectric substrate (the \emph{loading} phase) and subsequently retracted until the two surfaces separate (the \emph{unloading} phase). As established in Section~\ref{sec:flexo_theory}, the deformation at the contact interface produces strain gradients, which in turn induce flexoelectric polarization in the dielectric. This polarization terminates at the surface and creates a bound charge density $\rho_b = -\nabla \cdot \mathbf{p}$ in the bulk and $\sigma_b = \mathbf{p} \cdot \mathbf{n}$ at the surface. To reach electrostatic equilibrium, free charges must be supplied to screen (compensate) this bound polarization charge \cite{Silavnieks2025flexo}. The source of these free charges depends on the contact configuration: a conducting tip can supply electrons readily from the metal; when the tip is also a dielectric, surface states serve as the charge reservoir. We denote by $\sigfroz$ the free charge density accumulated at a point of the contact surface and by $\sigma_\mathrm{tar}$ the target charge density, i.e.\ the equilibrium value of $\sigfroz$ permitted by the available charge reservoir, which is specified for each contact scenario below.

\subsubsection{Tunneling transparency factor}
\label{sec:tunneling}

Charge transfer across a nanoscale interface gap occurs primarily through quantum mechanical tunneling, where the transmission probability depends exponentially on the barrier width. According to the Wentzel--Kramers--Brillouin (WKB) approximation, the tunneling probability through a rectangular barrier of width $d$ and height $U$ decays exponentially as $T \sim \exp(-2\kappa d)$, where $\kappa = \sqrt{2m(U-E)}/\hbar$ is the decay constant determined by the effective electron mass $m$ and the energy deficit below the barrier \cite{simmons1963generalized}.

A deficit of $\sigfroz$ relative to $\sigma_\mathrm{tar}$ produces a proportional potential difference across the gap, and in the low-bias limit the resulting tunneling current is proportional to this potential difference and to the barrier transmission~\cite{simmons1963generalized}. The surface charge therefore relaxes toward its target as
\begin{equation}
	\label{eq:kinetic}
	\frac{\mathrm{d}\sigfroz}{\mathrm{d}t} = -\frac{\sigfroz - \sigma_\mathrm{tar}}{\tau(g_N)}, \qquad \tau(g_N) = \tau_0 \exp\left(2\kappa g_N\right),
\end{equation}
where $\tau_0$ is the relaxation time at zero gap and the barrier width is identified with the normal gap. Integrating Eq.~\eqref{eq:kinetic} over a load increment of physical duration $\delta t$, with $\sigma_\mathrm{tar}$ and $g_N$ taken at the end of the increment, gives
\begin{equation}
	\label{eq:relax_update}
	\sigfroz^{(n)} = \sigfroz^{(n-1)} + \Transparency^{(n)} \left(\sigma_\mathrm{tar}^{(n)} - \sigfroz^{(n-1)}\right), \qquad \Transparency = 1 - \exp\left[-\frac{\delta t}{\tau(g_N)}\right],
\end{equation}
where the tunneling transparency factor $\Transparency$ is the fraction of the local charge deficit removed within one increment. Because $\tau$ grows exponentially with the gap, $\Transparency$ switches sharply from 1 to 0 around the critical gap at which $\tau$ becomes comparable to $\delta t$. The charge therefore equilibrates within one increment below this gap and is effectively frozen above it.

\begin{figure}[!htbp]
	\centering
	\includegraphics[width=0.6\textwidth]{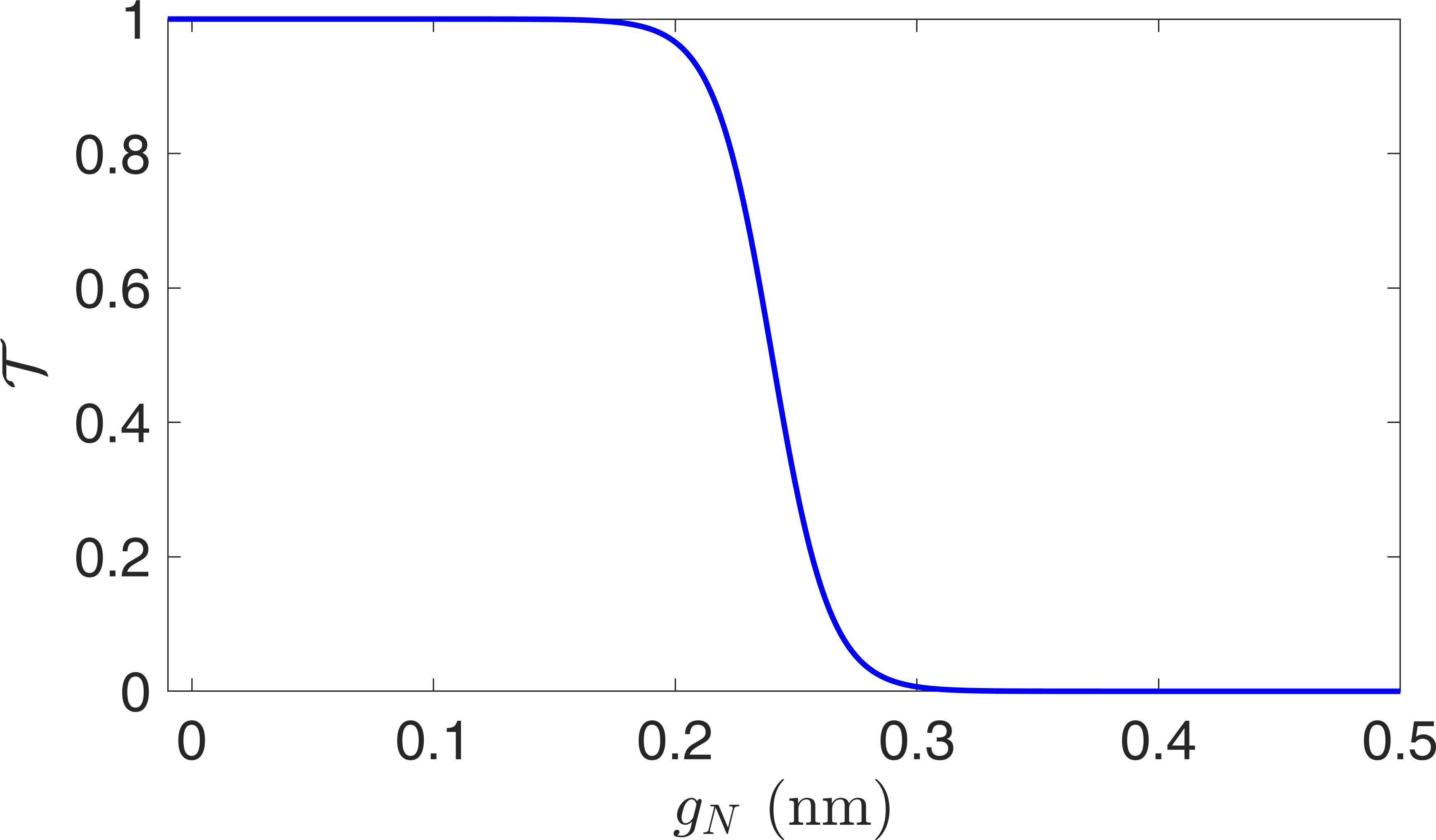}
	\caption{Tunneling transparency factor $\Transparency$ as a function of normal gap distance $g_N$, with $\ell_q = 0.24$~nm and $\Delta_g = 0.012$~nm.}
	\label{fig:transparency}
\end{figure}

Inspired by this switching behavior, we approximate the tunneling transparency factor by a phenomenological function that smoothly interpolates between fully open ($\Transparency = 1$) and fully closed ($\Transparency = 0$) tunneling channels as
\begin{equation}
	\label{eq:transparency}
	\Transparency = \left[1 + \exp\left(\frac{g_{N} - \ell_q}{\Delta_g}\right)\right]^{-1},
\end{equation}
where $g_{N}$ is the normal gap, $\ell_q$ is the characteristic tunneling length below which the channel is effectively open, and $\Delta_g$ controls the sharpness of the transition. This logistic form captures the rapid decay of transfer capability once the gap exceeds a critical distance while avoiding the numerical difficulties associated with a discontinuous step function. Based on the overlapped electron-cloud model \cite{Lin2020overlap,Xu2018electron}, we adopt $\ell_q = 0.24$~nm and $\Delta_g = 0.012$~nm in this work unless otherwise specified, corresponding to a transition zone of approximately 0.05~nm.
Figure~\ref{fig:transparency} illustrates the tunneling transparency factor as a function of the normal gap distance.

When $\Delta_g \to 0$ in Eq.~\eqref{eq:transparency}, the update in Eq.~\eqref{eq:relax_update} reduces to the sharp tunneling cut-off of the classical capacitor model for metal--metal and metal--insulator contacts~\cite{Harper1967,Lowell1979}, in which the charge remains in equilibrium as long as the separation is below a critical distance and is fixed once this distance is exceeded. The present formulation can be viewed as a spatially resolved continuum version of this picture, with the flexoelectric bound charge as the driving force.

Under the update rule of Eq.~\eqref{eq:relax_update}, charge freezing originates from the asymmetry between channel opening and closing. When a channel opens, the frozen charge jumps to the current target at once, whereas during unloading it follows the target until the channel closes, and the residual charge is therefore set by the local target at the instant of closure even if the elastic configuration path is retraced exactly.

During loading, as each point of the contact surface passes from a closed to an open channel, the frozen charge is approximated by the transparency-weighted equilibrium
\begin{equation}
	\label{eq:loading_eval}
	\sigfroz^{(n)} = \Transparency^{(n)}\,\sigma_\mathrm{tar}^{(n)}.
\end{equation}

\subsubsection{Metal-dielectric contact: Unbiased case}
\label{sec:unbiased}

In this scenario, a conducting tip contacts a dielectric surface without an applied bias. The flexoelectric polarization induced by contact deformation creates bound surface charge that must be screened by free charges. The conducting tip, acting as an ideal charge reservoir, supplies these free charges through the tunneling channels described in Section~\ref{sec:tunneling}. During loading, when the gap is closed or small and the transparency factor (Eq.~\eqref{eq:transparency}) is high, charge flows readily. During unloading, as the surfaces separate and gaps reopen at the contact periphery, the tunneling channels close progressively from the edge inward, ``freezing'' a portion of the transferred charge on the dielectric surface. To track this process, the accumulated frozen charge $\sigfroz$ records the free charge density at each surface point throughout the loading--unloading cycle, enabling prediction of the residual charge distribution after separation.

We extract the flexoelectric polarization in the current configuration as:
\begin{equation}
	\label{eq:p_flexo}
	\mathbf{p}_\mathrm{flexo} = \mathbf{F}^{-\mathrm{T}} \cdot \bd{\mu} \tripledots \bm{G}.
\end{equation}
This local flexoelectric polarization induces a \emph{charge demand} $\sigdemand$ that represents the equilibrium free charge density required to screen the bound polarization charge:
\begin{equation}
	\label{eq:sigma_unbiased}
	\sigdemand = -\mathbf{p}_\mathrm{flexo} \cdot \mathbf{n},
\end{equation}
with $\mathbf{n}$ being the outward surface normal. 

Because the conducting tip can supply charge of either sign, the target charge in the unbiased case is the full demand,
\begin{equation}
	\label{eq:charge_update}
	\sigma_\mathrm{tar}^{(n)} = \sigdemand^{(n)},
\end{equation}
which enters Eq.~\eqref{eq:loading_eval} during loading and Eq.~\eqref{eq:relax_update} during unloading.

The treatment of electrical boundary conditions at the interface reflects the physics of the tunneling channel. In regions where the channel is open ($\Transparency \to 1$), charge flows freely between the metal and dielectric. Numerically, this is enforced by imposing an electrical contact condition that ensures potential continuity:
\begin{equation}
	\label{eq:electrical_contact}
	\phi_{\mathrm{dielectric}} = \phi_{\mathrm{metal}} \quad \text{where } \Transparency \to 1.
\end{equation}
Under this constraint, charge transfer occurs implicitly through the coupled electromechanical solution, so that no explicit charge source term is required.
In contrast, where the tunneling channel is closed ($\Transparency \to 0$), the frozen charge $\sigfroz$ acts as an external surface charge contributing to the electric field. The charge contribution to the electrical Neumann boundary is:
\begin{equation}
	\label{eq:nodal_charge}
	\bar{\omega}^{(n)} = (1 - \Transparency) \cdot \sigfroz^{(n)},
\end{equation}
where the factor $(1 - \Transparency)$ ensures a smooth transition in intermediate regions.

\begin{figure}[!htbp]
	\centering
	\includegraphics[width=\textwidth]{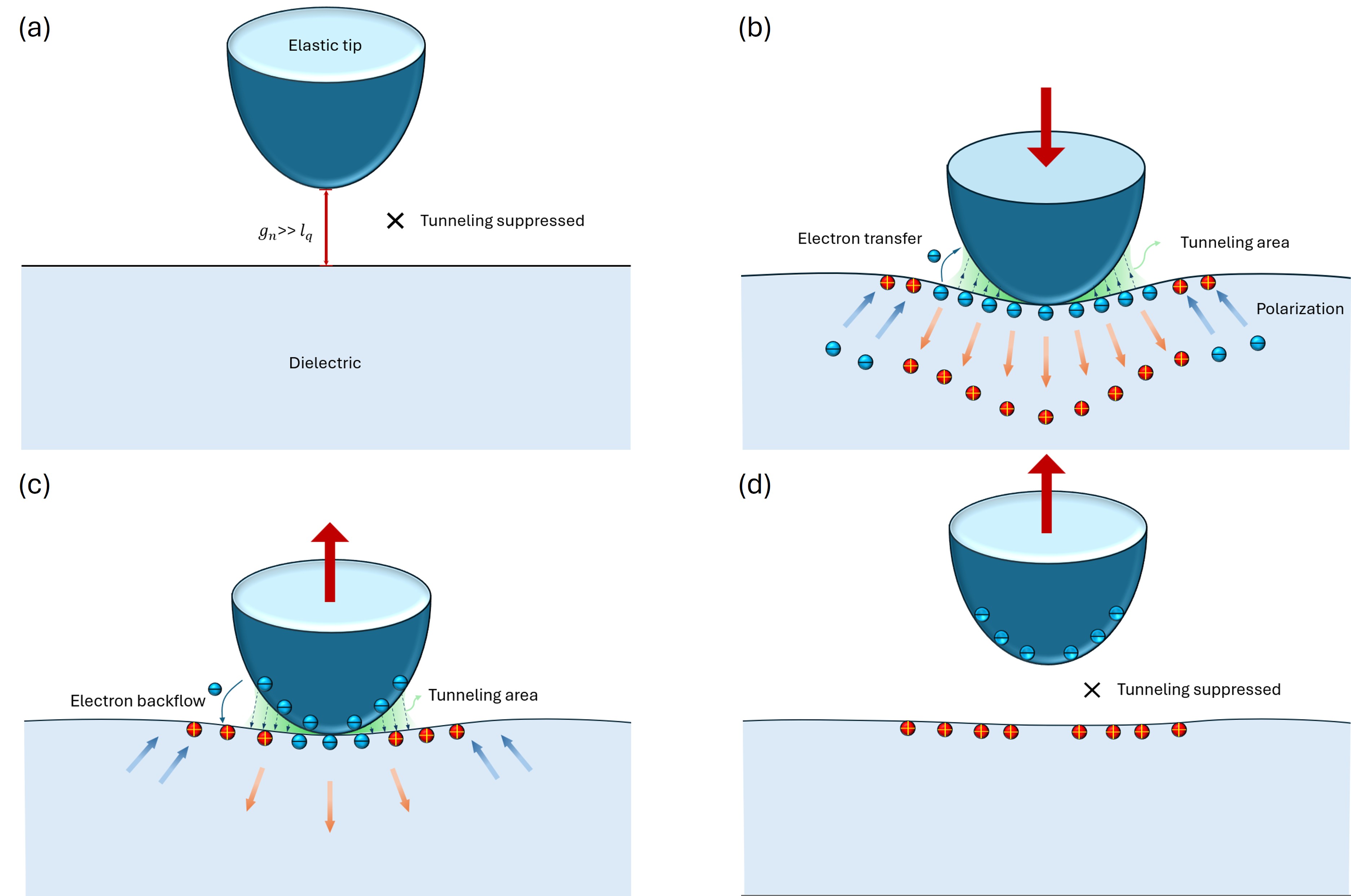}
	\caption{Schematic of charge transfer in unbiased metal-dielectric contact: (a) large gap with closed tunneling channel; (b) loading with flexoelectric polarization driving charge transfer; (c) unloading with partial charge backflow and edge freezing; (d) complete separation with frozen residual charge.}
	\label{fig:unbiased_schematic}
\end{figure}

Figure~\ref{fig:unbiased_schematic} illustrates the transparency factor's role throughout the loading--unloading cycle. When the surfaces are far apart (stage a), $\Transparency = 0$ and no charge can tunnel across the gap. As contact is established (stage b), tunneling channels open and free charges flow from the metal to meet the flexoelectric demand. During unloading (stage c), the gap opens first at the contact periphery, causing $\Transparency$ to decrease there while the center remains in close contact. The charge update rule permits only partial backflow through the narrowing channel, thereby yielding the irreversible ``edge-frozen'' charge distribution observed after complete separation (stage d).

\subsubsection{Metal-dielectric contact: Biased case}
\label{sec:biased}

When an external voltage $V_\mathrm{tip}$ is applied to the conducting tip, an additional polarity constraint governs the charge transfer. While an applied bias intrinsically generates a strong capacitive field and potential charge injection, this reversible capacitive charge vanishes upon separation. To systematically isolate the intrinsic electromechanical response, the present model considers only the flexoelectric bound charge as the driving force, treating the tip bias purely as a polarity-selective electron reservoir. Accordingly, the sign of the applied bias determines the polarity of the \emph{net} charge that can accumulate on the dielectric surface: a negatively biased tip preferentially injects electrons, while a positively biased tip preferentially extracts them from dielectric surface states. Although electrons may flow in either direction during any given load step, the net accumulated charge at each surface point is constrained to remain of the bias-determined sign. For instance, electrons injected under negative bias may partially return to the tip during unloading as the flexoelectric demand diminishes. This polarity constraint is the key distinction from the unbiased scenario, where charges of either sign can be transferred to satisfy the full flexoelectric demand.

As defined in Eq.~\eqref{eq:sigma_unbiased}, the flexoelectricity-induced polarized surface charge demand is given by $\sigdemand^{(n)} = -\mathbf{p}_\mathrm{flexo} \cdot \mathbf{n}$. 
The polarity constraint is then implemented through a projection operator as
\begin{equation}
	\label{eq:polarity_proj}
	\Pi_V(x) = \begin{cases}
		\min(x, 0), & V_\mathrm{tip} < 0, \\
		x, & V_\mathrm{tip} = 0, \\
		\max(x, 0), & V_\mathrm{tip} > 0.
	\end{cases}
\end{equation}
For a negatively biased tip, this projection retains only the negative portion of any quantity, naturally precluding positive charge demands that cannot be satisfied by the electron reservoir. The converse applies for positive bias.

The charge transfer process in the biased case is formulated by applying the polarity constraint to the flexoelectric demand before evaluating the charge update. Specifically, we define the target charge at step $n$ as
\begin{equation}
	\label{eq:sigma_tar_cur}
	\sigma_\mathrm{tar}^{(n)} = \Pi_V(\sigdemand^{(n)}),
\end{equation}
where $\sigma_\mathrm{tar}^{(n)}$ represents the theoretical maximum charge that the biased tip can supply to meet the local flexoelectric demand at the current configuration. Substituting $\sigma_\mathrm{tar}$ into the charge update rule in Eq.~\eqref{eq:relax_update}, which applies during both loading and unloading in the biased case, yields
\begin{equation}
	\label{eq:biased_update}
	\sigfroz^{(n)} = \sigfroz^{(n-1)} + \Transparency \cdot \left(\sigma_\mathrm{tar}^{(n)} - \sigfroz^{(n-1)}\right).
\end{equation}

The physical insight underlying this formulation becomes clear when comparing with the unbiased case. In unbiased contact, the charge transfer dynamically responds to the symmetric evolution of flexoelectricity. In biased contact, however, both loading and unloading are continuously governed by the asymmetric capacity boundary. When the target decreases during unloading, the frozen charge tracks this capacity contraction as long as the channel remains open. Consequently, this mechanism leads to greater charge recovery during unloading, yielding a residual charge that is nearly independent of the maximum loading force.

\begin{figure}[!htbp]
	\centering
	\includegraphics[width=\textwidth]{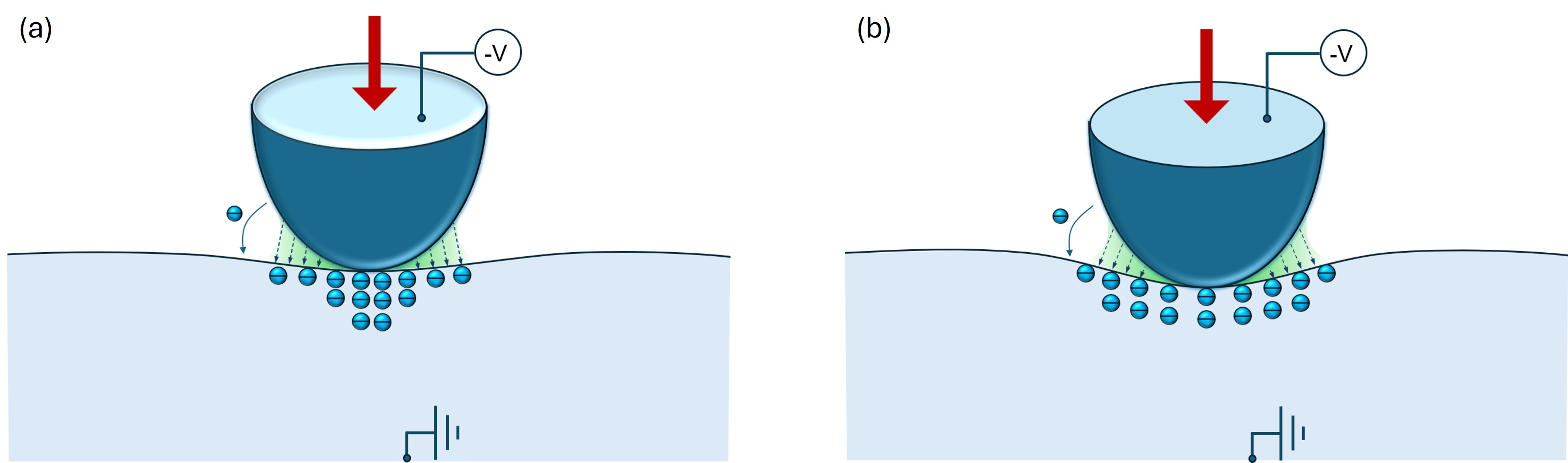}
	\caption{Schematic of charge transfer in biased metal-dielectric contact (negative tip bias): (a) initial contact with localized high charge density; (b) further contact with distributed lower charge density.}
	\label{fig:biased_schematic}
\end{figure}

Figure~\ref{fig:biased_schematic} illustrates two characteristic stages of biased contact. In the initial contact stage with small contact area, the localized flexoelectric demand leads to high local charge density. As contact progresses and the contact area increases, the local charge density decreases. Notably, for a negatively biased tip, all deposited charges are electrons. Even at the contact periphery where the flexoelectric polarization may reverse sign and create positive charge demand, no positive charges are deposited because of the polarity selectivity of the conducting tip. 

\subsubsection{Dielectric-dielectric contact}
\label{sec:dielectric_dielectric}

Contact between two dielectric bodies differs substantially from metal-dielectric contact because neither surface can supply charge freely. Instead, charge transfer relies on surface states: localized electronic levels within the bandgap that arise from lattice termination, defects, or surface contamination~\cite{Lowell1980}. These states act as electron traps with a finite areal density~\cite{Xu2018electron}. Several recent experiments support the surface-state mechanism for insulator-insulator electrification: temperature-dependent charge decay follows thermionic emission kinetics, indicating that charges reside in well-defined energy traps~\cite{Xu2018electron}; and identical polymers exchange charge when brought into contact with different curvatures, consistent with strain-induced shifts in surface state energy~\cite{Xu2019curvature}. In the model presented here, we characterize each surface by its maximum trappable charge density $\sigma_\mathrm{trap}$, representing the areal density of available surface states.

When two dielectric surfaces of identical material come into contact, the flexoelectric bound charges $\hat{\sigma}_{b,m}$ and $\hat{\sigma}_{b,s}$ on the master and slave surfaces generally differ due to geometric asymmetry in the strain gradients. This asymmetry alters the local electrostatic environment at the interface and bends the energy bands near each surface, shifting the surface states on the two sides relative to each other~\cite{Lin2022flexo}. The resulting electrochemical potential difference between the surface states drives electron tunneling until the potential imbalance is neutralized. To estimate the equilibrium transfer, we treat each interfacial point independently and require that the total surface charge be equal on both sides, i.e.\ $\hat{\sigma}_{b,m} - \Delta\sigma = \hat{\sigma}_{b,s} + \Delta\sigma$, which eliminates the local driving field across the gap: at this point the net unscreened charge is identical on both sides, so neither surface can extract further charge from the other. For contacts between dissimilar dielectrics with different permittivities, this simple halving rule no longer holds, and the equilibrium condition should account for the dielectric mismatch at the interface.
The idealized amount of charge transferred between the two surfaces is then
\begin{equation}
	\label{eq:delta_sigma}
	\Delta\sigma = \frac{\hat{\sigma}_{b,m} - \hat{\sigma}_{b,s}}{2},
\end{equation}
where $\hat{\sigma}_{b,m} = \mathbf{p}_{\mathrm{flexo},m} \cdot \mathbf{n}_m$ and $\hat{\sigma}_{b,s} = \mathbf{p}_{\mathrm{flexo},s} \cdot \mathbf{n}_s$ are the flexoelectric bound surface charge densities on the master and slave surfaces, with $\mathbf{n}_m$ and $\mathbf{n}_s$ being the respective outward surface normals. The actual transfer is limited by the surface state capacity, expressed as
\begin{equation}
	\label{eq:sigma_tar_dd}
	\sigma_\mathrm{tar}^{(n)} = \mathrm{sign}(\Delta\sigma^{(n)}) \cdot \min\left(\left|\Delta\sigma^{(n)}\right|, \sigma_\mathrm{trap}\right).
\end{equation}
In alignment with the framework established above, the charge evolution follows Eq.~\eqref{eq:loading_eval} during loading and the same charge update rule during unloading:
\begin{equation}
	\label{eq:dd_update}
	\sigfroz^{(n)} = \sigfroz^{(n-1)} + \Transparency \cdot \left(\sigma_\mathrm{tar}^{(n)} - \sigfroz^{(n-1)}\right),
\end{equation}
subject to the physical constraint of finite surface state capacity:
\begin{equation}
	\label{eq:dd_clamp}
	\left|\sigfroz^{(n)}\right| \le \sigma_\mathrm{trap}.
\end{equation}
The electrical Neumann contribution follows Eq.~\eqref{eq:nodal_charge}. This formulation is particularly relevant for contact between identical materials, where geometric asymmetry, such as surface curvature, creates differential strain gradients that drive charge separation despite identical material properties.

\subsubsection{Unified charge transfer algorithm}
\label{sec:unified_algorithm}

Algorithm~\ref{alg:unified_charge_transfer} collects the charge transfer rules developed in Sections~\ref{sec:unbiased}--\ref{sec:dielectric_dielectric} into a single unified procedure. All three contact scenarios share the same charge update structure, differing in the definition of the target charge $\sigma_\mathrm{tar}$. The transparency-weighted update rule, the Neumann boundary contribution, and the post-processing steps are common to all cases.

\begin{algorithm}[!htbp]
	\SetAlgoLined
	\DontPrintSemicolon
	\caption{Unified charge transfer algorithm for contact electrification}
	\label{alg:unified_charge_transfer}
	\KwIn{Converged fields at step $n$; previous frozen charge $\sigfroz^{(n-1)}$; contact type; loading or unloading phase}
	\KwOut{Updated frozen charge $\sigfroz^{(n)}$, Neumann contribution $\bar{\omega}^{(n)}$}
	\ForEach{integration point on the contact surface}{
		Compute normal gap $g_N$ from the deformed geometry\;
		Compute transparency: $\Transparency = \dfrac{1}{1 + \exp\bigl(\frac{g_N - \ell_q}{\Delta_g}\bigr)}$\;
		\BlankLine
		\Switch{contact scenario}{
			\uCase{Unbiased metal--dielectric}{
				$\sigdemand^{(n)} \leftarrow -\mathbf{p}_\mathrm{flexo} \cdot \mathbf{n}$\;
				$\sigma_\mathrm{tar}^{(n)} \leftarrow \sigdemand^{(n)}$\;
			}
			\uCase{Biased metal--dielectric}{
				$\sigdemand^{(n)} \leftarrow -\mathbf{p}_\mathrm{flexo} \cdot \mathbf{n}$\;
				$\sigma_\mathrm{tar}^{(n)} \leftarrow \Pi_V(\sigdemand^{(n)})$\;
			}
			\Case{Dielectric--dielectric}{
				$\Delta\sigma^{(n)} \leftarrow \dfrac{\hat{\sigma}_{b,m}^{(n)} - \hat{\sigma}_{b,s}^{(n)}}{2}$\;
				$\sigma_\mathrm{tar}^{(n)} \leftarrow \mathrm{sign}(\Delta\sigma^{(n)}) \cdot \min\left(|\Delta\sigma^{(n)}|,\, \sigma_\mathrm{trap}\right)$\;
			}
		}
		\BlankLine
		\eIf{loading phase \textbf{and} scenario $\neq$ biased}{
			Update frozen charge: $\sigfroz^{(n)} \leftarrow \Transparency \cdot \sigma_\mathrm{tar}^{(n)}$\;
		}{
			Update frozen charge: $\sigfroz^{(n)} \leftarrow \sigfroz^{(n-1)} + \Transparency \cdot \left(\sigma_\mathrm{tar}^{(n)} - \sigfroz^{(n-1)}\right)$\;
		}
		\BlankLine
		Compute Neumann contribution: $\bar{\omega}^{(n)} \leftarrow (1 - \Transparency) \cdot \sigfroz^{(n)}$\;
	}
\end{algorithm}

\subsection{Variational formulation and discretization}
\label{sec:numerical}

The contact contribution to the weak form arises from the virtual work performed by contact tractions and surface charges at the interface:
\begin{equation}
	\label{eq:contact_virtual_work}
	\delta\Pi_C = \int_{\partial\mathcal{B}_t^C} \left[ t_N \, \delta g_N + q_C \, \delta g_\phi \right] \mathrm{d}\gamma,
\end{equation}
where $t_N$ is the normal contact traction, $q_C$ is the contact surface charge density, and $g_\phi = \phi_s - \bar{\phi}_m$ is the potential gap across the interface, defined analogously to the normal gap $g_N$. For metal-dielectric contact, $\bar{\phi}_m$ reduces to the prescribed metal potential; for dielectric-dielectric contact, it corresponds to the interpolated master surface potential.

The normal contact traction comprises penalty regularization for contact~\cite{wriggers2006computational} and the adhesive cohesive zone model introduced in Section~\ref{sec:contact_mech}:
\begin{equation}
	\label{eq:contact_traction}
	t_N = \begin{cases}
		\varepsilon_N g_N & \text{if } g_N \leq 0, \\[3pt]
		\displaystyle\frac{\phi_N}{(g_N^\mathrm{max})^2}\,g_N \exp\left(-\frac{g_N}{g_N^\mathrm{max}}\right) & \text{if } g_N > 0,
	\end{cases}
\end{equation}
where the first case approximates the impenetrability constraint~\eqref{eq:KKT} via a large positive penalty parameter $\varepsilon_N$, permitting slight interpenetration ($g_N < 0$) as a trade-off for computational efficiency. The second case applies the adhesive traction~\eqref{eq:adhesion} during separation.

Analogously, the surface charge density arising from the electrical contact condition~\eqref{eq:electrical_contact} takes the penalty form
\begin{equation}
	\label{eq:penalty_elec}
	q_C = \varepsilon_\phi \, \Transparency \, g_\phi,
\end{equation}
where $\varepsilon_\phi$ is the electrical penalty parameter, which physically assumes the role of an infinitely large interfacial capacitance per unit area. This constitutive relation strongly penalizes electrical potential jumps, driving the potential gap $g_\phi \to 0$ at locations where the tunneling channel is open ($\Transparency \approx 1$), thereby enforcing the required local electrostatic equilibrium for transferring compensating charges. It is worth noting that while macroscopic capacitive charge $q_C$ is invoked to equilibrate the potential constraint during contact, it is fully reversible and inherently vanishes upon separation as $\Transparency \to 0$. The final irreversible surface charge remaining after separation is governed exclusively by the dynamically tracked fraction $\sigfroz$ subject to polarity and trap limits.

The contact integrals in~\eqref{eq:contact_virtual_work} are evaluated using isogeometric analysis with NURBS basis functions~\cite{Hughes2005}. The mortar method is employed to project contact quantities onto nodal values via weighted averaging, ensuring smooth transfer of contact forces between non-matching meshes. The resulting coupled electromechanical system is solved using a monolithic Newton--Raphson scheme. The detailed formulation and consistent linearization are provided in~\ref{app:contact}.

\section{Metal-Dielectric AFM Contact}
\label{sec:validation}

\subsection{Problem setup}
\label{sec:setup}

\begin{figure}[!htbp]
	\centering
	\includegraphics[width=\textwidth]{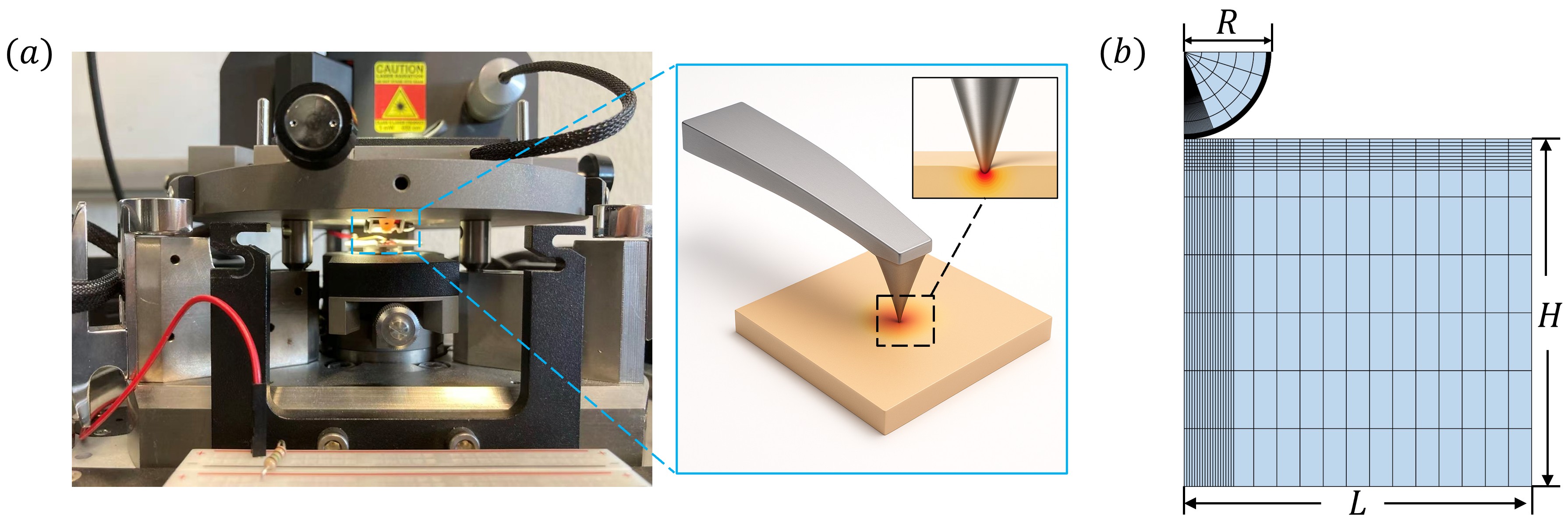}
	\caption{AFM contact configuration: (a) experimental setup; (b) computational mesh.}
	\label{fig:afm_schematic}
\end{figure}

This section evaluates the model by comparing its predictions with experimental observations in atomic force microscopy (AFM) experiments on polymer substrates. We consider poly(meth\-yl meth\-acryl\-ate) (PMMA) and poly(di\-allyl phthal\-ate) (PDAP), and contact scenarios including an AFM tip with and without an applied bias. Figure~\ref{fig:afm_schematic} shows the experimental setup and a schematic of the tip-substrate contact.
The AFM contact geometry is modeled as a two-dimensional plane-strain problem, with a metal-coated silicon tip (radius $R = 25$~nm) indenting a finite dielectric substrate. The computational domain is set to $L\times H = 100 \times 100$~nm$^2$, which is sufficiently large compared to the contact radius ($\sim$10~nm) to avoid boundary effects. Complete separation is defined when the minimum normal gap exceeds $2\ell_q$, at which point the tunneling transparency is negligible across the entire interface.

The material parameters are summarized in Table~\ref{tab:material}. The mechanical and dielectric properties are obtained from the literature. Flexoelectric coefficients for polymers are not yet routinely measured, and the reported values can span orders of magnitude depending on the measurement technique and sample preparation~\cite{lu2019temperature,deng2014flexoelectricity}. Reasonable estimates are therefore adopted based on available measurements and typical polymer values. In particular, the sign of the transverse coefficient $\mu_T$ may differ between polymers because it depends on the local molecular polarizability and symmetry of the repeat unit. The signs of the transverse coefficient $\mu_T$ are phenomenologically inferred to be opposite for PMMA and PDAP to interpret their contrasting electrification polarities observed in experiments, which serves as a working assumption in our simulations. A characteristic length of $\ell_s = 0.1$~nm is used for both materials to account for strain gradient elasticity effects.
\begin{table}[!htbp]
	\centering
	\caption{Material and model parameters. }
	\label{tab:material}
	\begin{tabular}{lcccl}
		\toprule
		Parameter & PMMA \cite{ishiyama2002effects,harper2000modern,tang2022mechanoelectric} & PDAP~\cite{harper2000modern,guo2017thermosets} & Tip & Unit \\
		\midrule
		Young's modulus $E$ & 3.5 & 13.0 & 170 & GPa \\
		Poisson's ratio $\nu$ & 0.36 & 0.358 & 0.22 & -- \\
		Relative permittivity $\epsilon/\epsilon_0$ & 3.6 & 5.2 & -- & -- \\
		Flexoelectric coeff. $\mu_L$ & $4.67 \times 10^{-11}$ & $3.62 \times 10^{-11}$ & -- & C/m \\
		Flexoelectric coeff. $\mu_T$ & $-1.65 \times 10^{-11}$ & $7.95 \times 10^{-11}$ & -- & C/m \\
		Flexoelectric coeff. $\mu_S$ & 0 & $1.20 \times 10^{-11}$ & -- & C/m \\
		Adhesion work $\phi_N$ & 5 & 10 & -- & J/m$^2$ \\
		Max adhesive traction $p_N^\mathrm{max}$ & 10 & 25 & -- & MPa \\
		\bottomrule
	\end{tabular}
\end{table}
The mechanical penalty parameter is set to $\varepsilon_N = 10^{10}\,E$~m$^{-1}$, where $E$ is the Young's modulus of the substrate, and the electrical penalty parameter is evaluated element-wise as $\varepsilon_\phi = \theta\,\epsilon / h_e$ with $\theta = 0.3$ and $h_e$ the local element size. The resulting maximum interpenetration remains below 0.01~nm, negligible relative to both the contact dimensions and the tunneling length $\ell_q$. The AFM simulations use a contact-zone element size $h = 0.29$~nm and a pseudo-time increment $\Delta t = 0.05$. A convergence study confirming the adequacy of these choices is provided in~\ref{app:convergence}.

\subsection{Unbiased AFM contact}
\label{sec:unbiased_results}

\begin{figure}[!htbp]
	\centering
	\includegraphics[width=0.6\textwidth]{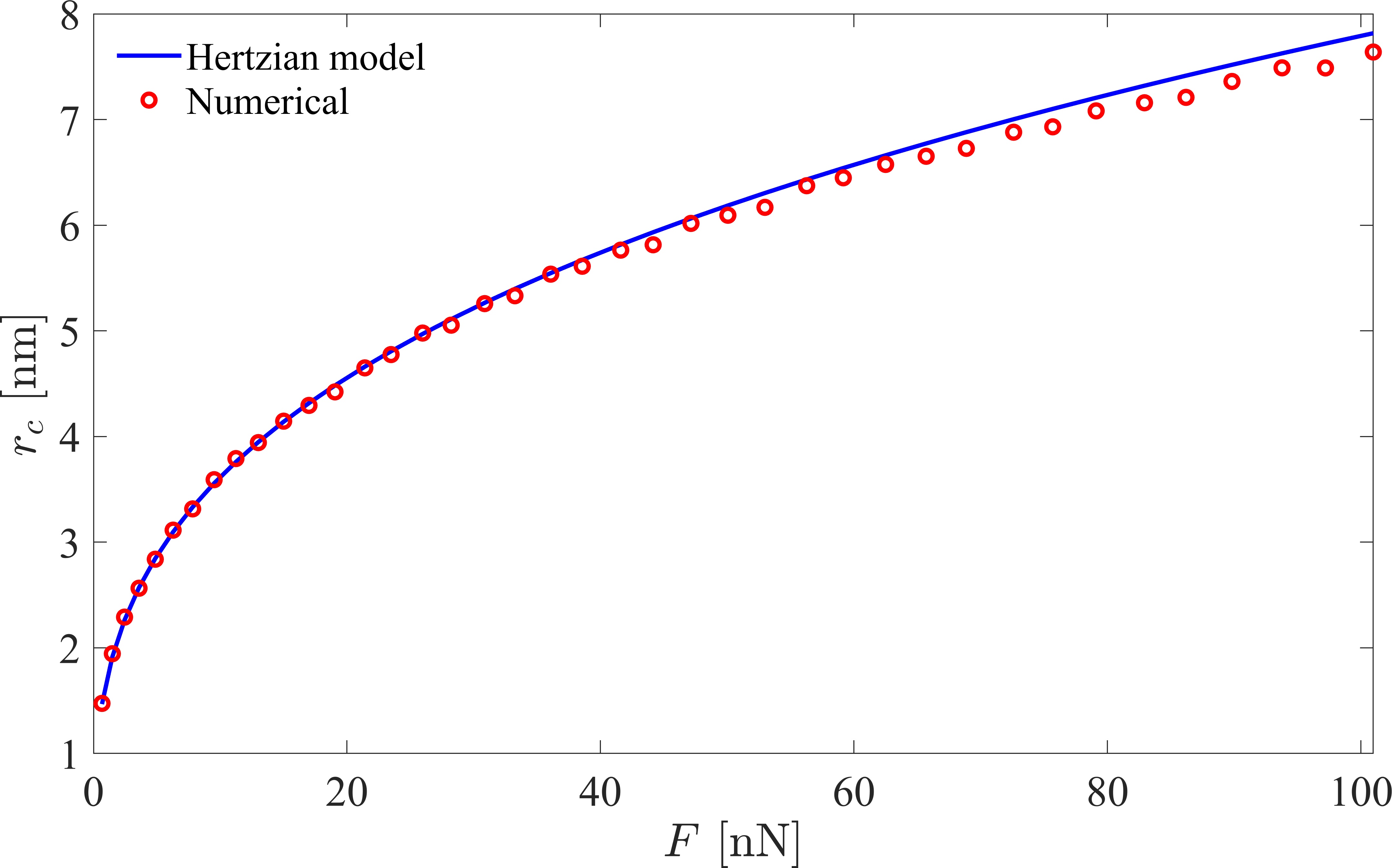}
	\caption{Contact radius as a function of contact force for PMMA substrate, compared with classical Hertz theory. The contact force is the equivalent three-dimensional force.}
	\label{fig:hertz_validation}
\end{figure}

To verify the mechanical response of the model, we compare the computed contact radius with the classical Hertz theory, as shown in Fig.~\ref{fig:hertz_validation}. The figure shows good agreement between the numerical results and the analytical Hertz solution, confirming the accuracy of the contact mechanics implementation. The slight discrepancy above 60~nN is attributed to finite deformation effects and the nonlinear flexoelectric response at large strains.

\begin{figure}[!htbp]
	\centering
	\begin{subfigure}{0.48\textwidth}
		\includegraphics[width=\textwidth]{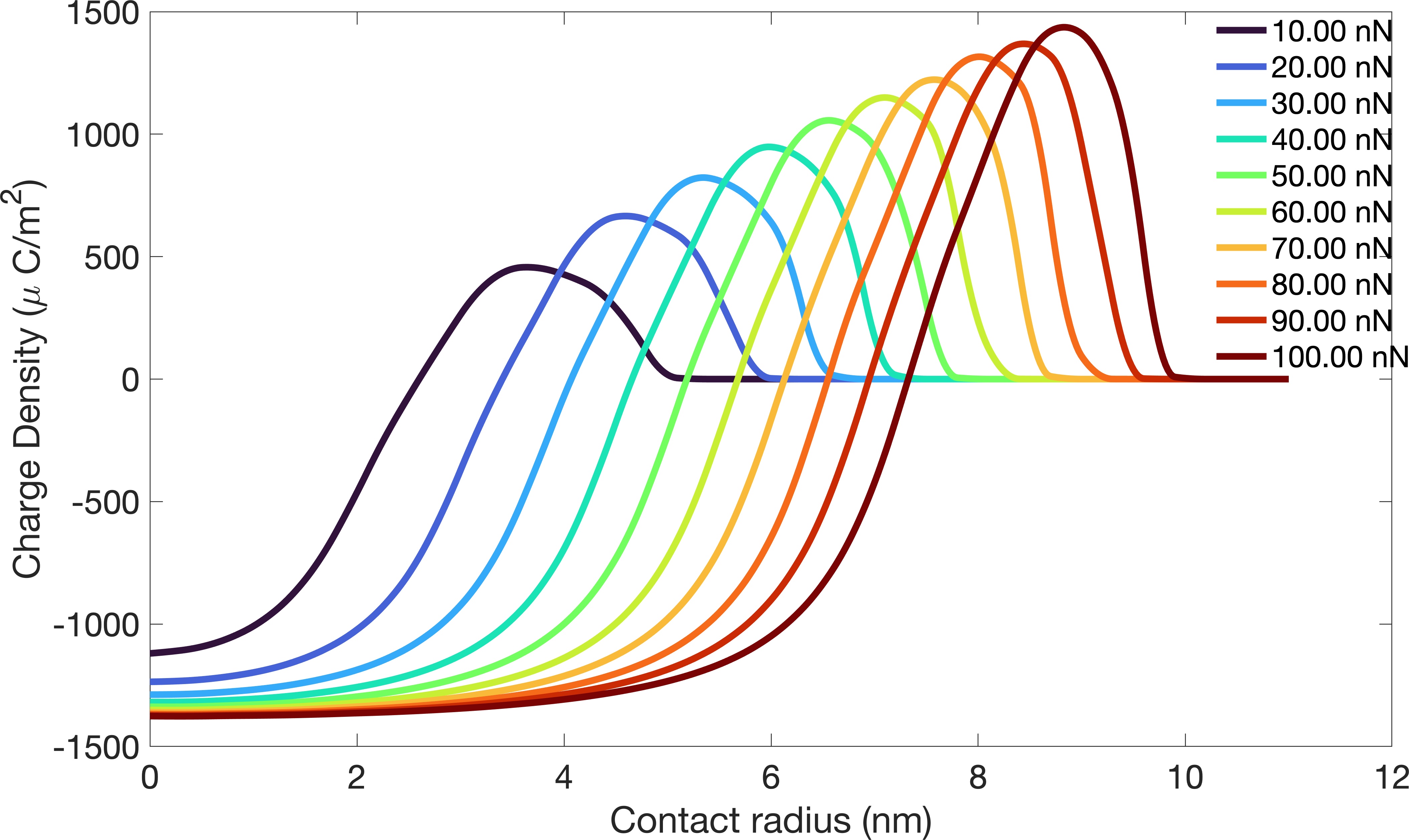}
		\caption{PMMA}
	\end{subfigure}
	\hfill
	\begin{subfigure}{0.48\textwidth}
		\includegraphics[width=\textwidth]{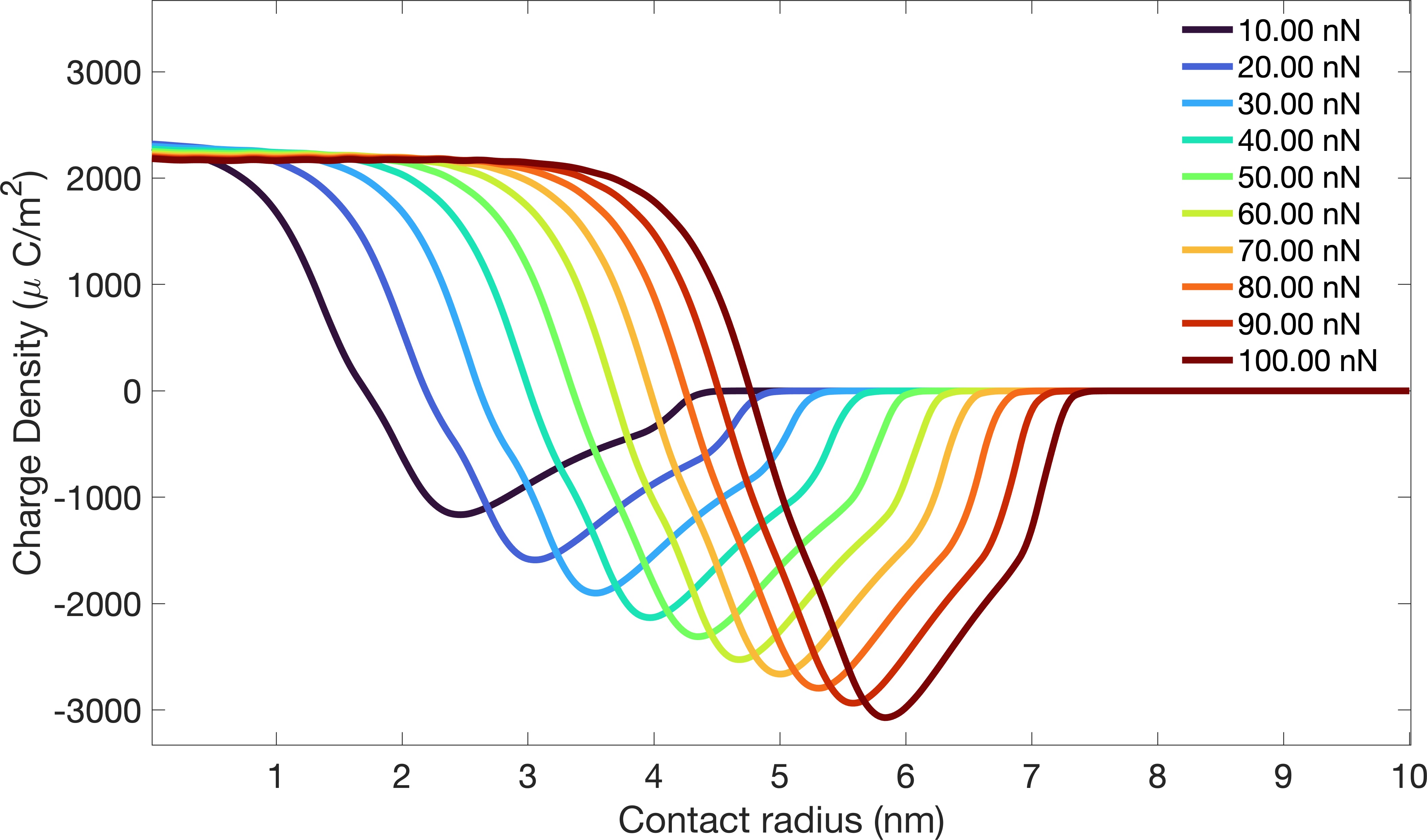}
		\caption{PDAP}
	\end{subfigure}
	\caption{Surface free charge density distribution at the end of loading for different contact forces: (a) PMMA; (b) PDAP.}
	\label{fig:loading_charge}
\end{figure}

Figure~\ref{fig:loading_charge} presents the surface free charge density distribution at the end of the loading phase for different contact forces. It is observed that the two materials exhibit opposite charge patterns due to their different flexoelectric properties.
For PMMA, as shown in Fig.~\ref{fig:loading_charge}(a), the contact center exhibits negative charge density while the contact edge shows positive charge. At the contact center, the strain gradient induces positive flexoelectric polarization, attracting negative charges from the metal tip. At the edge, the strain gradient reverses sign due to the local strain field transition from compression to tension, leading to positive charge accumulation. As the contact force increases, the negative charge density at the center initially increases but then saturates when the strain gradient at the center stops growing. Meanwhile, the positive charge density at the contact edge continues to increase, and the charged region extends outward with the expanding contact area.
For PDAP, as shown in Fig.~\ref{fig:loading_charge}(b), the charge polarity is reversed: the contact center shows positive charge density while the edge exhibits negative charge. Within the present model, this opposite pattern arises from the assumed sign of the transverse flexoelectric coefficient $\mu_T$, which governs the polarization response to the dominant strain gradient component.

\begin{figure}[!htbp]
	\centering
	\begin{subfigure}{0.48\textwidth}
		\includegraphics[width=\textwidth]{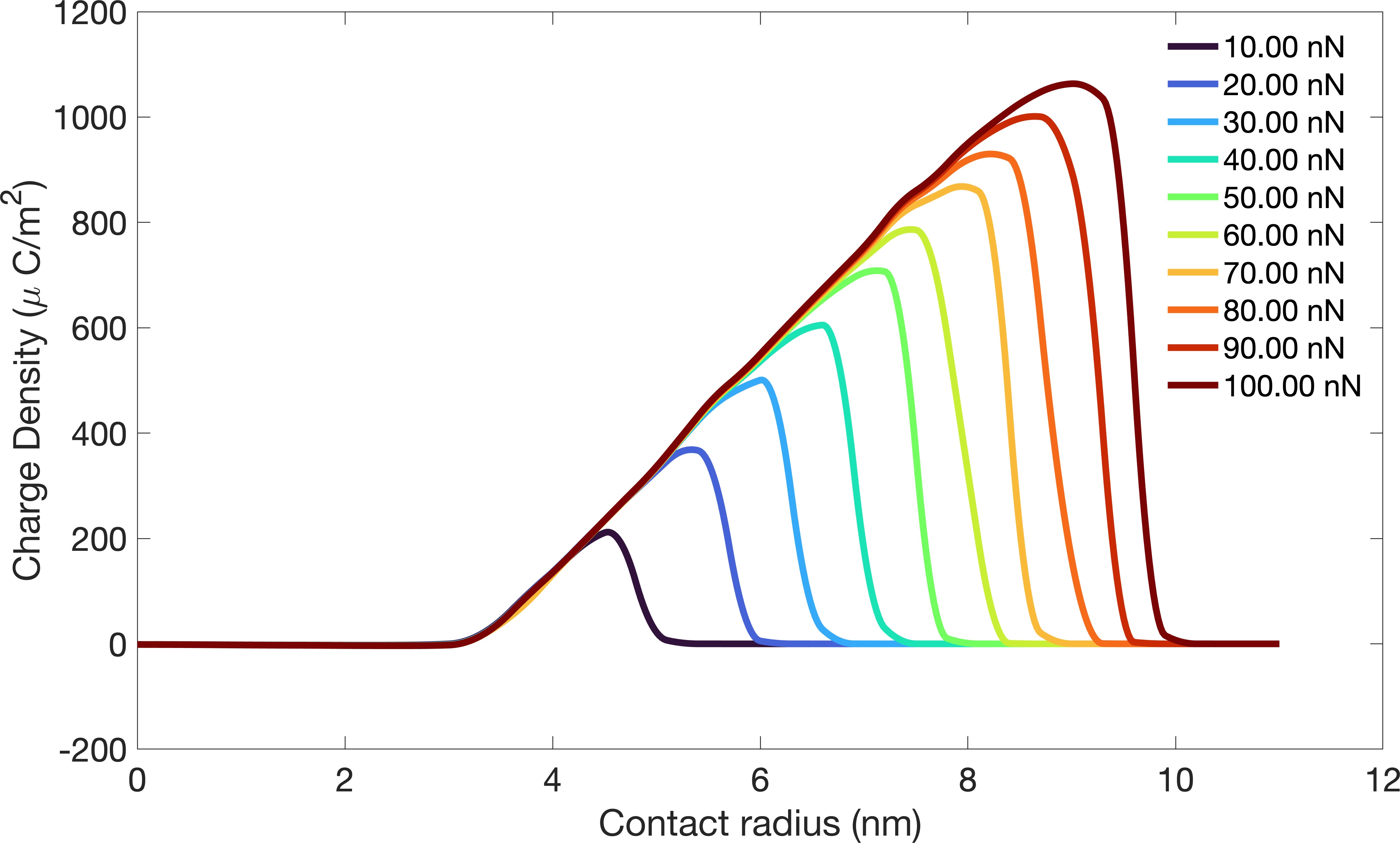}
		\caption{PMMA}
	\end{subfigure}
	\hfill
	\begin{subfigure}{0.48\textwidth}
		\includegraphics[width=\textwidth]{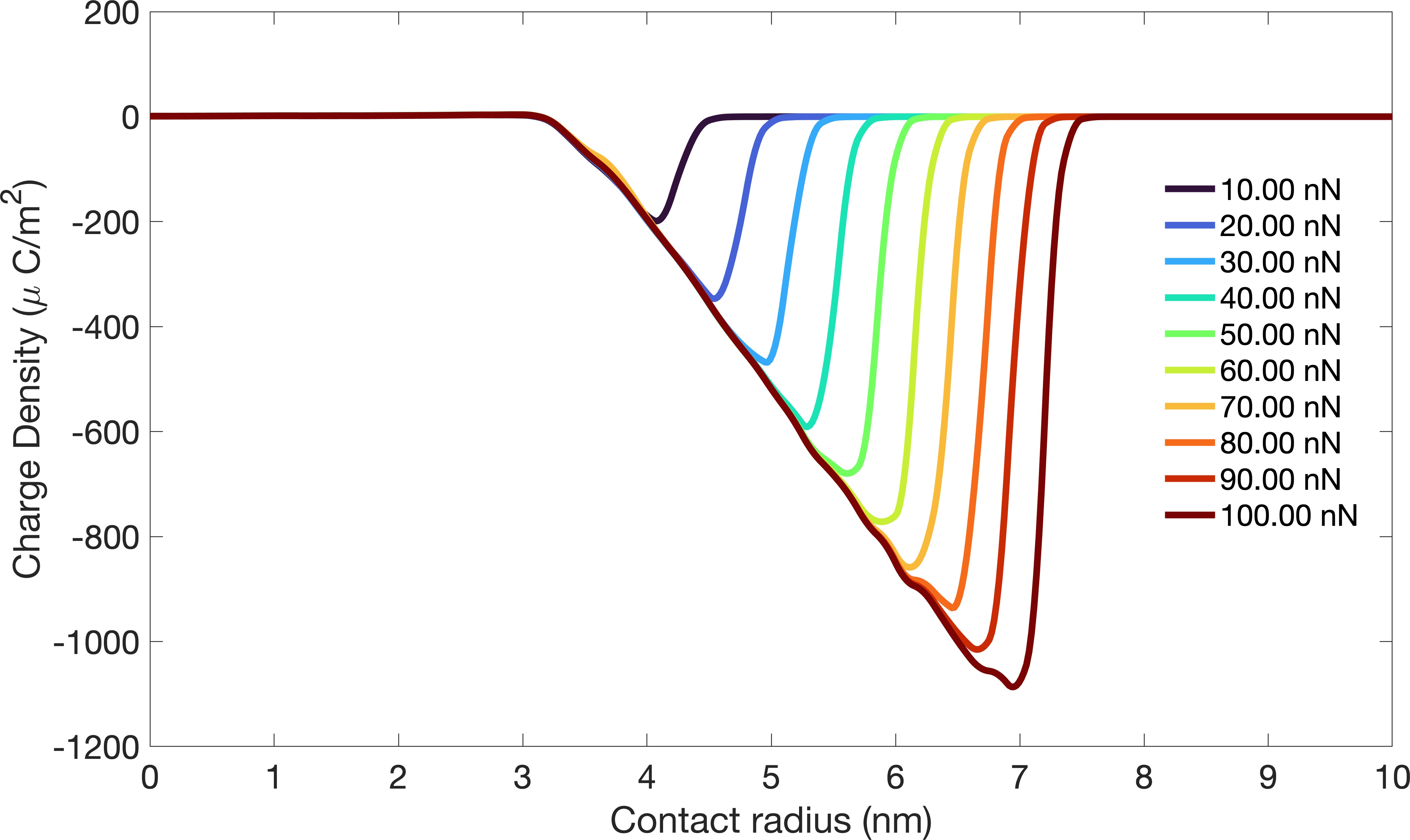}
		\caption{PDAP}
	\end{subfigure}
	\caption{Residual surface charge density distribution after complete separation for different maximum loading forces: (a) PMMA; (b) PDAP.}
	\label{fig:unloading_charge}
\end{figure}

Figure~\ref{fig:unloading_charge} shows the residual charge distribution after complete separation. In contrast to the loading phase, the contact center region shows nearly zero residual charge, while the original contact edge retains significant charge. 
This characteristic distribution follows directly from the charge transfer mechanism and the elastic recovery of the substrate. As the mechanical load is removed and the substrate returns to its undeformed shape, the flexoelectric demand at the center diminishes. During unloading, the tunneling channel closes first at the contact periphery where the gap opens earliest, freezing the accumulated charge before it can flow back to the tip. Conversely, the channel at the contact center remains open longest, allowing the charge there to flow back to the metal tip to maintain local equilibrium. The final state is thus a ring of frozen charge at the periphery with an uncharged center, which agrees with the theoretical predictions of Silavnieks et al.~\cite{Silavnieks2025flexo}. It is also worth noting that under the current cohesive parameters, the adhesive effects during unloading do not visibly alter this central discharge behavior.

\begin{figure}[!htbp]
	\centering
	\begin{subfigure}{0.48\textwidth}
		\includegraphics[width=\textwidth]{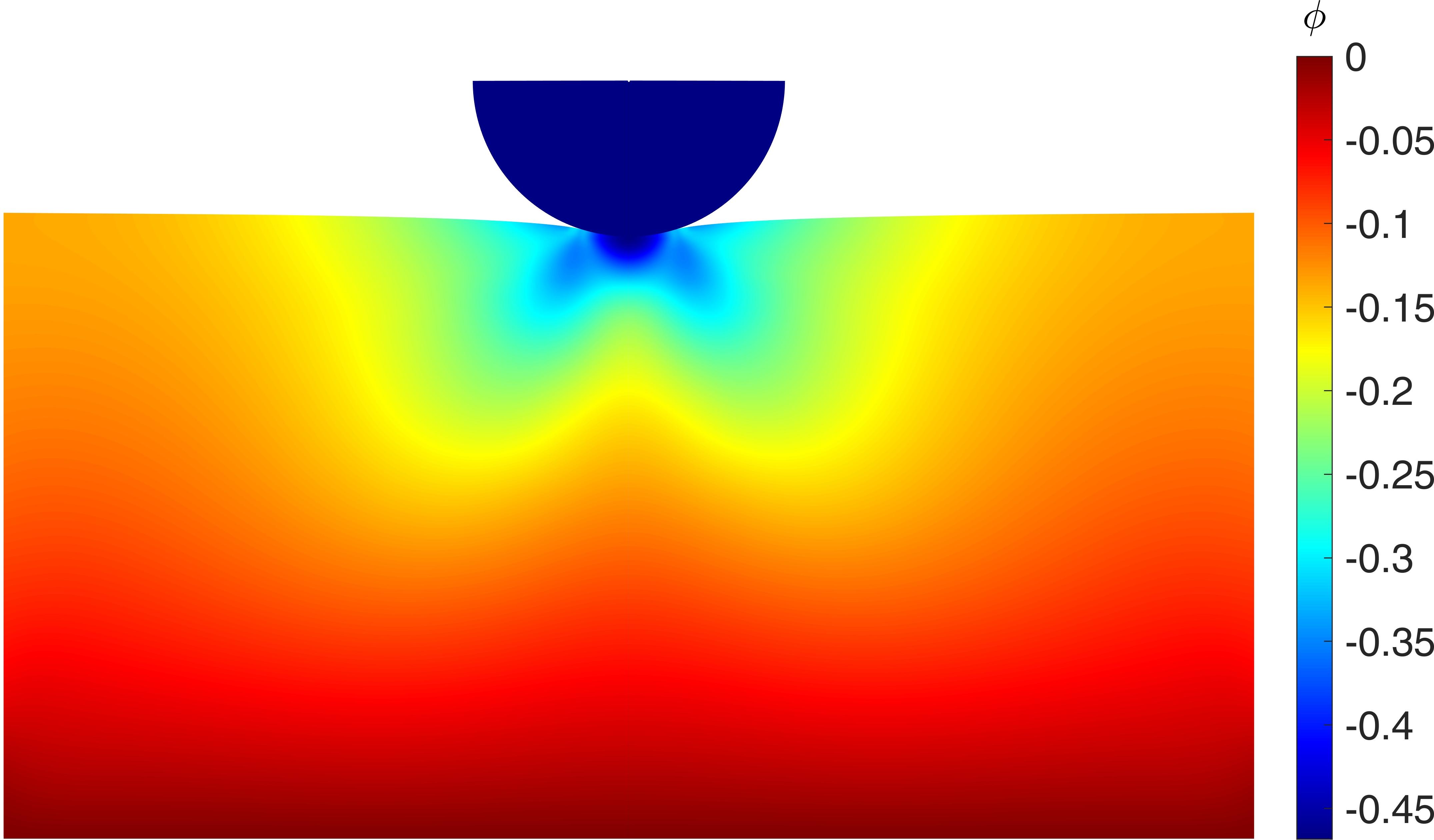}
		\caption{PMMA, loading}
	\end{subfigure}
	\hfill
	\begin{subfigure}{0.48\textwidth}
		\includegraphics[width=\textwidth]{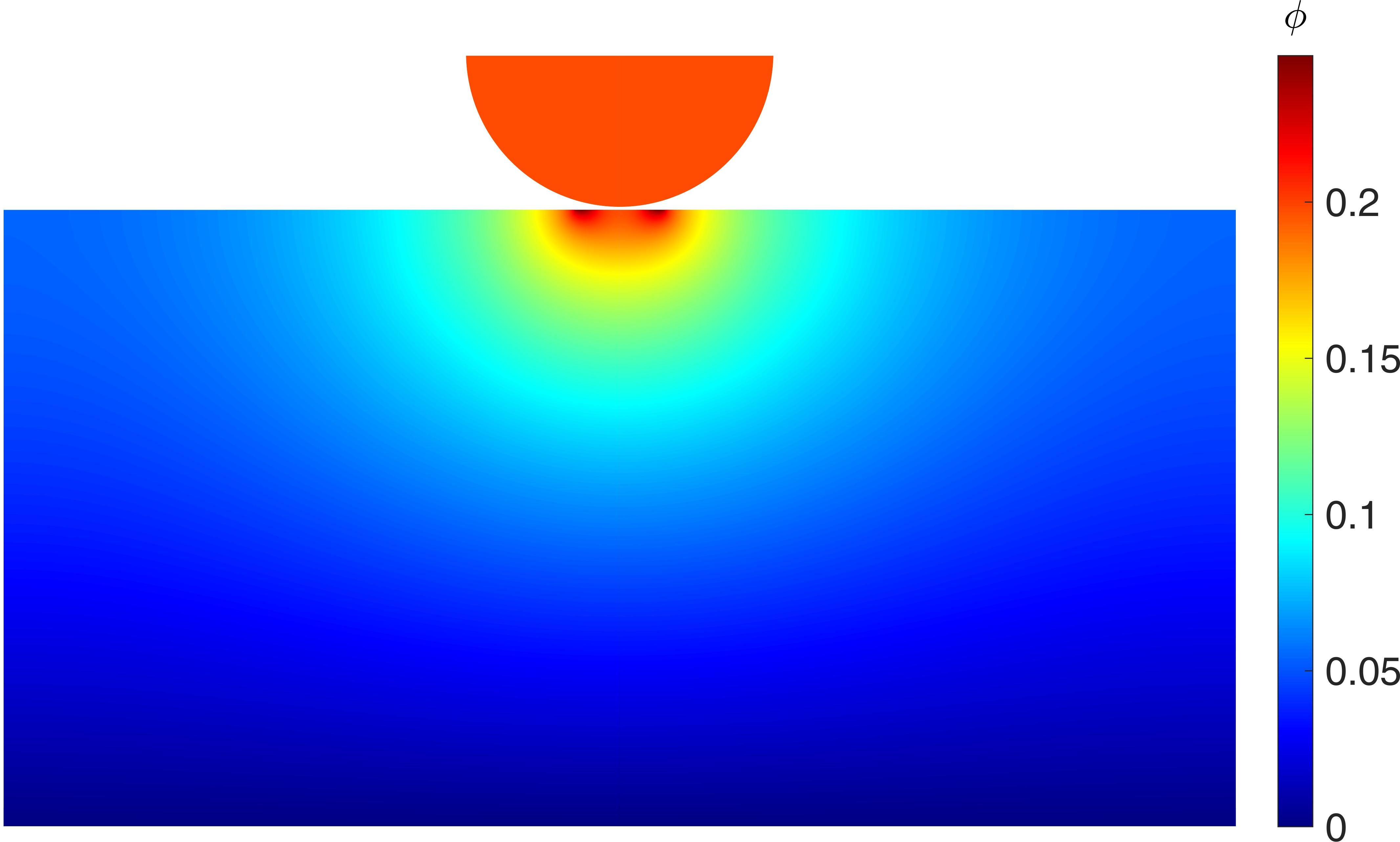}
		\caption{PMMA, separation}
	\end{subfigure}
	\\
	\begin{subfigure}{0.48\textwidth}
		\includegraphics[width=\textwidth]{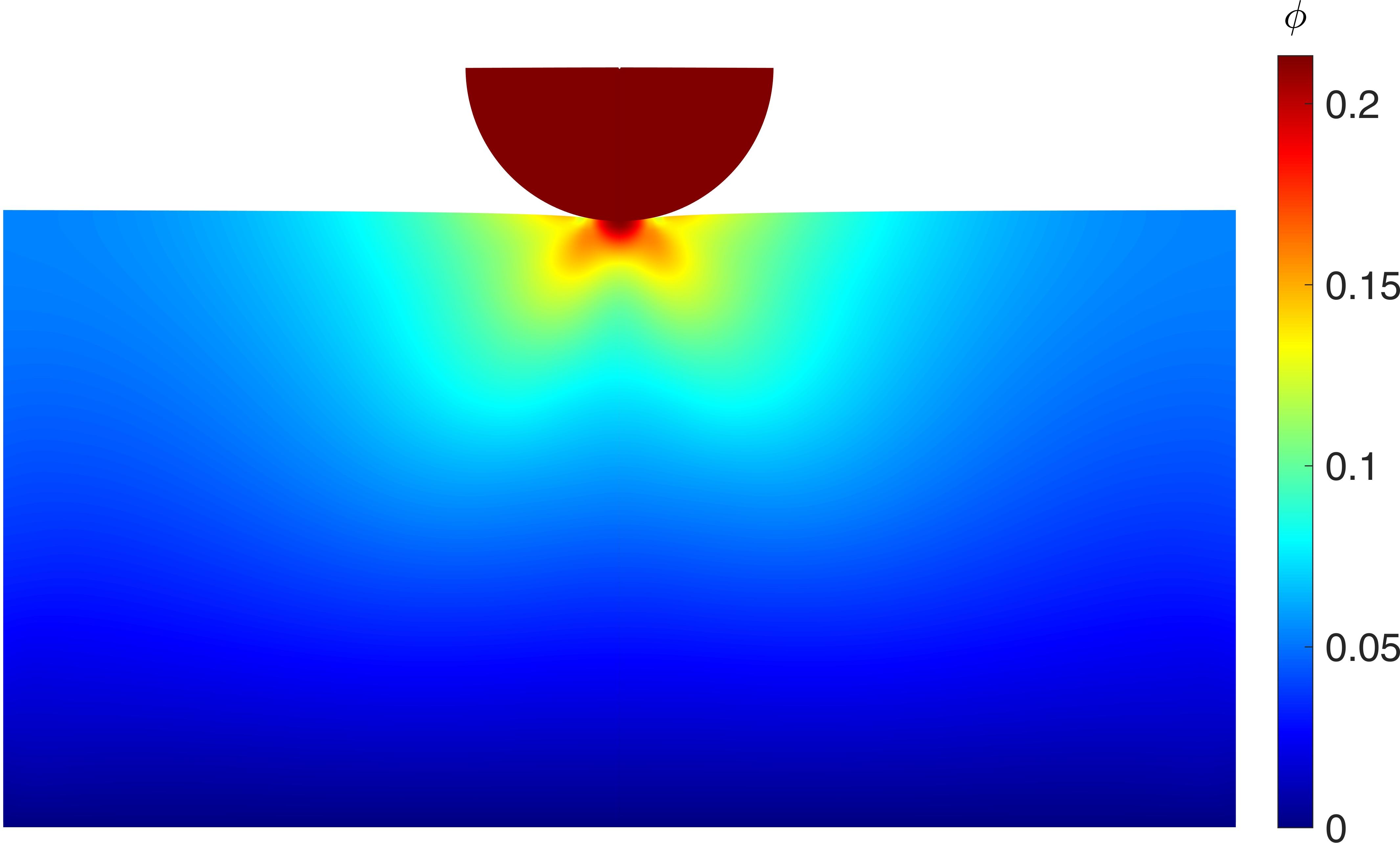}
		\caption{PDAP, loading}
	\end{subfigure}
	\hfill
	\begin{subfigure}{0.48\textwidth}
		\includegraphics[width=\textwidth]{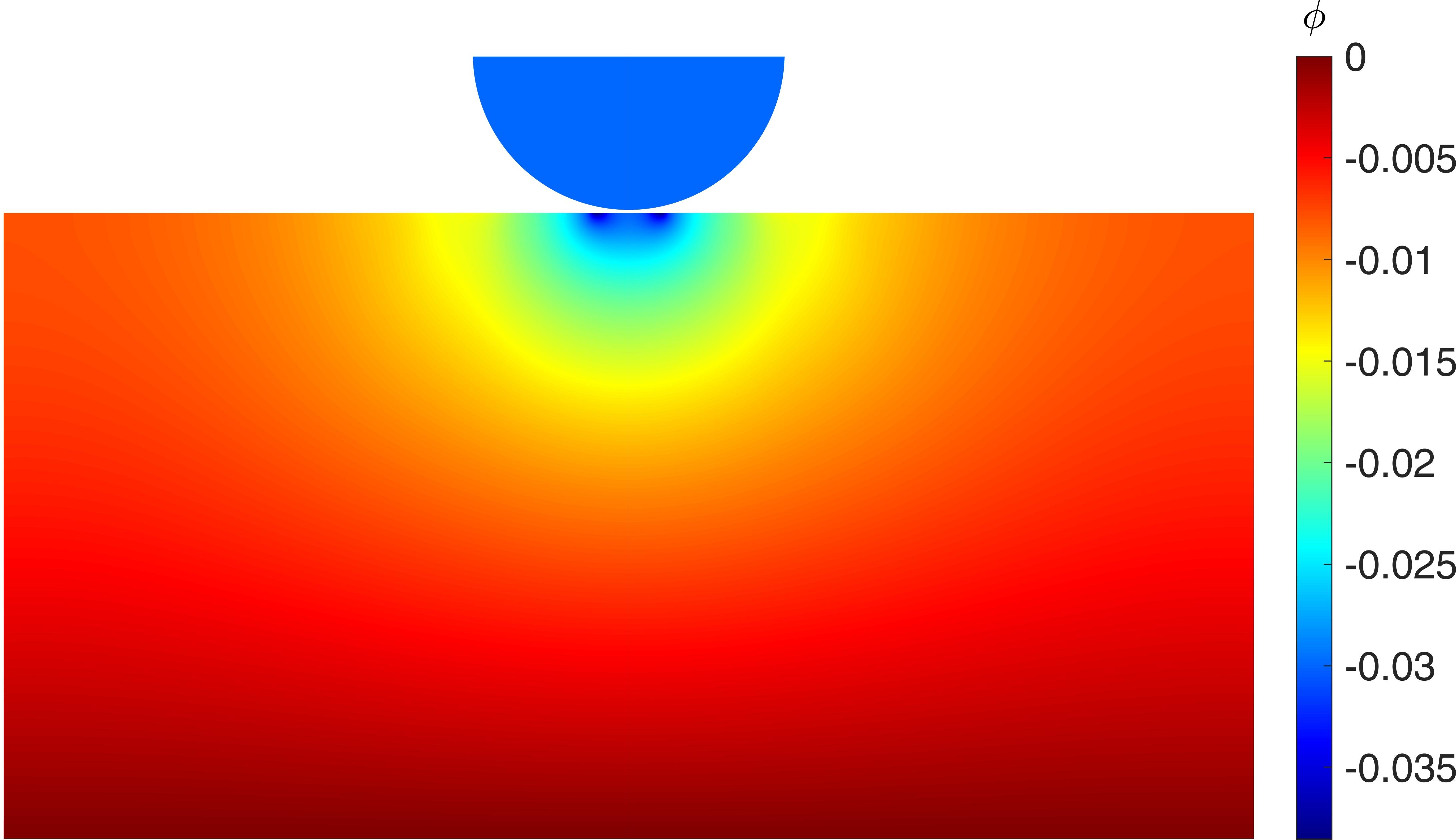}
		\caption{PDAP, separation}
	\end{subfigure}
	\caption{Electric potential distribution at 50~nN contact force: (a,c) at the end of loading; (b,d) after complete separation.}
	\label{fig:potential_unbiased}
\end{figure}

Notably, the residual charge distributions for different loading forces follow the same spatial profile, with only the extent varying with the maximum contact area. Since PMMA and PDAP exhibit opposite edge charging during loading, the frozen residual charge after separation also displays opposite polarity, as shown in Figs.~\ref{fig:unloading_charge}(a) and (b).
Figure~\ref{fig:potential_unbiased} displays the electric potential distribution at the end of loading and after complete separation for a loading force of 50~nN. The potential field reflects the coupled electromechanical response, with the flexoelectric polarization creating localized potential variations near the contact interface. The potential contours reveal that the field perturbation penetrates approximately 20--30~nm into the substrate. It can be observed that the potential magnitude after separation is significantly reduced.

\begin{figure}[!htbp]
	\centering
	\begin{subfigure}{0.48\textwidth}
		\includegraphics[width=\textwidth]{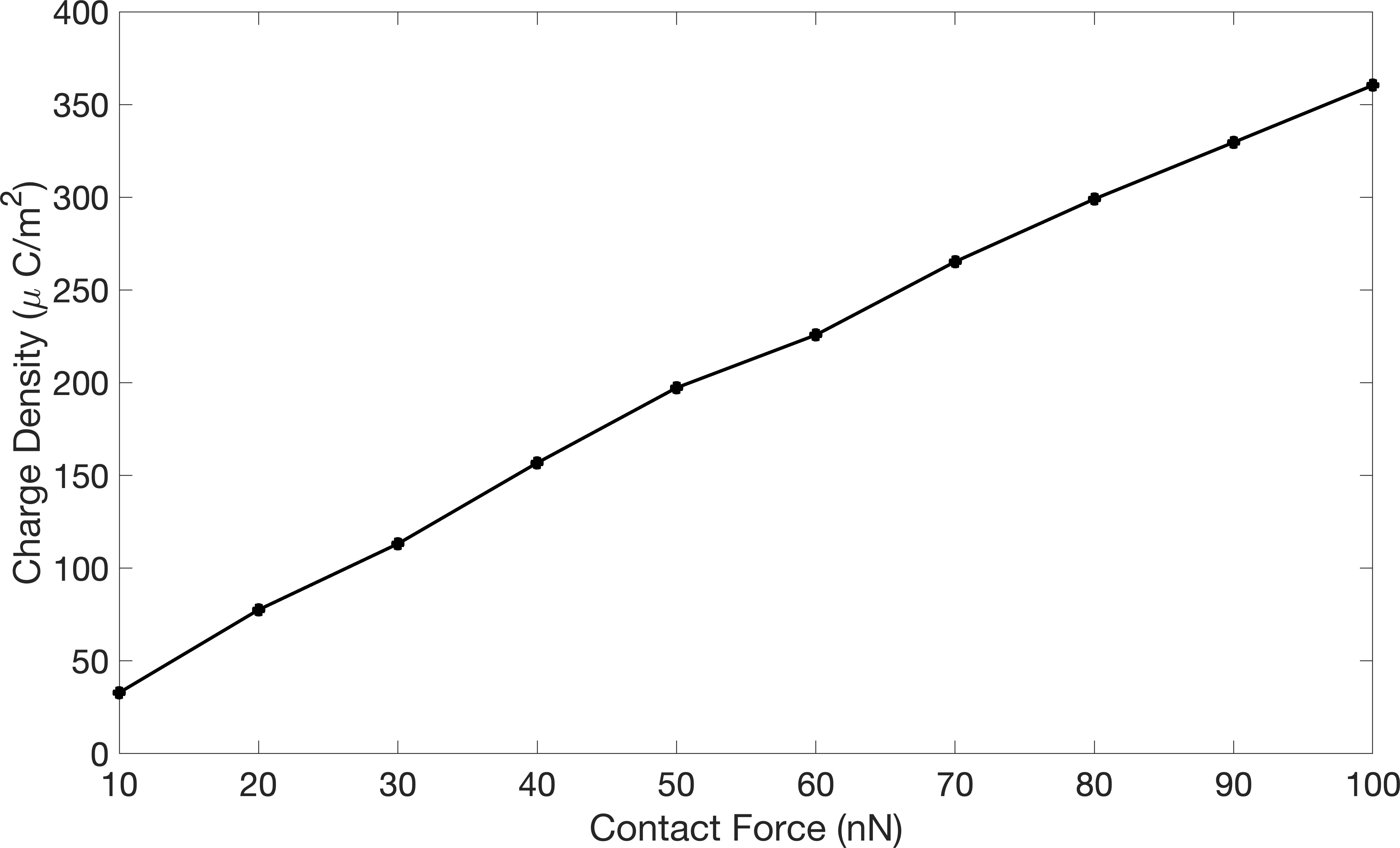}
		\caption{}
	\end{subfigure}
	\hfill
	\begin{subfigure}{0.45\textwidth}
		\includegraphics[width=\textwidth]{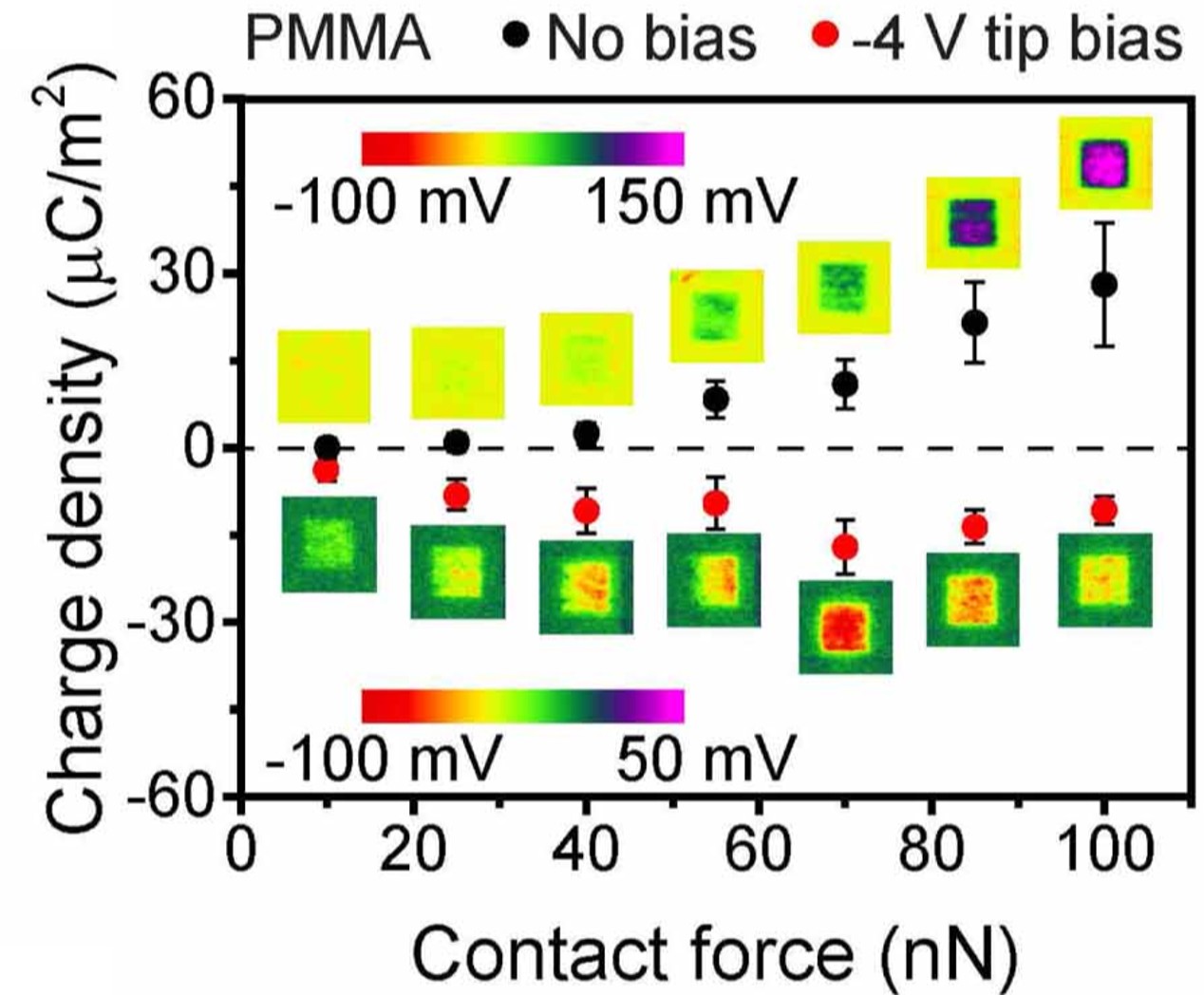}
		\caption{}
	\end{subfigure}
	\\
	\begin{subfigure}{0.48\textwidth}
		\includegraphics[width=\textwidth]{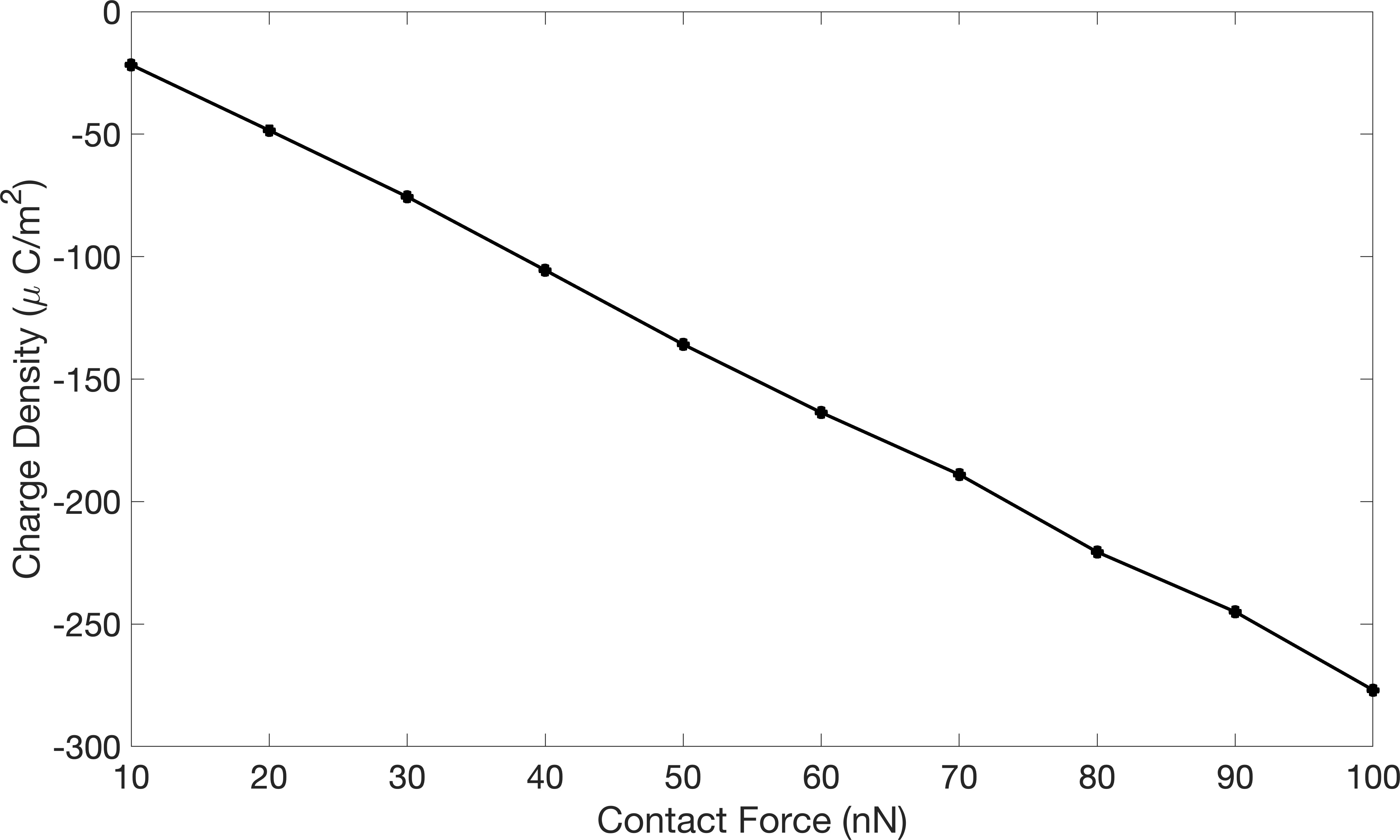}
		\caption{}
	\end{subfigure}
	\hfill
	\begin{subfigure}{0.45\textwidth}
		\includegraphics[width=\textwidth]{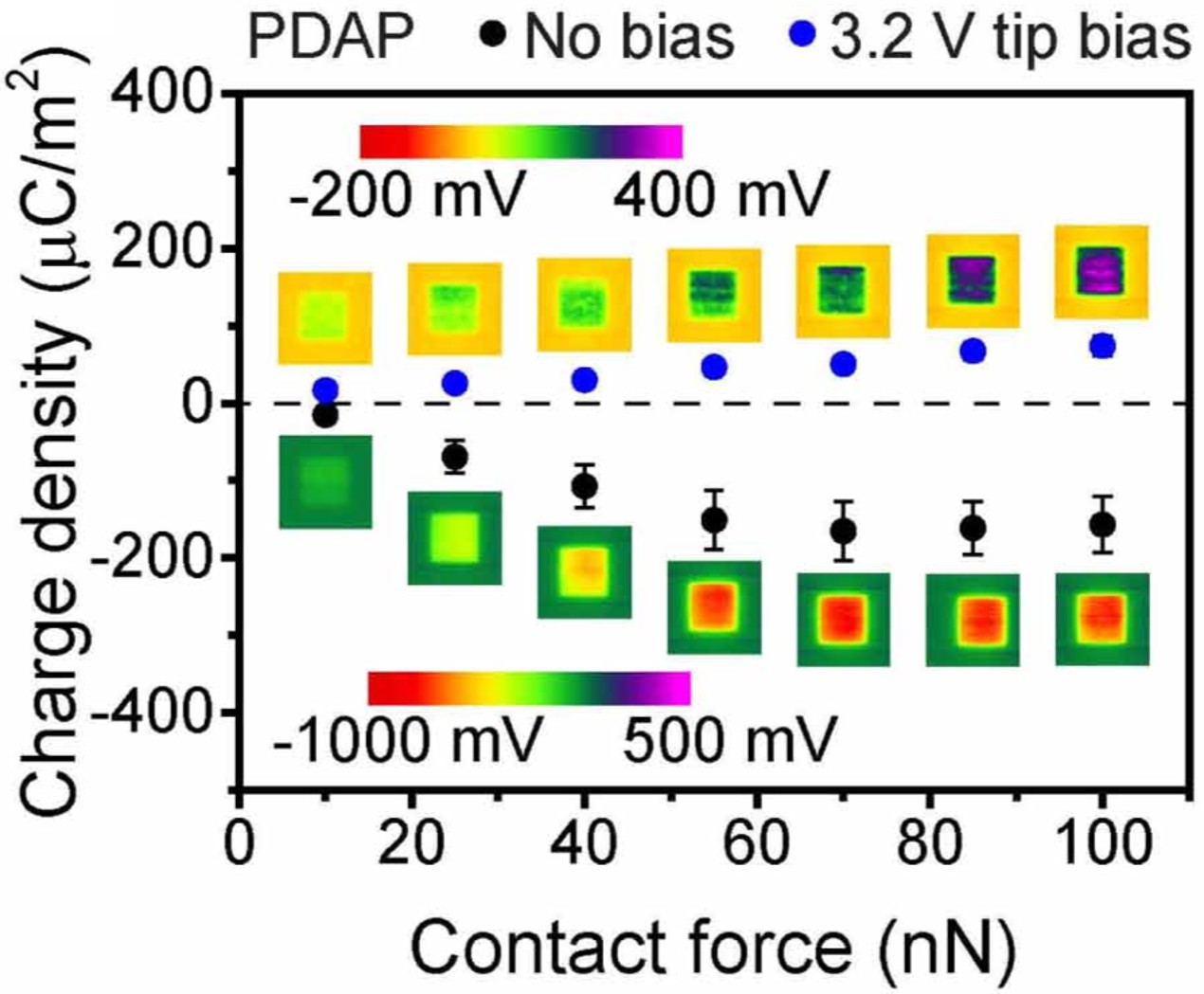}
		\caption{}
	\end{subfigure}
	\caption{Comparison of average surface charge density between simulation and experiment: (a,c) simulations for PMMA and PDAP; (b,d) experimental data adapted from Lin et al.~\cite{Lin2022flexo}. \textcopyright~IOP Publishing. Reproduced with permission. All rights reserved.}
	\label{fig:exp_comparison_unbiased}
\end{figure}

Figure~\ref{fig:exp_comparison_unbiased} compares the computed average surface charge density ($\bar{\sigma}=\int_{\gamma_c} \sigfroz\mathrm{d}\gamma/\int_{\gamma_c} \mathrm{d}\gamma$, where $\gamma_c$ is the total charged area) with experimental data from Lin et al.\ \cite{Lin2022flexo}. The simulation results capture the qualitative trend of increasing charge density with contact force for both PMMA and PDAP, but a quantitative gap is evident. For PDAP at 100~nN, the predicted residual charge density is approximately $280$~$\mu$C/m$^2$, compared with the measured value of roughly $160$~$\mu$C/m$^2$. For PMMA, the model overestimates by roughly one order of magnitude.
Because the charge demand scales linearly with the flexoelectric coefficients, the predicted charge density inherits this linear dependence. Matching the PMMA measurements would therefore require reducing the assumed flexoelectric coefficients by a factor of about 10, placing them at approximately $5\times10^{-12}$~C/m. Given that measured flexoelectric coefficients for polymers span several orders of magnitude depending on the measurement technique, sample preparation, and loading conditions~\cite{lu2019temperature,deng2014flexoelectricity}, values in this range remain physically plausible.
Additionally, the predicted charge magnitude is related to the strain gradient characteristic length $\ell_s$, as larger values of $\ell_s$ would smooth the localized strain gradients and further reduce the theoretical charge demand.
Beyond coefficient uncertainty, other factors may contribute to the discrepancy. At small contact forces, the charge transfer may be partly driven by the intrinsic work function difference between the tip and substrate~\cite{Qiao2021mixed,Olson2022band}, which is not included in the present model. The Hertzian contact model also assumes an idealized tip geometry that may differ from the actual AFM tip-sample interaction.

\subsection{Biased AFM contact}
\label{sec:biased_results}
\begin{figure}[!htbp]
	\centering
	\begin{subfigure}{0.48\textwidth}
		\includegraphics[width=\textwidth]{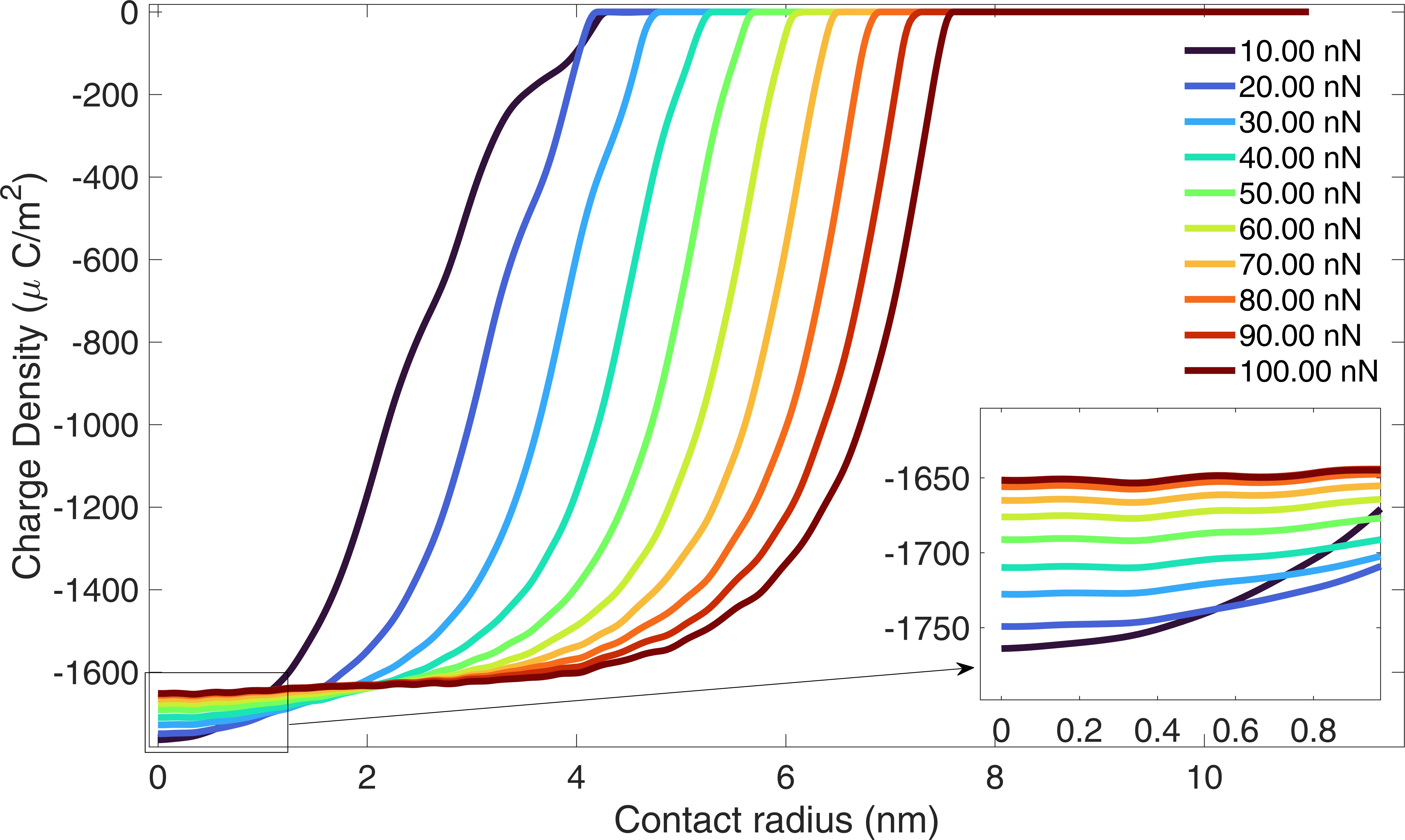}
		\caption{PMMA (negative bias)}
	\end{subfigure}
	\hfill
	\begin{subfigure}{0.48\textwidth}
		\includegraphics[width=\textwidth]{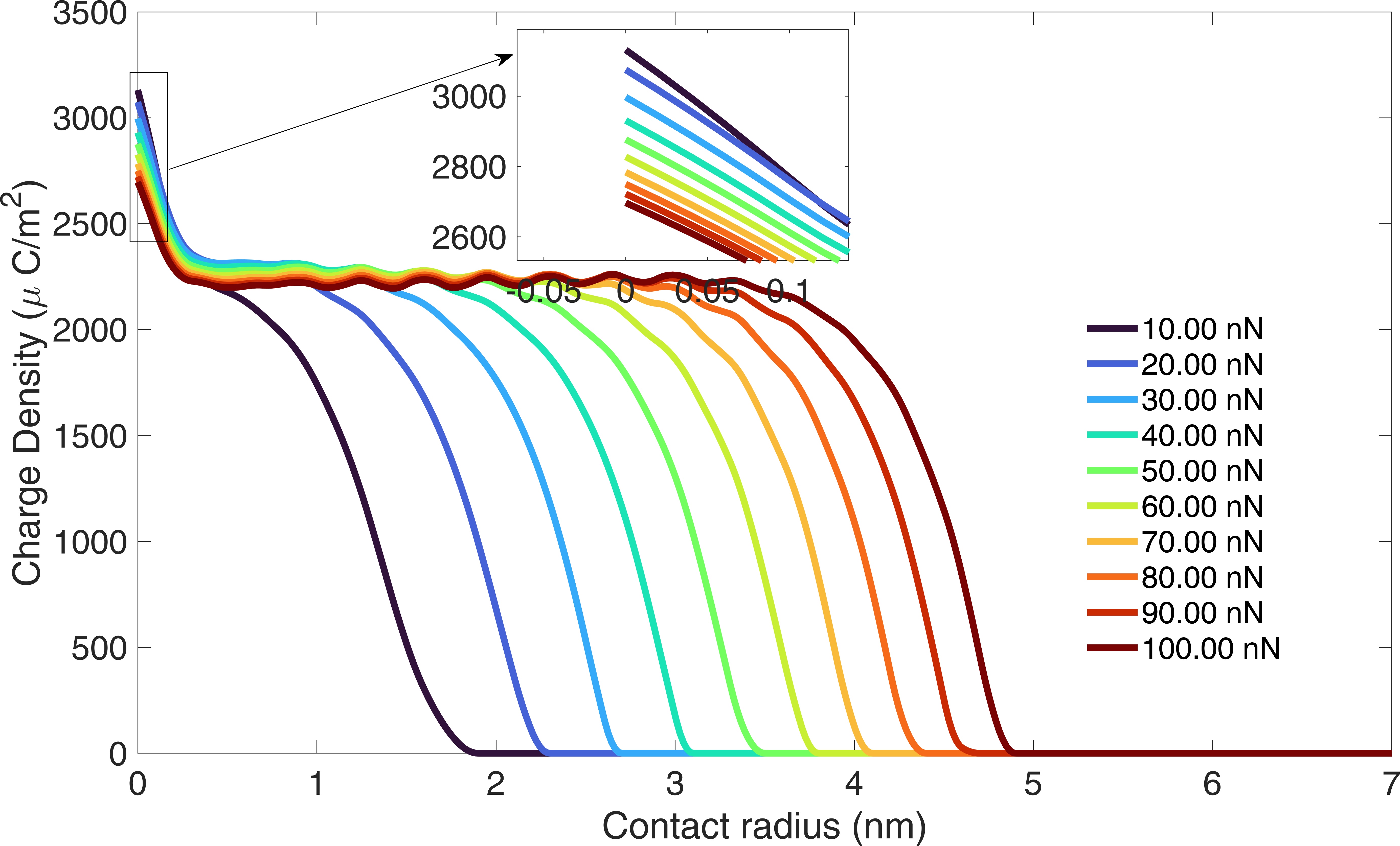}
		\caption{PDAP (positive bias)}
	\end{subfigure}
	\caption{Surface charge density distribution at the end of loading under biased conditions: (a) PMMA with negative tip bias; (b) PDAP with positive tip bias.}
	\label{fig:loading_charge_biased}
\end{figure}

In this section, we investigate the effect of external tip bias on the charge transfer process. The basic setup is the same as in Section~\ref{sec:unbiased_results}, except that a $-4$~V tip bias is applied for PMMA and a $+3.2$~V bias for PDAP, matching the experimental conditions of Lin et al.~\cite{Lin2022flexo}.
Figure~\ref{fig:loading_charge_biased} shows the surface charge distribution at the end of the loading phase. For PMMA, as shown in Fig.~\ref{fig:loading_charge_biased}(a), only negative charges are transferred, with an approximately uniform charge density within the contact region that drops quickly to zero at the contact edge. For PDAP, as shown in Fig.~\ref{fig:loading_charge_biased}(b), only positive charges are transferred, exhibiting a similar plateau distribution. Notably, as shown in the insets, the charge density at the contact center is higher for smaller loading forces, consistent with the two stages illustrated in Fig.~\ref{fig:biased_schematic}.

\begin{figure}[!htbp]
	\centering
	\begin{subfigure}{0.48\textwidth}
		\includegraphics[width=\textwidth]{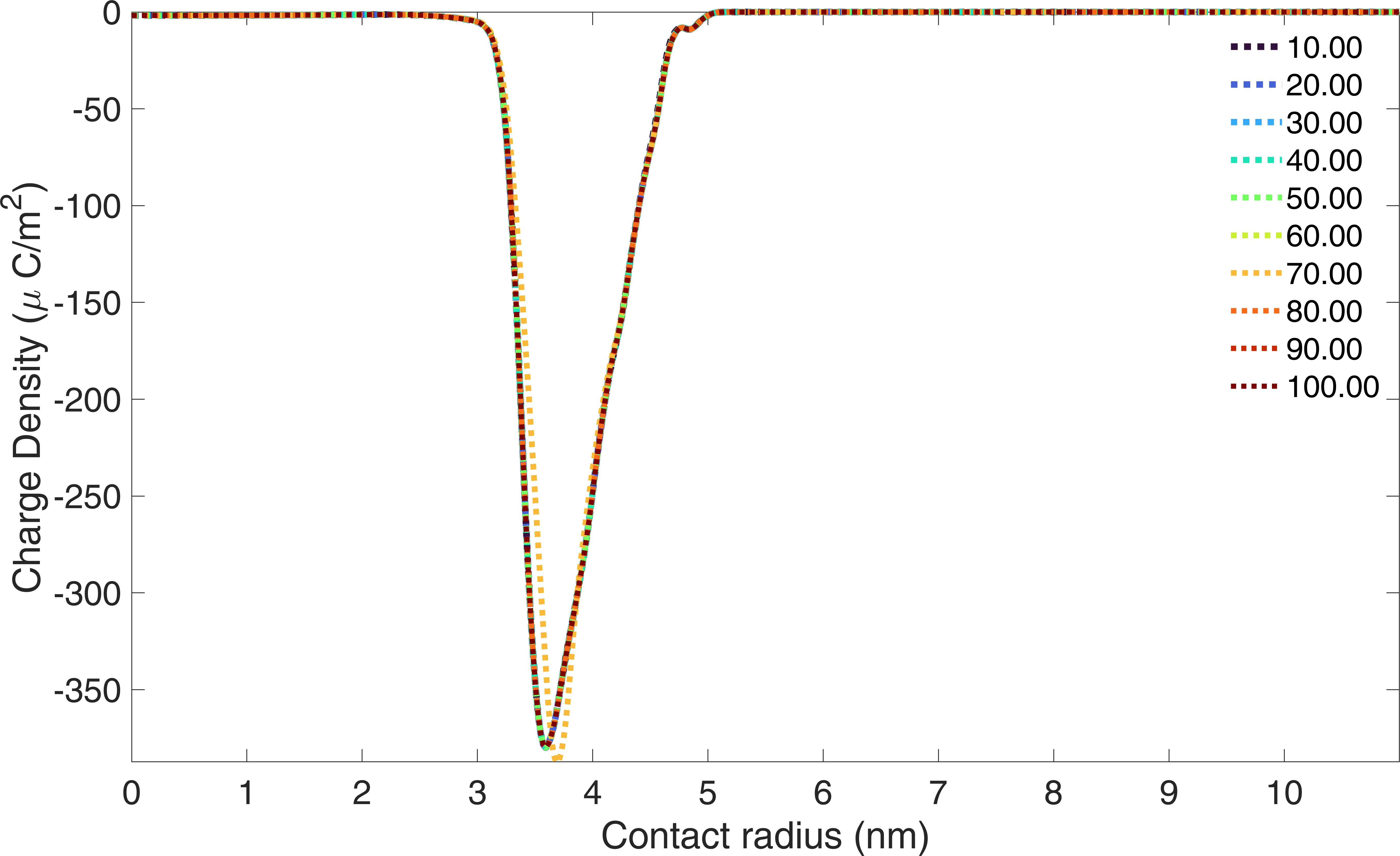}
		\caption{PMMA (negative bias)}
	\end{subfigure}
	\hfill
	\begin{subfigure}{0.48\textwidth}
		\includegraphics[width=\textwidth]{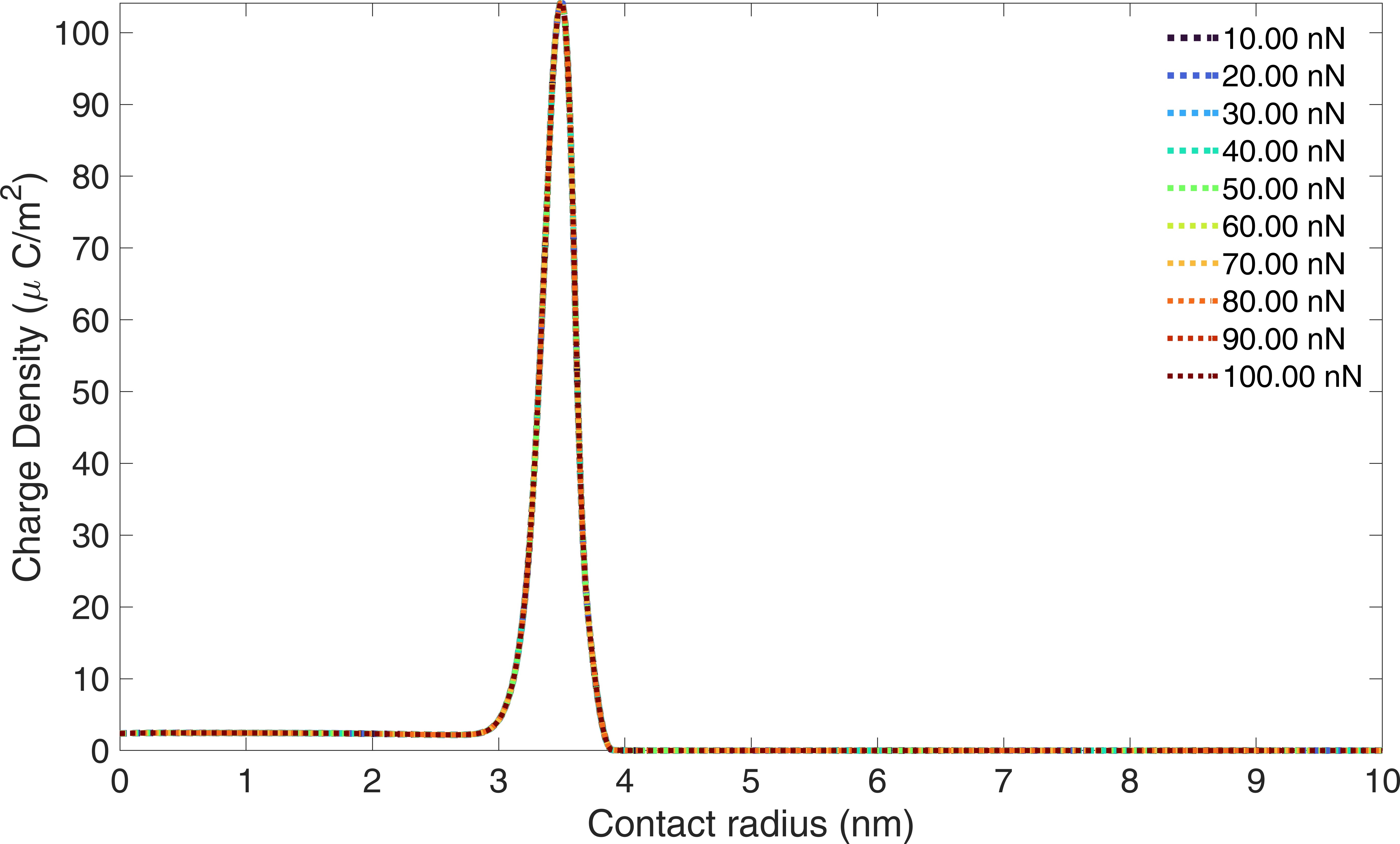}
		\caption{PDAP (positive bias)}
	\end{subfigure}
	\caption{Residual surface charge density distribution after complete separation under biased conditions: (a) PMMA; (b) PDAP. Note the near-overlap of curves for different loading forces.}
	\label{fig:unloading_charge_biased}
\end{figure}

\begin{figure}[!htbp]
	\centering
	\begin{subfigure}{0.48\textwidth}
		\includegraphics[width=\textwidth]{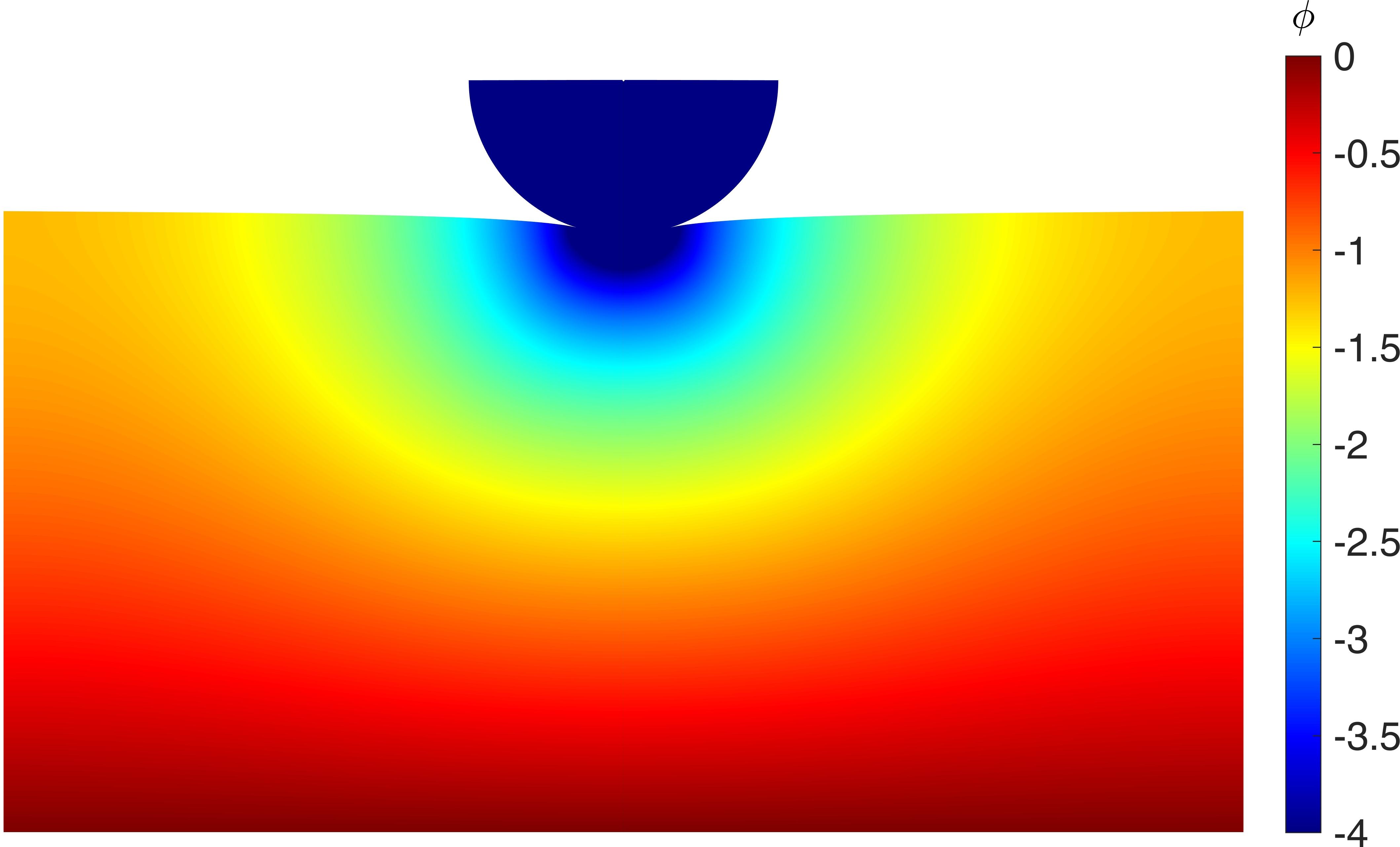}
		\caption{PMMA, loading}
	\end{subfigure}
	\hfill
	\begin{subfigure}{0.48\textwidth}
		\includegraphics[width=\textwidth]{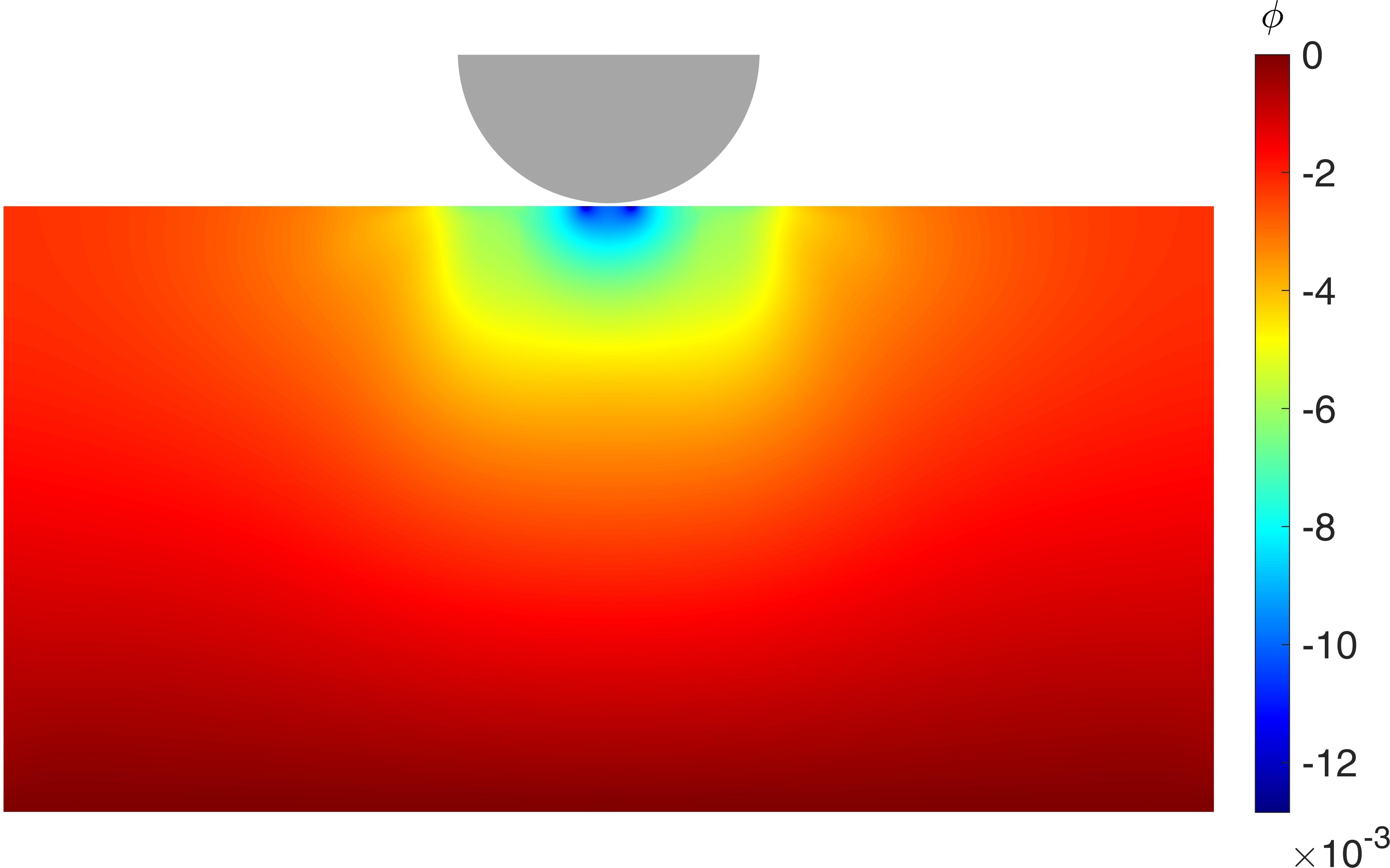}
		\caption{PMMA, separation}
	\end{subfigure}
	\\
	\begin{subfigure}{0.48\textwidth}
		\includegraphics[width=\textwidth]{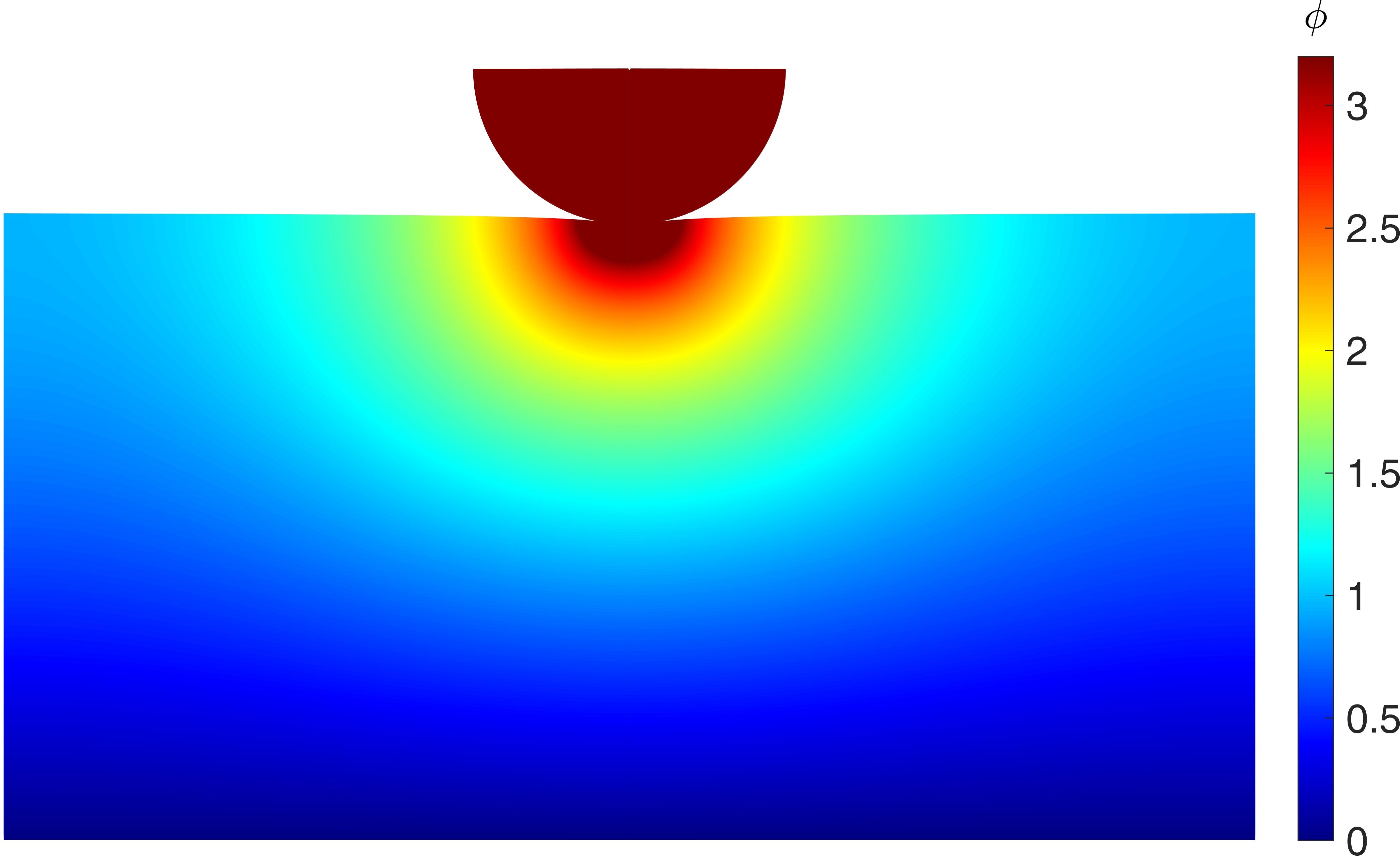}
		\caption{PDAP, loading}
	\end{subfigure}
	\hfill
	\begin{subfigure}{0.48\textwidth}
		\includegraphics[width=\textwidth]{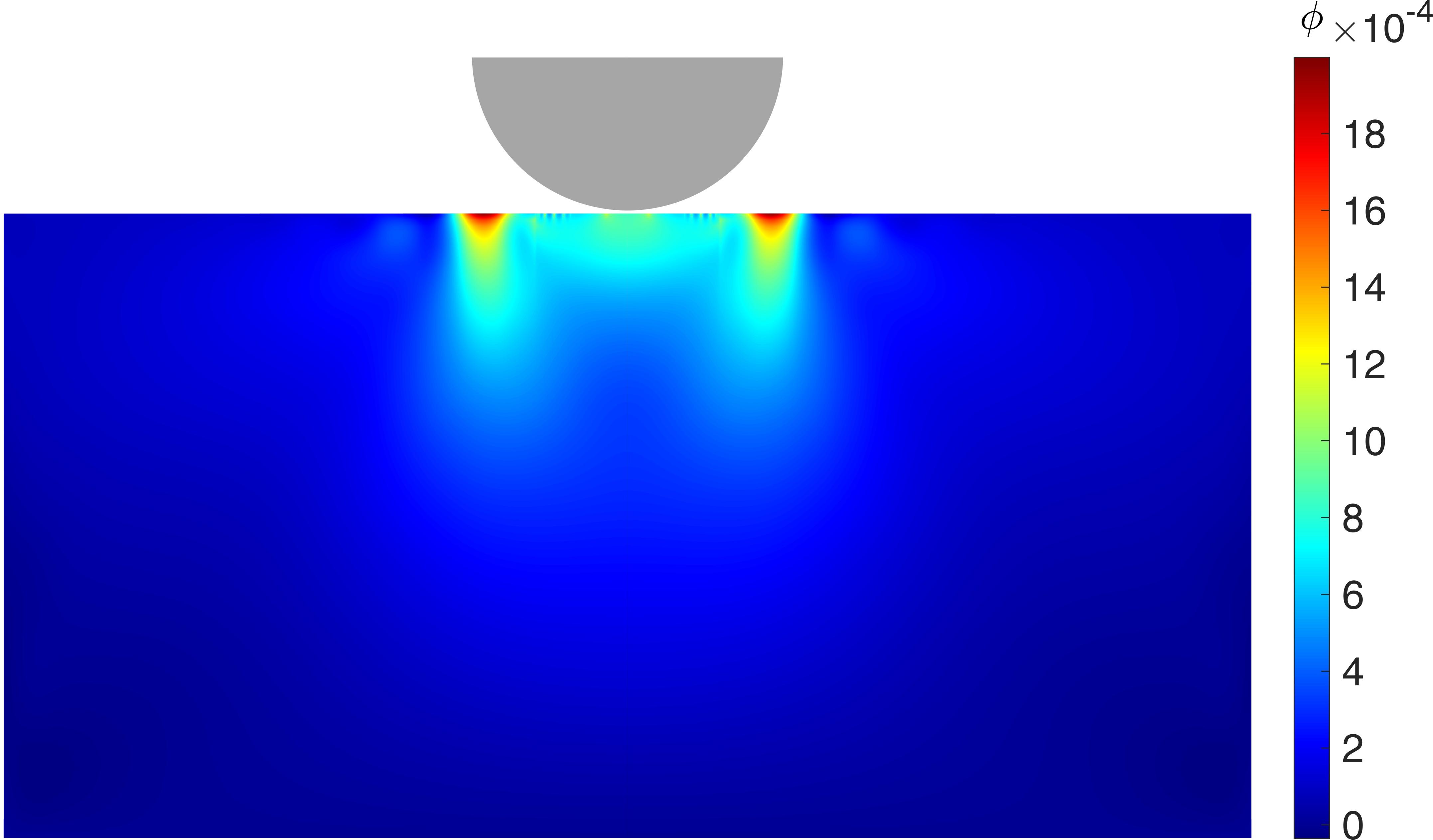}
		\caption{PDAP, separation}
	\end{subfigure}
	\caption{Electric potential distribution at 100~nN contact force under biased conditions: (a,c) at the end of loading; (b,d) after complete separation.}
	\label{fig:potential_biased}
\end{figure}

The residual charge distribution after complete separation is shown in Fig.~\ref{fig:unloading_charge_biased}. An important feature is that the charge distributions for different loading forces nearly overlap, indicating that the residual charge is almost independent of the maximum loading force. This behavior contrasts with the unbiased case and can be attributed to the single-polarity constraint. Because only one polarity of charge can be transferred, the extended contact edge region remains uncharged during loading, and the residual charge is significantly lower than in the unbiased case. In the unbiased case, the unrestricted charge supply fully satisfies the bipolar flexoelectric demand, allowing significant peripheral charge of the opposite polarity to freeze upon separation.
Figure~\ref{fig:potential_biased} shows the electric potential distribution at the end of the loading phase and after complete separation for a maximum loading force of 100~nN. The small residual potential after separation, in contrast to the applied bias magnitude during loading, reflects the limited charge transfer through the polarity-restricted pathway.

\begin{figure}[!htbp]
	\centering
	\begin{subfigure}{0.48\textwidth}
		\includegraphics[width=\textwidth]{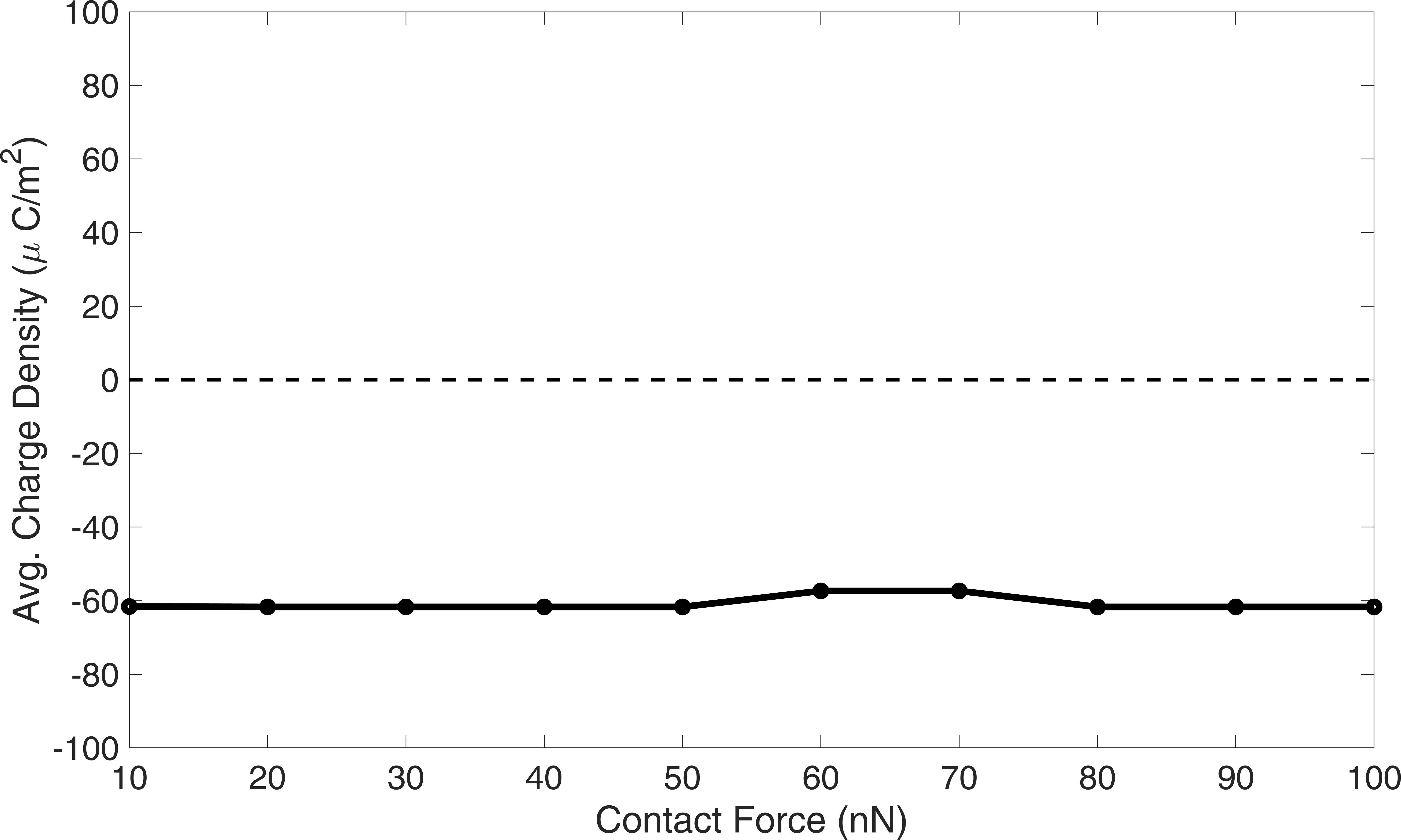}
		\caption{PMMA simulation}
	\end{subfigure}
	\hfill
	\begin{subfigure}{0.48\textwidth}
		\includegraphics[width=\textwidth]{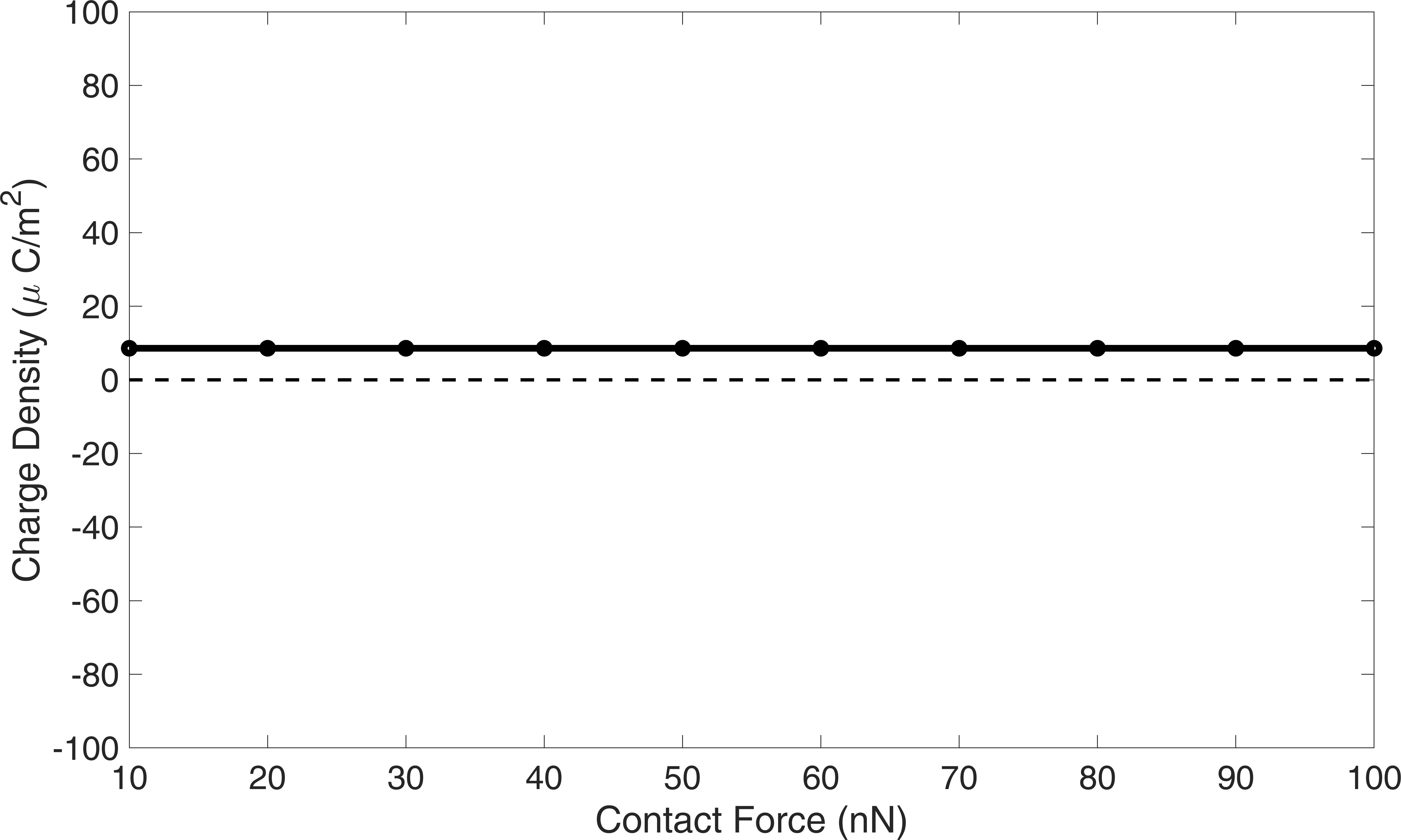}
		\caption{PDAP simulation}
	\end{subfigure}
	\caption{Average surface charge density as a function of contact force under biased conditions: (a) PMMA; (b) PDAP.}
	\label{fig:exp_comparison_biased}
\end{figure}

The computed average surface charge density for biased contact is shown in Fig.~\ref{fig:exp_comparison_biased}, and the corresponding experimental data are given in Fig.~\ref{fig:exp_comparison_unbiased}(b,d). 
The simulation results agree with experimental observations that the residual charge follows the polarity of the applied bias, opposite to that in the unbiased case.

\subsection{Effect of tip radius and characteristic tunneling length}
\label{sec:parametric}

We further investigate the effect of tip radius $R$ and characteristic tunneling length $\ell_q$ on the charge distribution, using PDAP in the unbiased contact scenario as a representative case.

\begin{figure}[!htbp]
	\centering
	\includegraphics[width=0.65\textwidth]{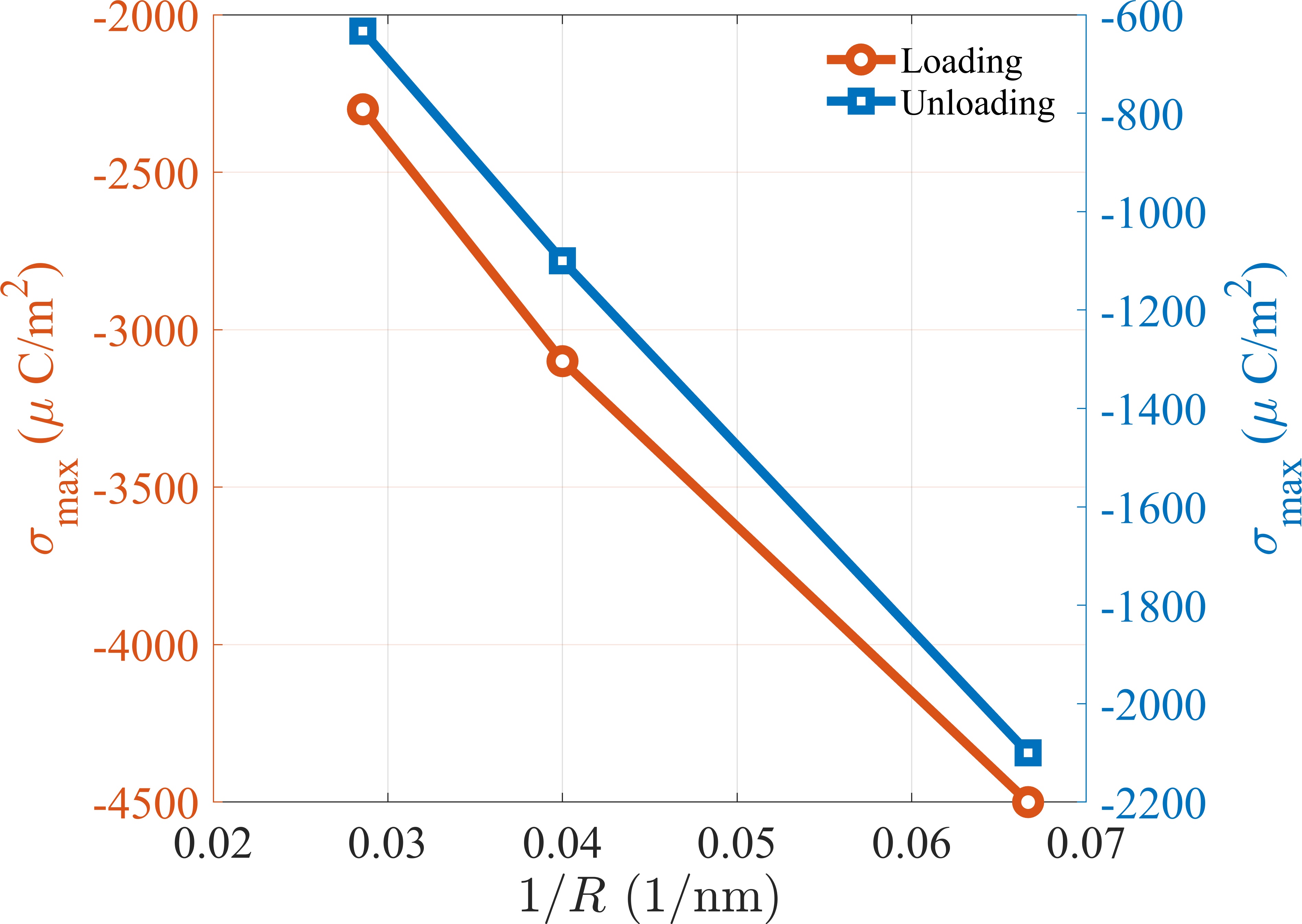}
	\caption{Peak charge density as a function of inverse tip radius $1/R$ for PDAP in unbiased contact at loading force of 100~nN.}
	\label{fig:R_effect_peak}
\end{figure}

\begin{figure}[!htbp]
	\centering
	\begin{subfigure}{0.48\textwidth}
		\includegraphics[width=\textwidth]{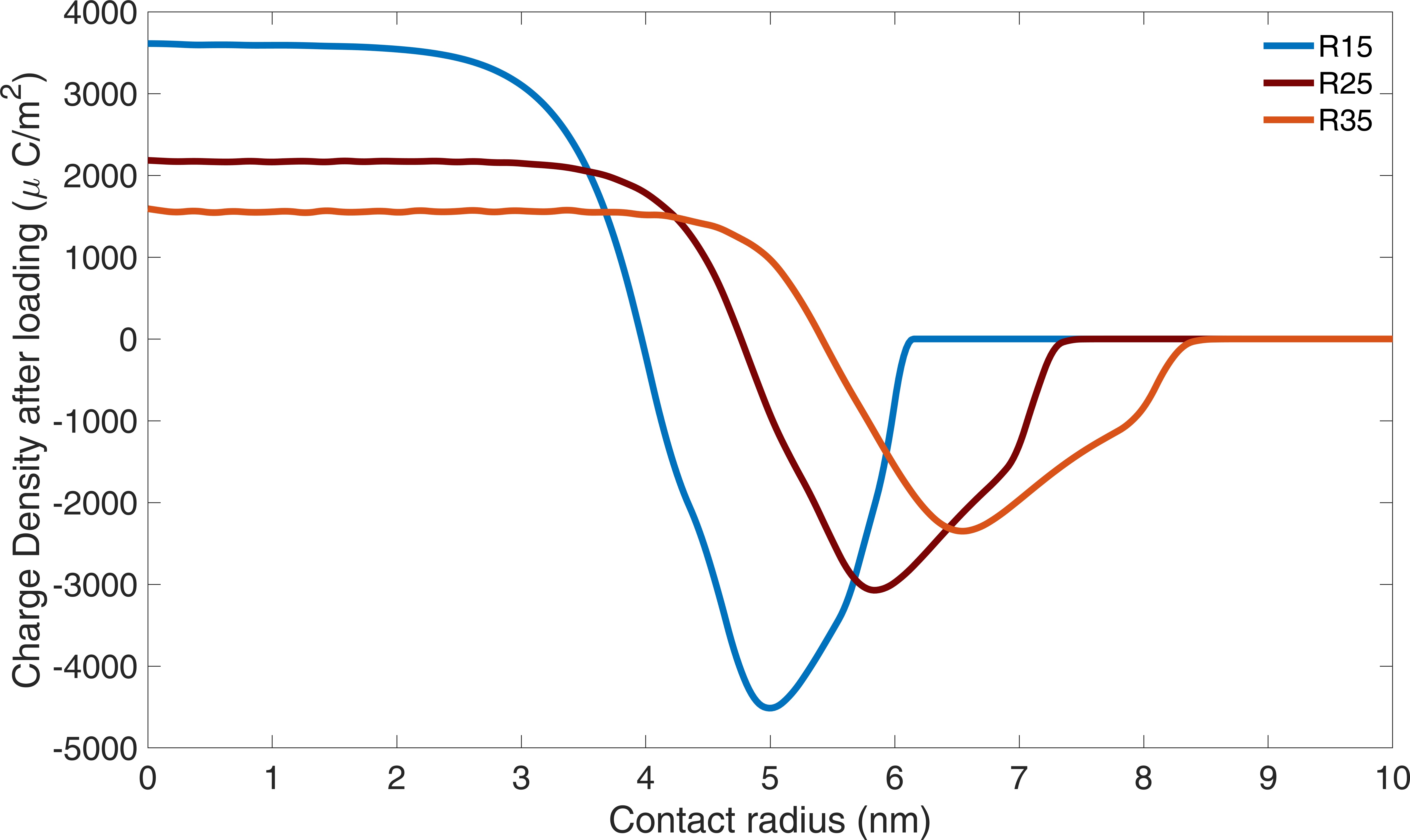}
		\caption{Charge distribution at end of loading}
	\end{subfigure}
	\hfill
	\begin{subfigure}{0.48\textwidth}
		\includegraphics[width=\textwidth]{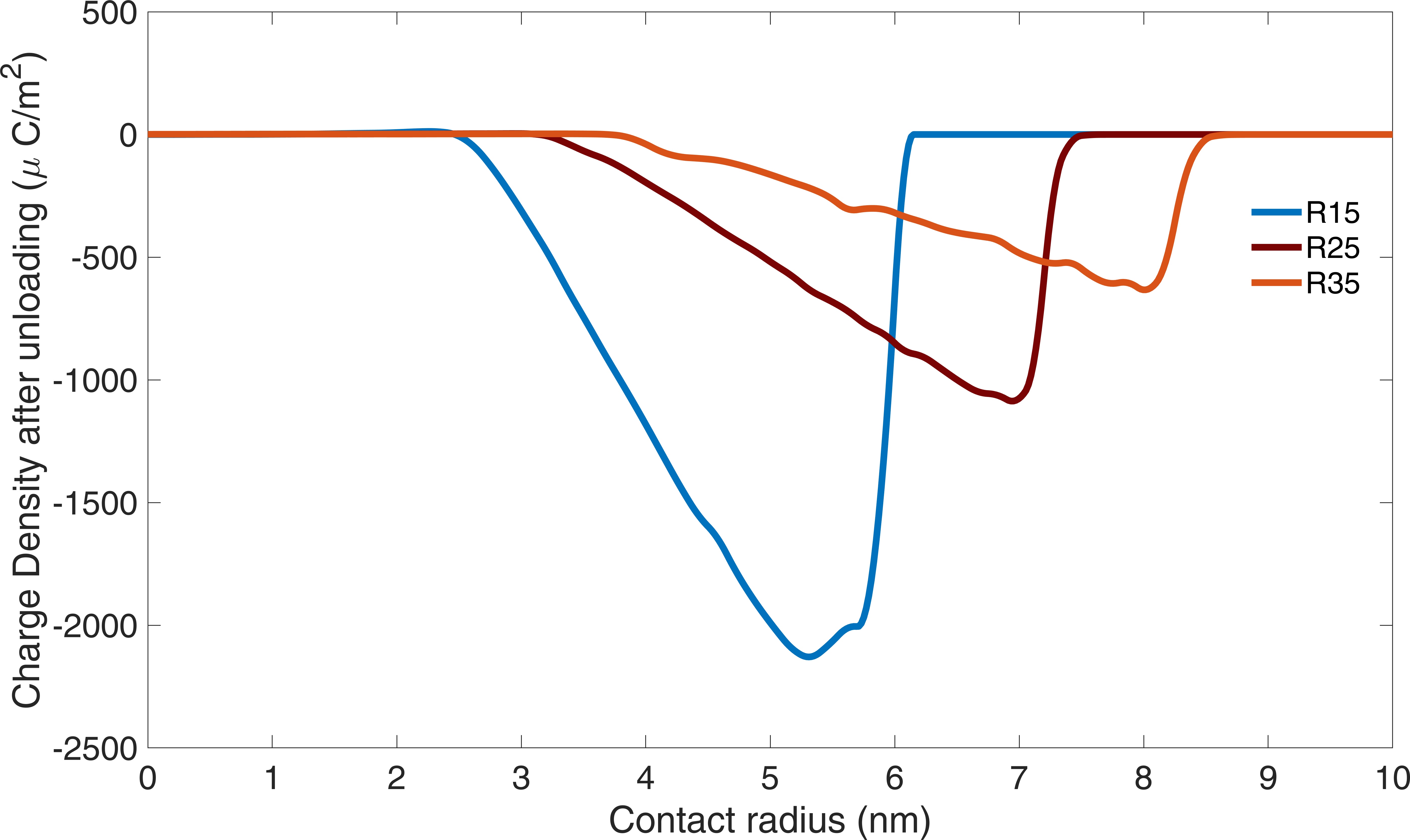}
		\caption{Residual charge distribution after separation}
	\end{subfigure}
	\caption{Spatial charge distributions for varying tip radius $R$ (unit: nm): (a) at maximum indentation; (b) after complete unloading, for PDAP in unbiased contact at loading force of 100~nN.}
	\label{fig:R_effect_dist}
\end{figure}

Figure~\ref{fig:R_effect_peak} shows the peak charge density when varying the inverse tip radius $1/R$. A clear positive correlation is observed: the peak charge density increases approximately linearly with $1/R$, suggesting that tip sharpness directly correlates with charge density. This trend arises because, for a fixed contact force, a sharper tip not only concentrates the deformation into a smaller contact area but also induces a significantly steeper spatial profile of displacement beneath the indenter. Consequently, local strain gradients are strongly amplified. Since the flexoelectric polarization scales with strain gradient, the driving force for charge transfer is enhanced for sharper tips.
The corresponding charge distributions are presented in Fig.~\ref{fig:R_effect_dist}. During the loading phase (Fig.~\ref{fig:R_effect_dist}(a)), the charge profiles exhibit characteristic bipolar patterns for all tip radii, with sharper peaks occurring at smaller $R$ due to the more concentrated strain gradients. After separation (Fig.~\ref{fig:R_effect_dist}(b)), the residual charge distributions retain similar shapes but with reduced magnitudes, as partial charge backflow occurs during unloading. Notably, the residual charge for larger tips is more spread out laterally due to the larger contact area.

\begin{figure}[!htbp]
	\centering
	\includegraphics[width=0.65\textwidth]{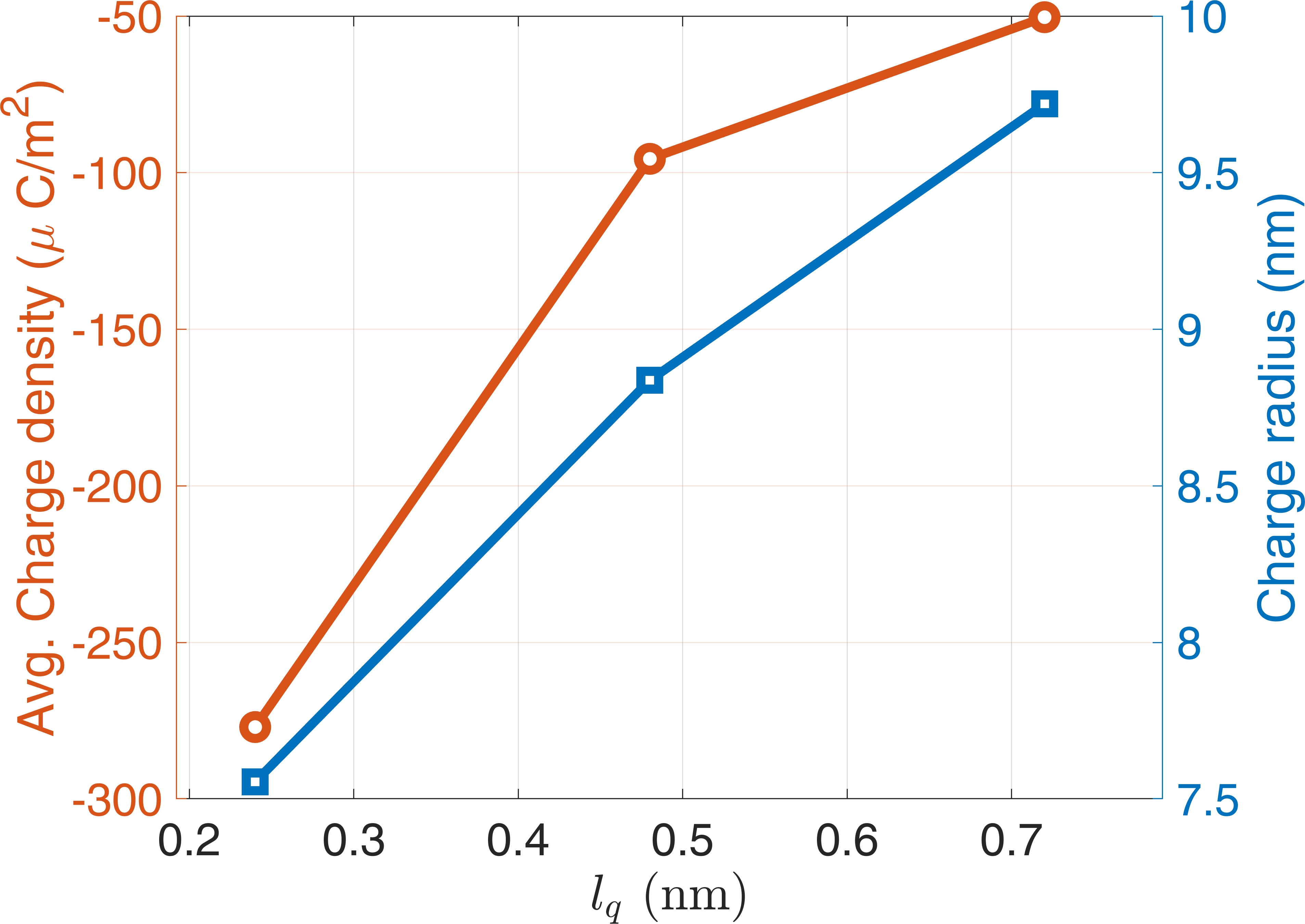}
	\caption{Average charge density and charged radius as functions of characteristic tunneling length $\ell_q$ for PDAP at 100~nN contact force.}
	\label{fig:lq_effect_trend}
\end{figure}

The tunneling length $\ell_q$ governs how rapidly the charge transfer rate decays with gap distance, as given by Eq.~\eqref{eq:transparency}. Figure~\ref{fig:lq_effect_trend} presents the average charge density and charged radius as functions of $\ell_q$. The charged radius increases monotonically with $\ell_q$, as a larger characteristic length extends the effective range of charge transfer at the contact periphery. In contrast, the magnitude of the average charge density decreases monotonically with increasing $\ell_q$. This reduction is attributed to the unloading phase: a larger $\ell_q$ maintains a conductive tunneling channel over larger separation distances, thereby allowing more accumulated charge to flow back before the channel effectively closes. Consequently, a smaller $\ell_q$ traps surface charge more effectively.

\begin{figure}[!htbp]
	\centering
	\begin{subfigure}{0.48\textwidth}
		\includegraphics[width=\textwidth]{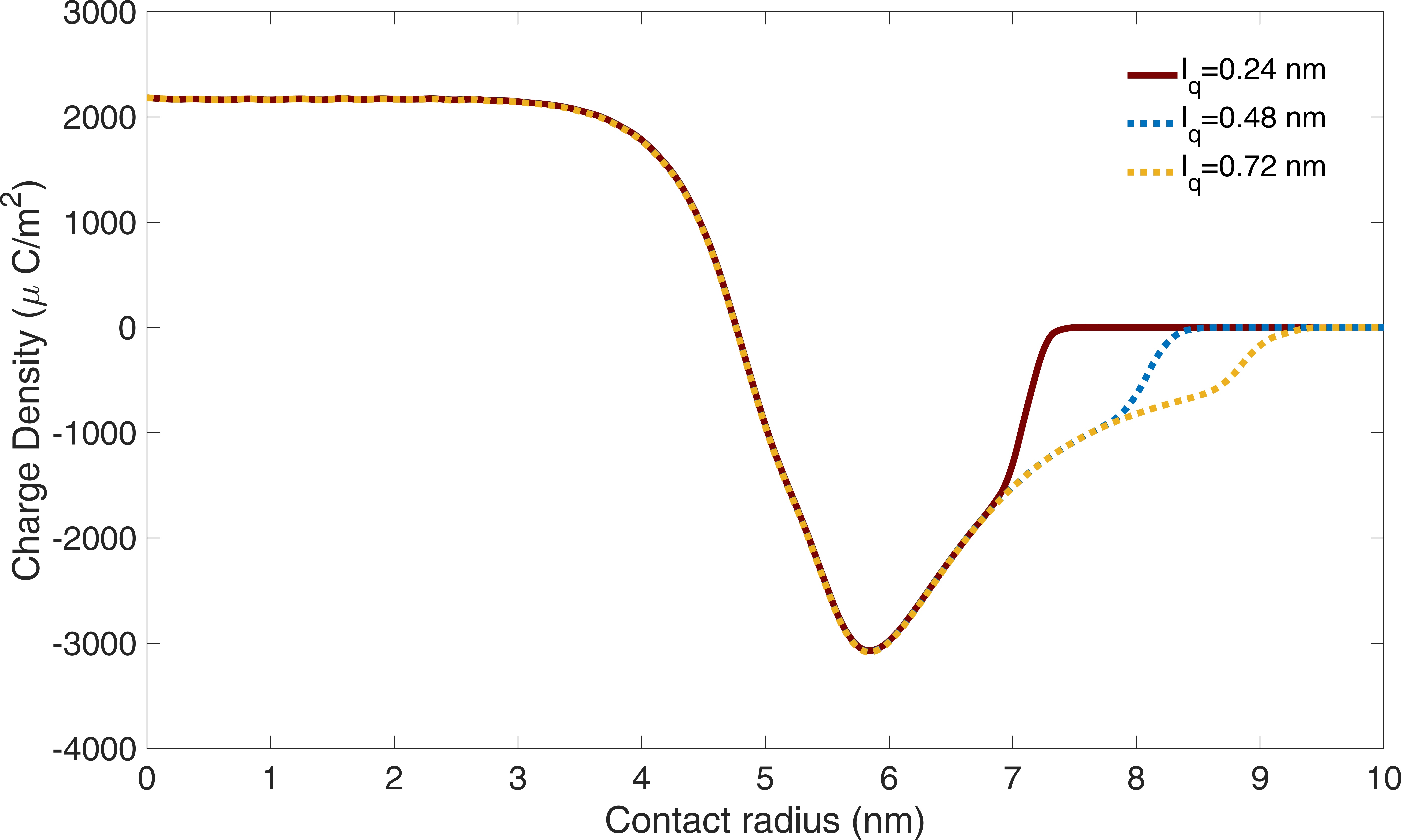}
		\caption{}
	\end{subfigure}
	\hfill
	\begin{subfigure}{0.48\textwidth}
		\includegraphics[width=\textwidth]{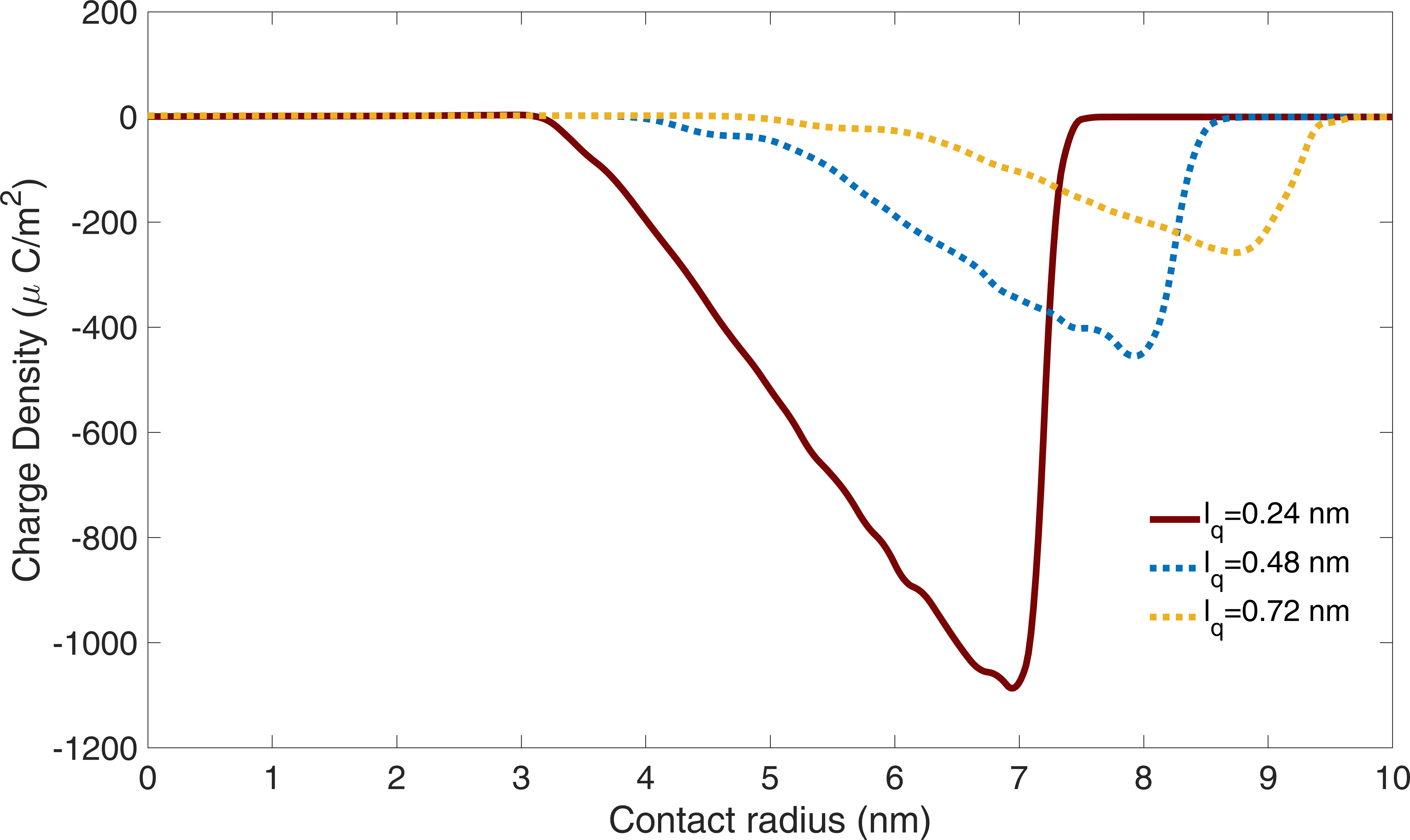}
		\caption{}
	\end{subfigure}
	\caption{Spatial charge distributions for varying tunneling length $\ell_q$: (a) at maximum indentation; (b) after complete separation for PDAP at 100~nN contact force.}
	\label{fig:lq_effect_dist}
\end{figure}

The spatial profiles in Fig.~\ref{fig:lq_effect_dist} further illustrate these effects. At maximum loading (Fig.~\ref{fig:lq_effect_dist}(a)), all curves show similar central charge densities, but larger $\ell_q$ values result in charge distributions that extend further beyond the nominal contact edge. After separation (Fig.~\ref{fig:lq_effect_dist}(b)), smaller $\ell_q$ values yield higher residual charge because the tunneling channel closes earlier upon unloading, thereby ``trapping'' the charge. In contrast, larger $\ell_q$ maintains an open channel longer, allowing more charge to flow back before separation is complete.

\section{Contact Electrification between Identical Dielectrics}
\label{sec:application}
Contact electrification between identical materials has long posed a challenge to classical theories because standard electron transfer models rely on intrinsic differences in work function or chemical potential, which are absent across symmetric interfaces \cite{Lacks2011,Xu2019curvature}. While this phenomenon has been observed for more than a century \cite{jamieson1910electrification} and extensively studied \cite{Lowell1986, apodaca2010contact, Baytekin2011mosaic}, its underlying mechanism remains debated. Recently, experimental studies \cite{Xu2019curvature} have linked charge polarity to local surface curvature, suggesting that the symmetry breaking required for electrification may originate from the physical state of the contact itself. In this section, we apply our computational model to dielectric-dielectric contact electrification and show that geometric asymmetry and flexoelectric effects together can induce charge separation even between two contacting bodies with identical material properties.

\subsection{Two-dimensional wavy surface contact}
\label{sec:wavy_contact}
\begin{figure}[!htbp]
	\centering
	\includegraphics[width=0.5\textwidth]{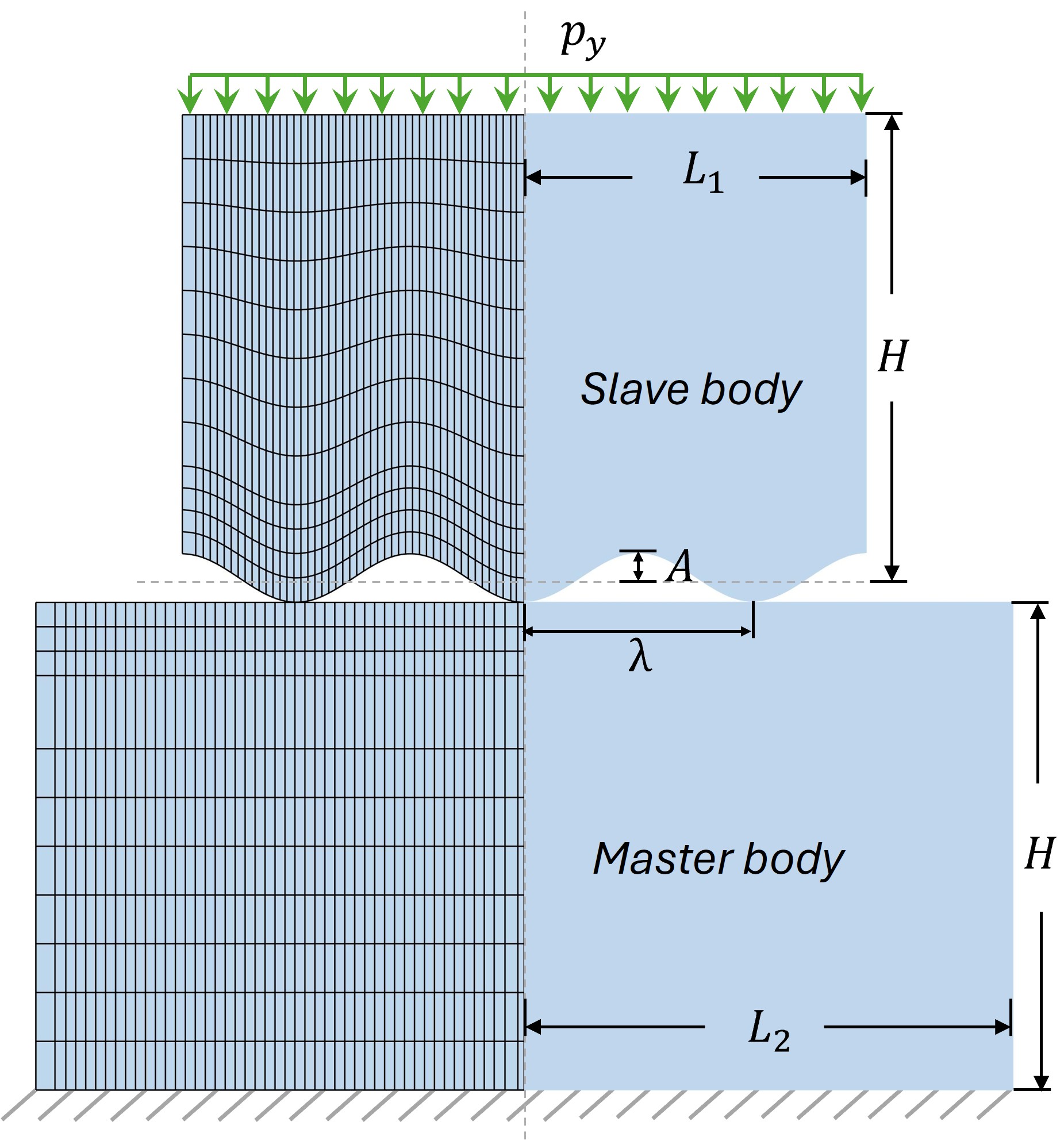}
	\caption{Schematic of wavy surface contact geometry and mesh (wavenumber $k = L_1/\lambda = 1.5$). The slave body has a wavy surface, while the master body has a flat surface.}
	\label{fig:wavy_schematic}
\end{figure}

We consider contact between a flat lower PMMA body and a wavy upper PMMA body characterized by a wavenumber parameter $k$ that controls the density of asperities. The geometric parameters are set to $L_1 = 70$~nm, $L_2 = 100$~nm, $H = 100$~nm, $A = 5$~nm, with an applied vertical pressure $p_y = 500$~MPa. Figure~\ref{fig:wavy_schematic} illustrates the geometry and mesh for a representative wavenumber $k = 1.5$. The simulation employs the same PMMA material parameters as in Section~\ref{sec:validation}, but introduces a surface state capacity parameter $\sigma_\mathrm{trap} = 100$~$\mu$C/m$^2$ to impose a physical limit on the maximum transferable charge density at dielectric-dielectric interfaces. This value is conservative relative to the experimentally reported surface state densities of $10^{12}$--$10^{13}$~cm$^{-2}$~\cite{Xu2018electron}, and is chosen so that the saturation regime of charge transfer can be reached within the simulated loading conditions.

\begin{figure}[!htbp]
	\centering
	\begin{subfigure}{\textwidth}
		\includegraphics[width=\textwidth]{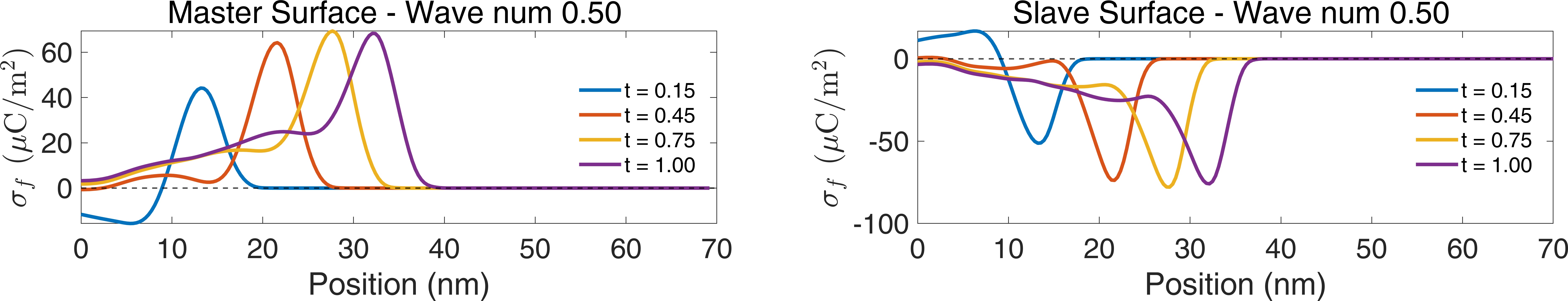}
		\caption{}
	\end{subfigure}
	\hfill
	\begin{subfigure}{\textwidth}
		\includegraphics[width=\textwidth]{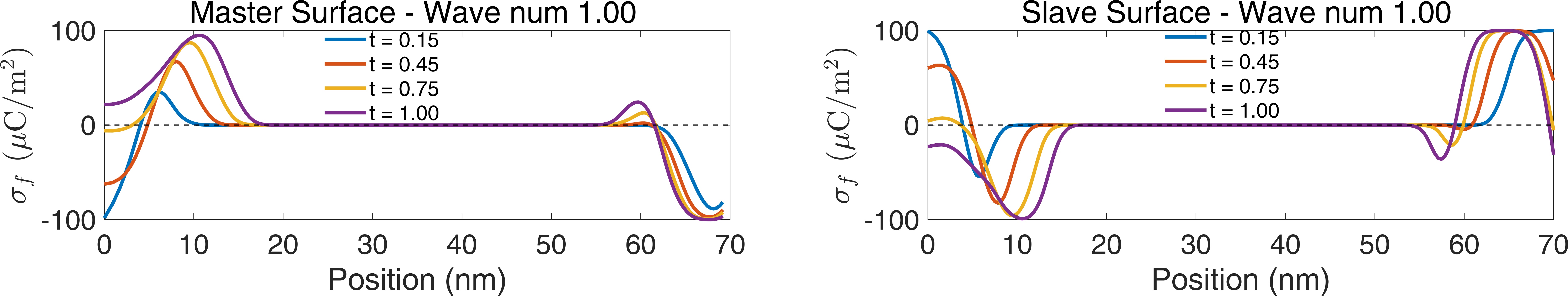}
		\caption{}
	\end{subfigure}
	\hfill
	\begin{subfigure}{\textwidth}
		\includegraphics[width=\textwidth]{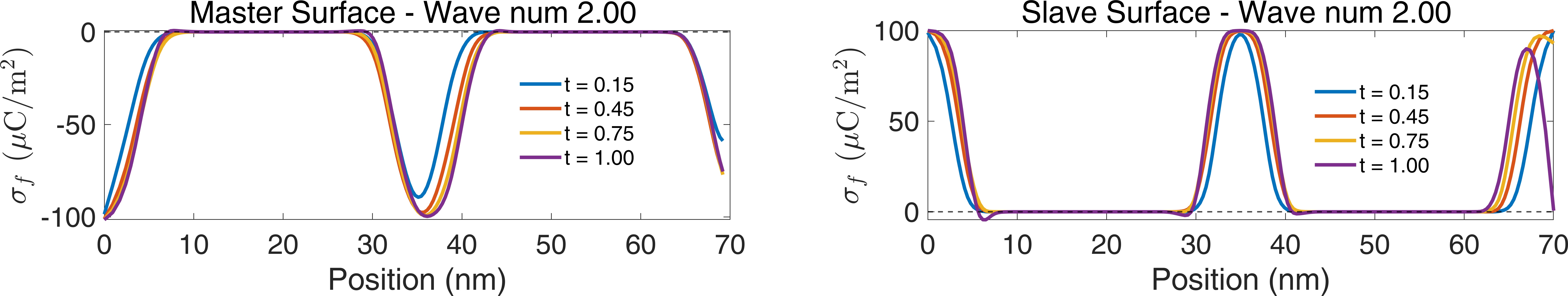}
		\caption{}
	\end{subfigure}
	\caption{Surface charge density distribution during loading for different wavenumbers: (a) $k = 0.5$; (b) $k = 1.0$; (c) $k = 2.0$.}
	\label{fig:wavy_loading}
\end{figure}

The evolution of surface charge distribution during loading reveals how local geometry dictates polarity. Figure~\ref{fig:wavy_loading} presents the charge density profiles at different loading stages for three distinct wavenumbers. For a low wavenumber ($k = 0.5$, Fig.~\ref{fig:wavy_loading}(a)), contact is localized at the center. Here, the convex slave surface approaches the flat master surface, inducing a negative charge on the slave and a positive charge on the master. This polarity aligns with the experimental ``curvature rule'' reported by Xu et al. \cite{Xu2019curvature}, where they found that convex surfaces in contact acquire negative charge while concave surfaces become positively charged. However, as the wavenumber increases to $k = 1.0$ (Fig.~\ref{fig:wavy_loading}(b)), a second contact region emerges at the edge (around 70~nm). The polarity at this edge contact is opposite to that at the center: the master surface gains negative charge while the center remains positive. This spatial variation indicates that charge transfer direction is governed locally by the specific geometric conditions at each contact point rather than by a global material property. At a high wavenumber ($k = 2.0$, Fig.~\ref{fig:wavy_loading}(c)), multiple contact points develop, and the charge density rapidly saturates to the trap limit $\sigma_\mathrm{trap}$. Notably, the global polarity pattern reverses compared to the low wavenumber case, with the master surface becoming predominantly negatively charged at the central and interior asperities as well as at the edge of the slave body.

\begin{figure}[!htbp]
	\centering
	\includegraphics[width=\textwidth]{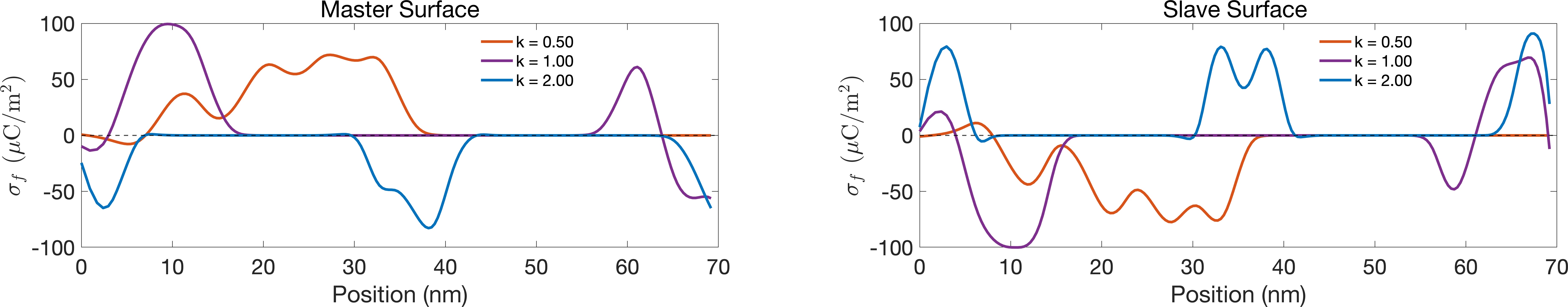}
	\caption{Residual surface charge density distribution after complete separation for different wavenumbers.}
	\label{fig:wavy_unloading}
\end{figure}

\begin{figure}[!htbp]
	\centering
	\begin{subfigure}{0.40\textwidth}
		\includegraphics[width=\textwidth]{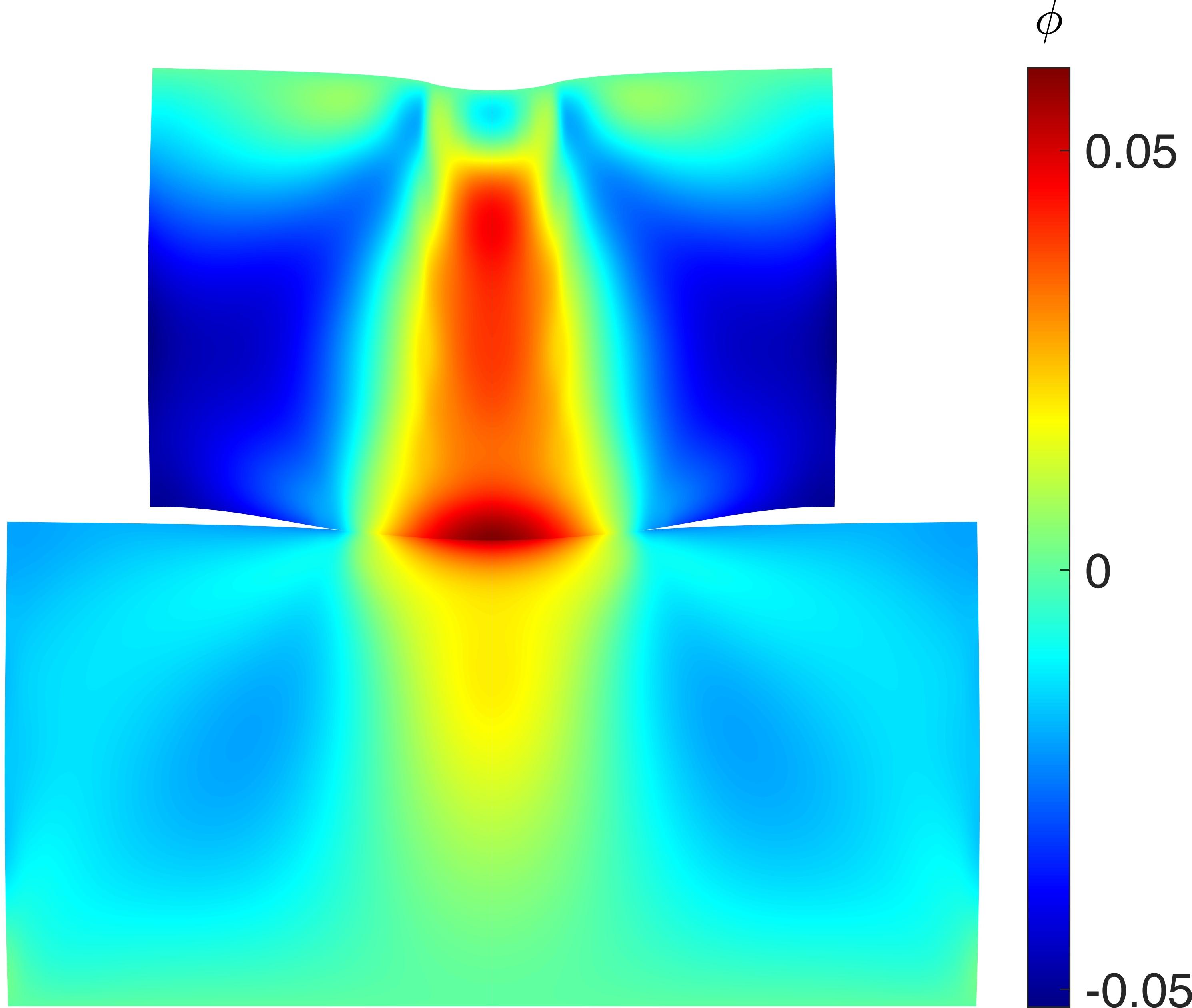}
		\caption{$k = 0.5$, loading}
	\end{subfigure}
	\hfill
	\begin{subfigure}{0.40\textwidth}
		\includegraphics[width=\textwidth]{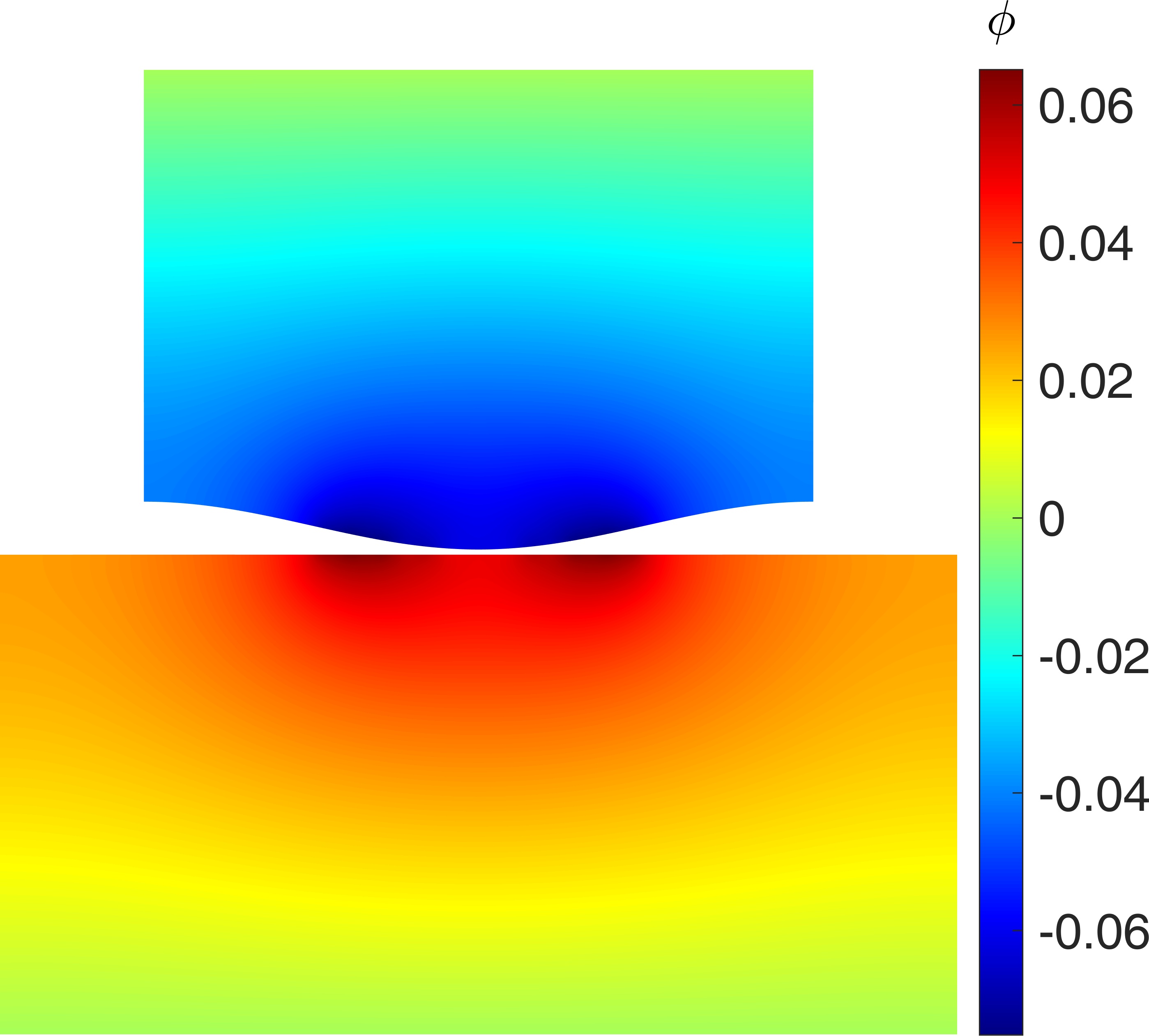}
		\caption{$k = 0.5$, unloading}
	\end{subfigure}
	\\[0.5em]
	\begin{subfigure}{0.40\textwidth}
		\includegraphics[width=\textwidth]{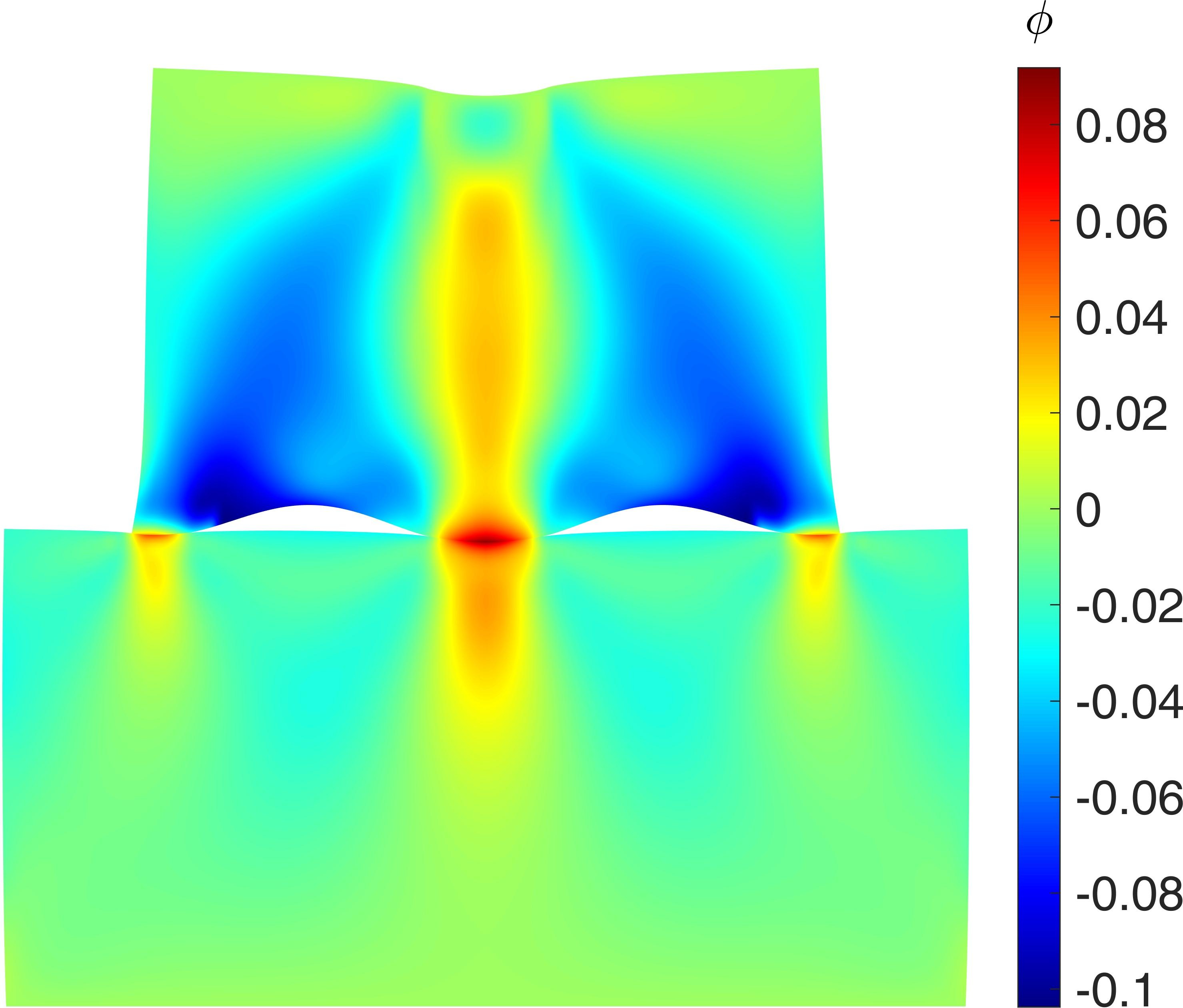}
		\caption{$k = 1.0$, loading}
	\end{subfigure}
	\hfill
	\begin{subfigure}{0.40\textwidth}
		\includegraphics[width=\textwidth]{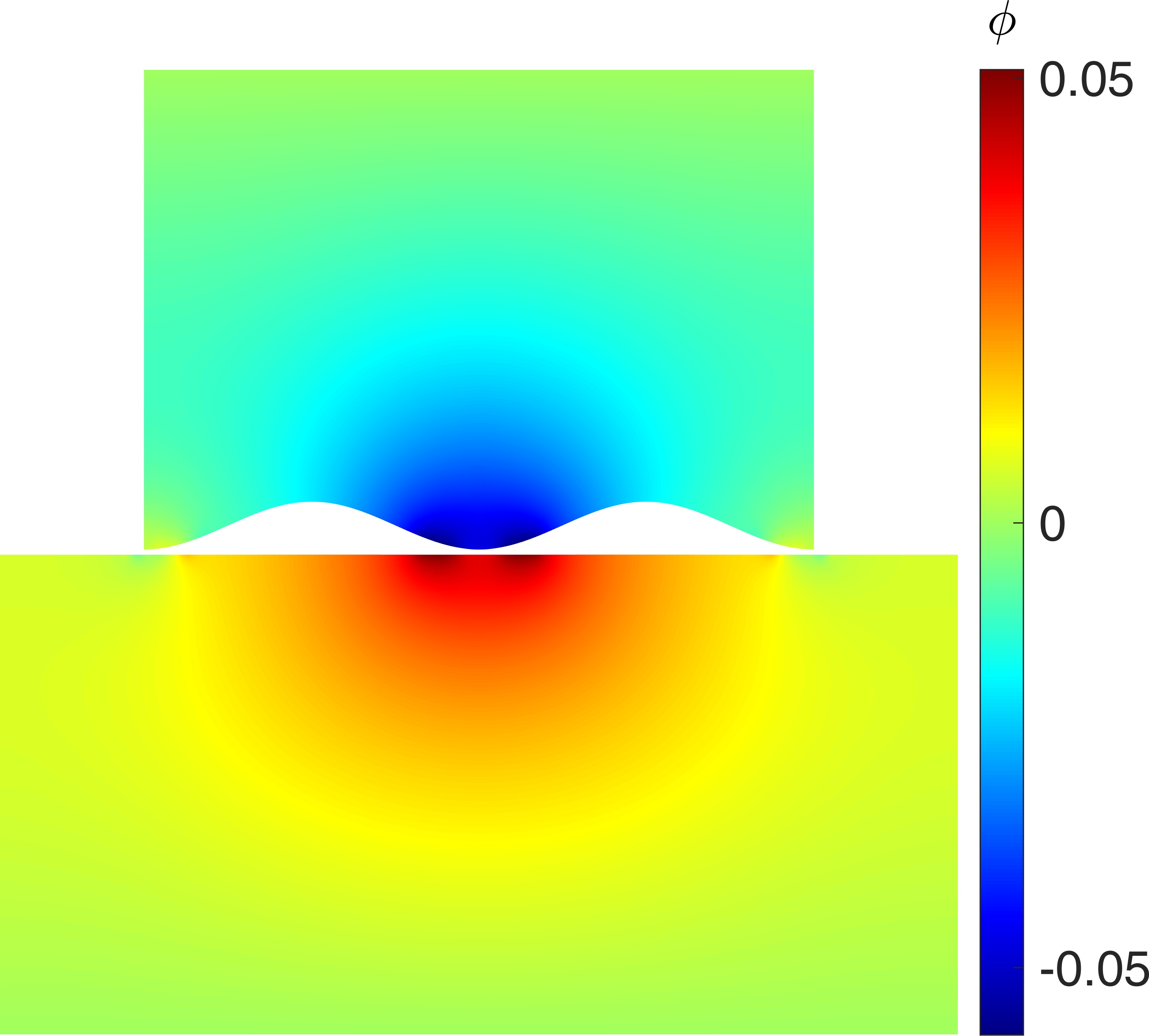}
		\caption{$k = 1.0$, unloading}
	\end{subfigure}
	\\[0.5em]
	\begin{subfigure}{0.40\textwidth}
		\includegraphics[width=\textwidth]{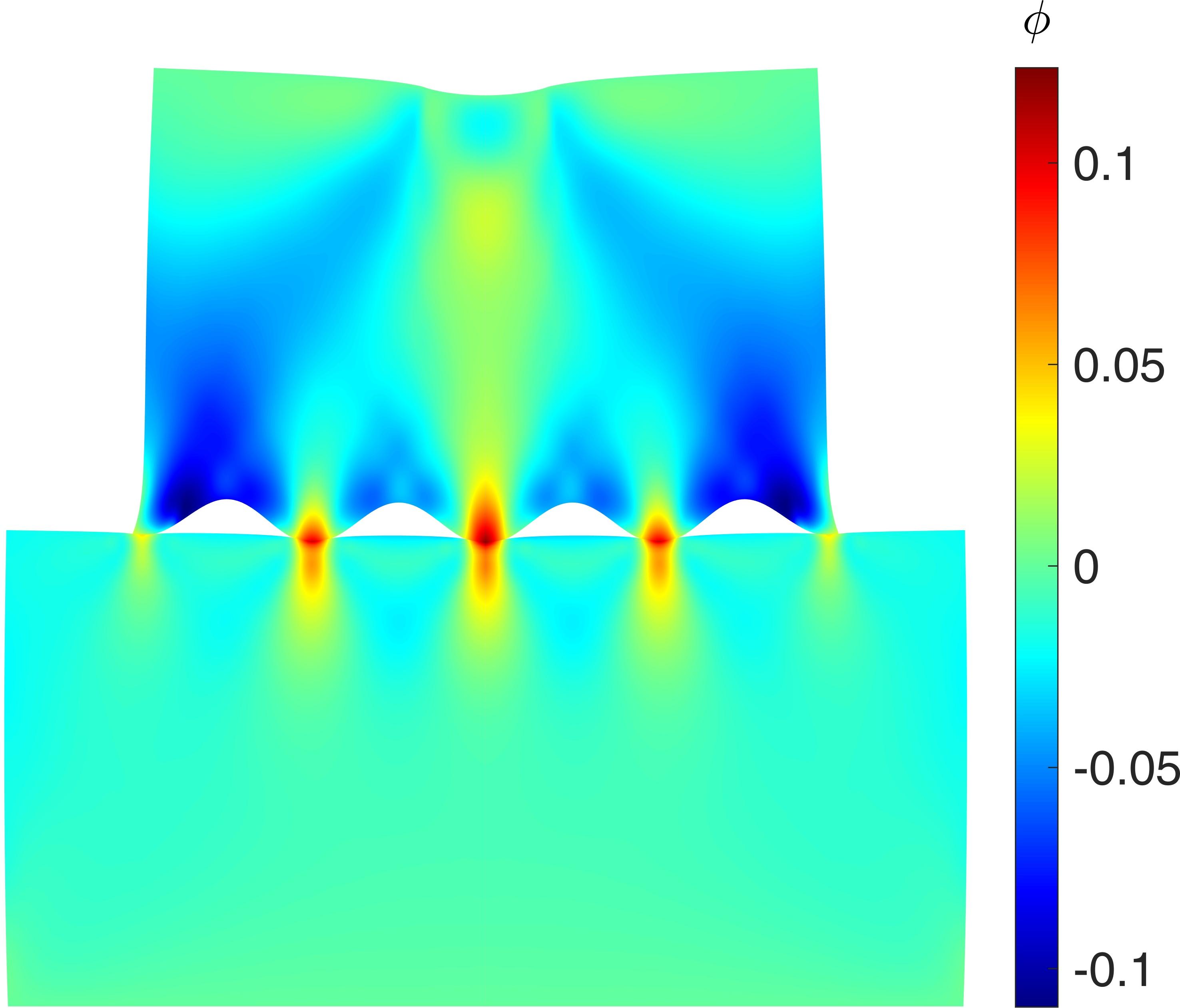}
		\caption{$k = 2.0$, loading}
	\end{subfigure}
	\hfill
	\begin{subfigure}{0.40\textwidth}
		\includegraphics[width=\textwidth]{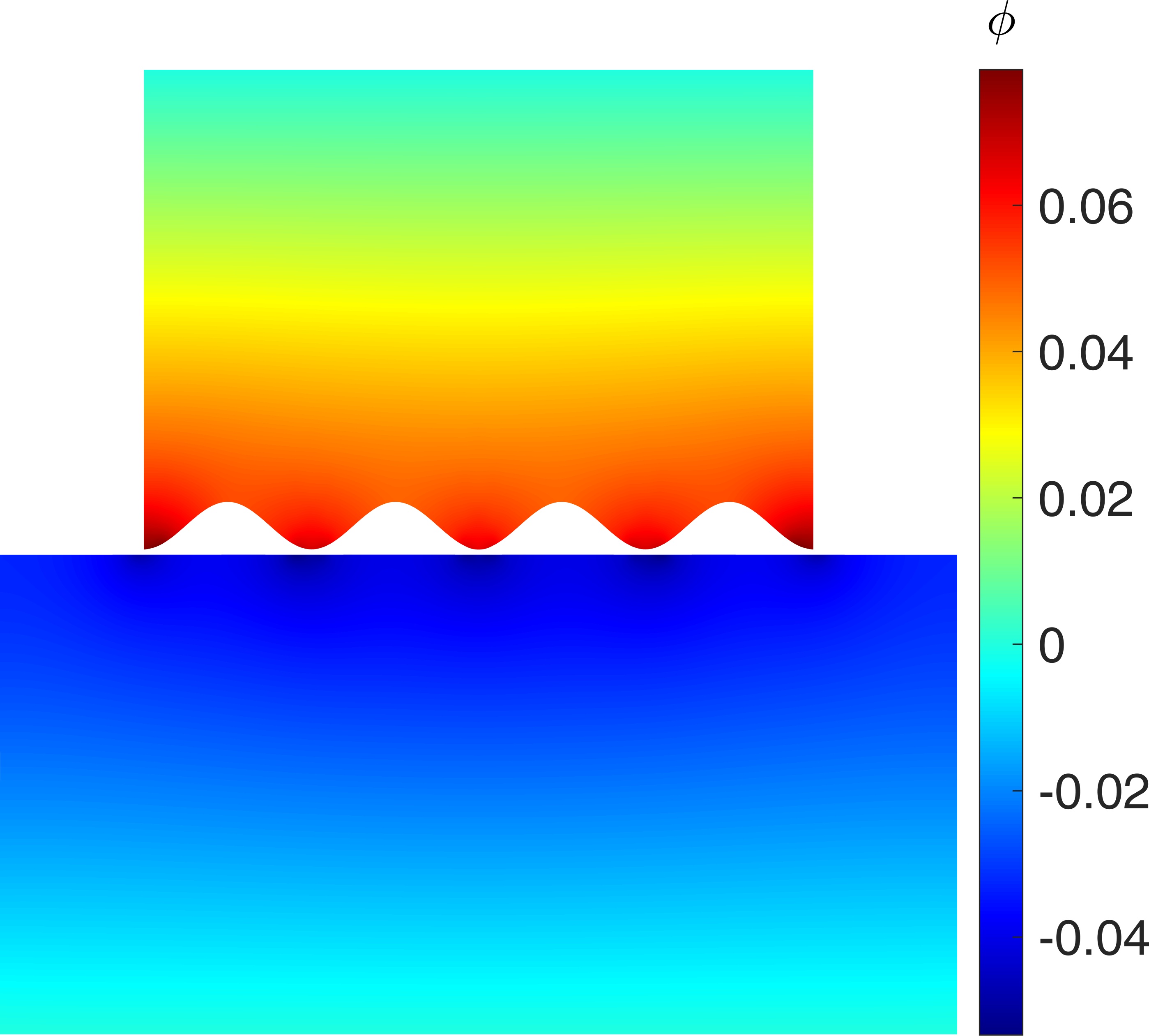}
		\caption{$k = 2.0$, unloading}
	\end{subfigure}
	\caption{Electric potential distribution for different wavenumbers: (a,b) $k = 0.5$; (c,d) $k = 1.0$; (e,f) $k = 2.0$. Left column: end of loading; right column: after separation.}
	\label{fig:wavy_potential}
\end{figure}

Upon separation, these charge patterns are largely preserved due to the freezing mechanism. Figure~\ref{fig:wavy_unloading} shows the residual charge distributions, where the polarity reversal between the low ($k = 0.5$) and high ($k = 2.0$) wavenumber cases is clearly visible. The corresponding electric potential distributions, shown in Fig.~\ref{fig:wavy_potential}, further illustrate the transition in the master body from a positive-dominant potential field at low wavenumbers to a negative-dominant field at high wavenumbers.

\begin{figure}[!htbp]
	\centering
	\begin{subfigure}{0.48\textwidth}
		\includegraphics[width=\textwidth]{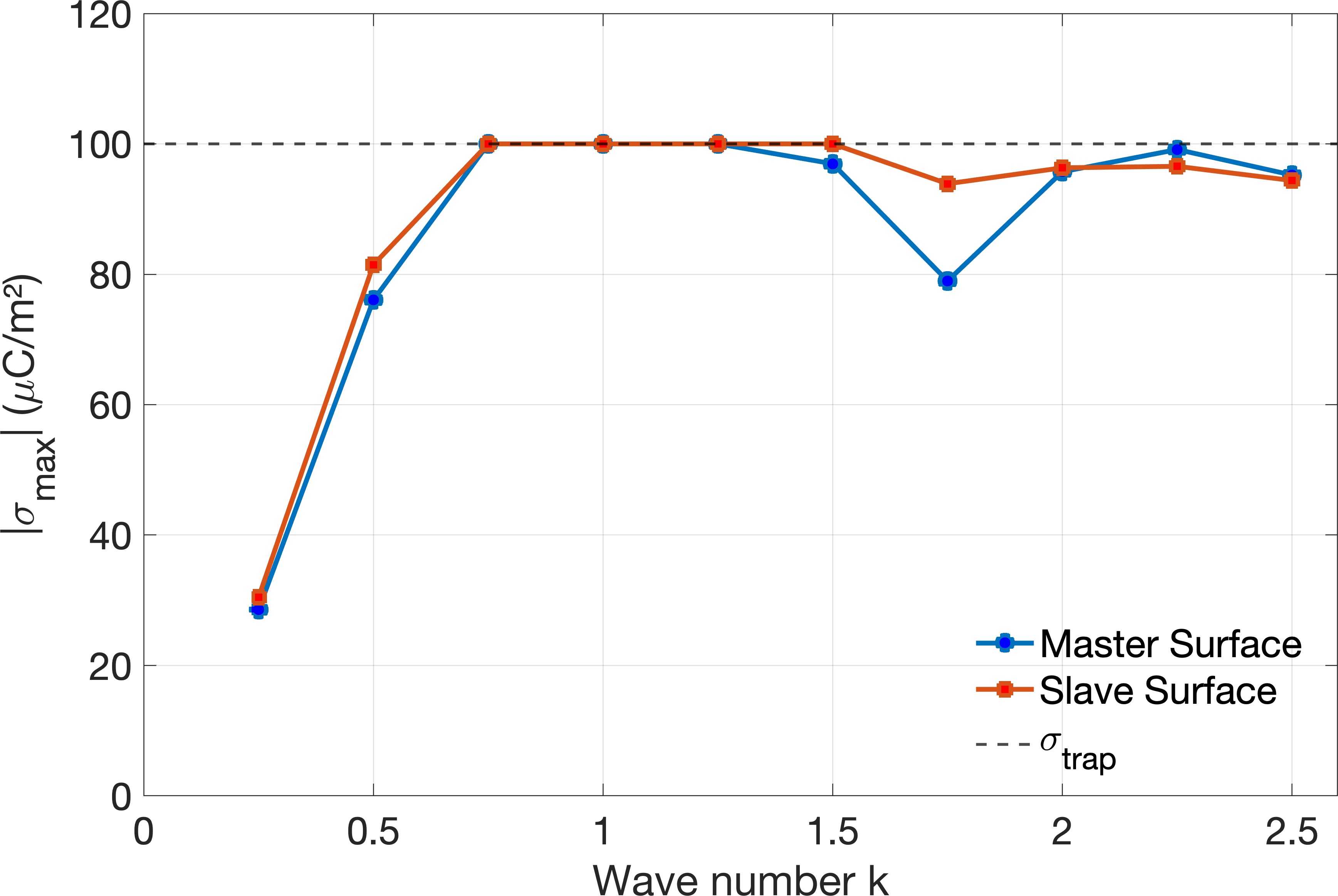}
		\caption{Peak charge density vs. wavenumber}
	\end{subfigure}
	\hfill
	\begin{subfigure}{0.48\textwidth}
		\includegraphics[width=\textwidth]{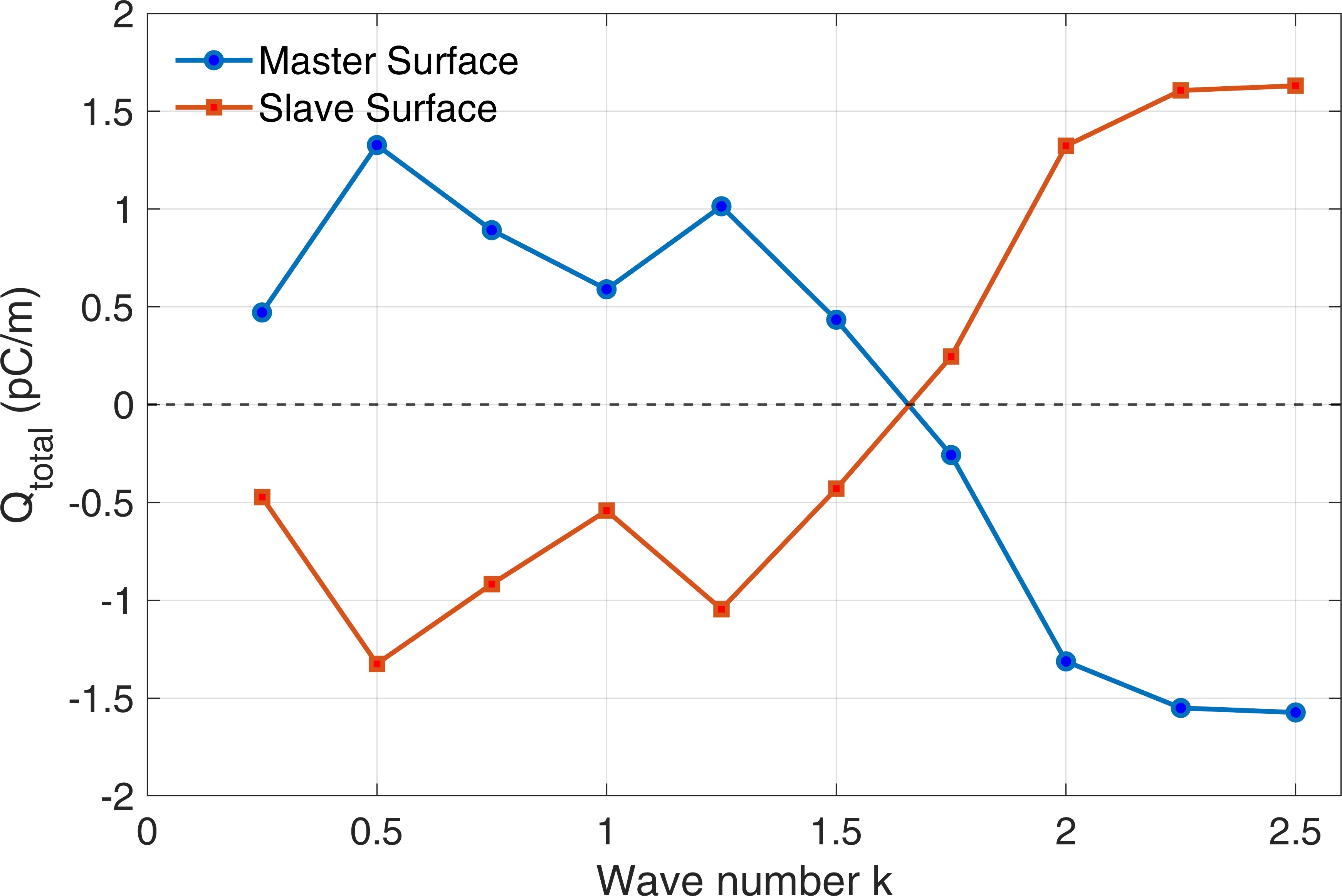}
		\caption{Total charge vs. wavenumber}
	\end{subfigure}
	\caption{Quantitative analysis of charge transfer vs. wavenumber: (a) peak charge density on master and slave surfaces; (b) total charge on each surface. Note the polarity reversal around $k \approx 1.5$--$2$.}
	\label{fig:wavy_quantitative}
\end{figure}

A systematic analysis of the peak charge density and total charge transferred between surfaces is summarized in Fig.~\ref{fig:wavy_quantitative}. For low wavenumbers ($k < 0.75$), the peak charge density increases monotonically with $k$, driven by the sharpening of local contact curvature. In the intermediate range ($0.75 < k < 1.5$), the local charge density saturates at the trap limit $\sigma_\mathrm{trap}$. The most significant feature appears in the total charge plot (Fig.~\ref{fig:wavy_quantitative}(b)): a critical transition occurs around $k \approx 1.5$--$2$, where the net charge on the master surface reverses from positive to negative. The net charge also reverses within the subset of wavenumbers for which the edge of the slave body does not come into contact with the master surface, namely $k = 0.5$, $1.5$ and $2.5$.
This reversal can be understood through the competition between the flexoelectric polarization on the two contacting surfaces. For identical materials, the net charge transfer direction is determined by whichever surface develops a stronger local flexoelectric response. At low wavenumbers, contact occurs at isolated smooth crests where the slave (wavy) surface undergoes larger local deformation, producing a stronger flexoelectric polarization of the appropriate sign that attracts electrons from the master surface. As the wavenumber increases, the asperities become sharper and more densely packed, which intensifies the local deformation on the initially flat master surface. When the master surface's flexoelectric response exceeds that of the slave, the direction of charge transfer reverses, and the master surface becomes the electron acceptor.
This finding offers insight into the pressure-dependent polarity reversal observed experimentally~\cite{Xu2019curvature, Sow2012reversal}. While such reversal has previously been attributed to material transfer or surface damage at high loads~\cite{Baytekin2012}, our results suggest that changes in the local contact conditions, such as the asperity curvature and the true local contact pressure, can by themselves shift the balance of flexoelectric response between the two surfaces.

\subsection{Three-dimensional random rough surface contact}
The wavy surface in the preceding section, while useful for isolating the effect of a single geometric parameter, is an idealization. Real surfaces are randomly rough at the nanoscale, and the resulting charge distributions are correspondingly irregular. To bridge this gap, we now extend the framework to three dimensions and replace the deterministic sinusoid with a stochastic surface profile. The rough surface is generated as a Gaussian random field with prescribed root-mean-square (RMS) amplitude and correlation length $l_\mathrm{corr}$, following the methodology in \cite{hu2022isogeometric}.

\begin{figure}[!htbp]
	\centering
	\includegraphics[width=0.85\textwidth]{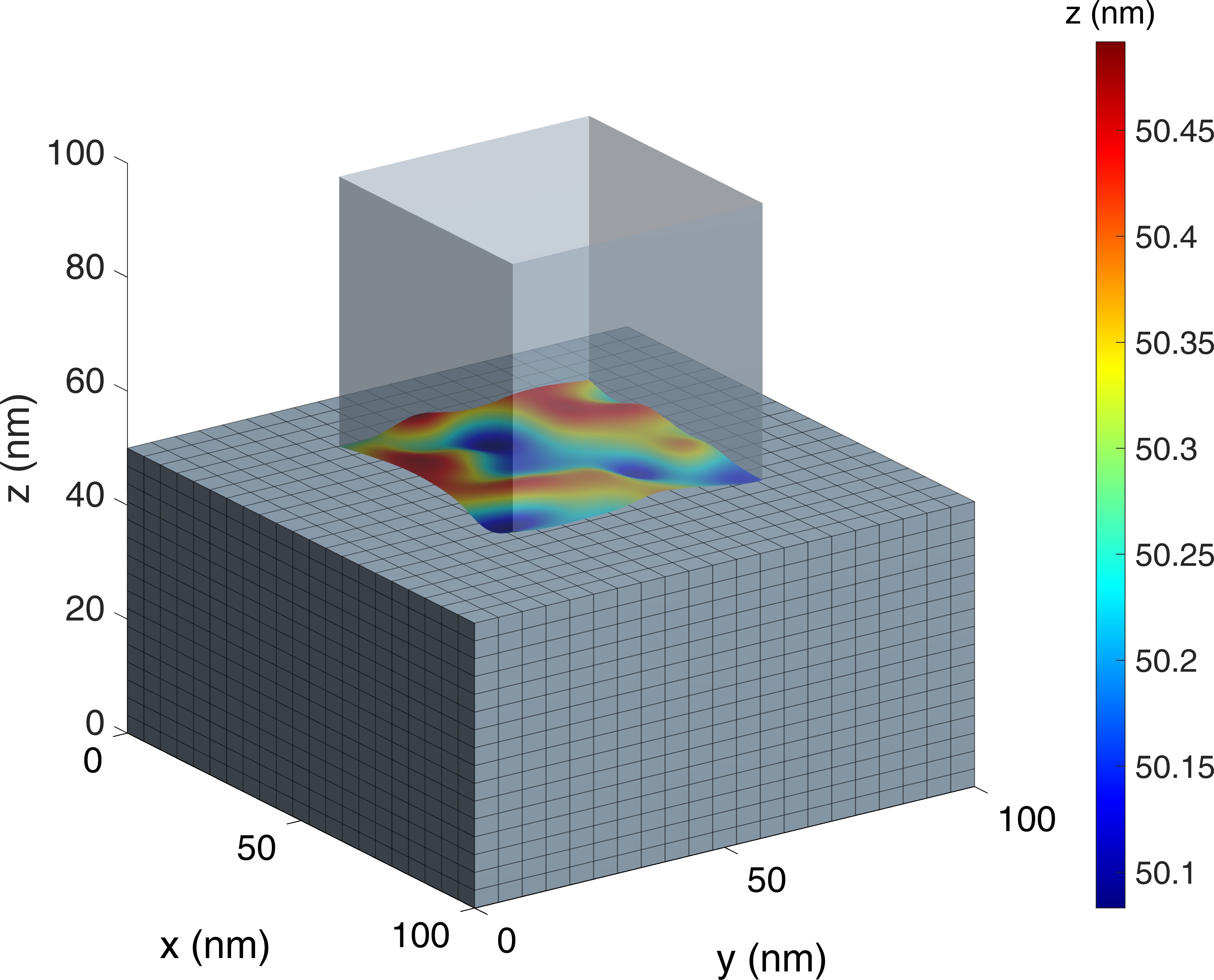}
	\caption{Geometry of 3D random rough surface contact. The upper block ($50 \times 50 \times 50$~nm) has a randomly rough bottom surface and is pressed into the flat lower block ($100 \times 100 \times 50$~nm). Coloring on the surface indicates the local height variation.}
	\label{fig:rough_geometry}
\end{figure}

The computational domain is shown in Fig.~\ref{fig:rough_geometry}. A flat PMMA block of dimensions $100 \times 100 \times 50$~nm is in contact with a smaller PMMA block of dimensions $50 \times 50 \times 50$~nm, whose bottom surface carries the random roughness. The RMS amplitude of the roughness is $0.1$~nm. The base of the lower block is fully clamped, while a vertical displacement of $\delta = 8$~nm is prescribed on the top face of the upper block. Both the bottom surface of the lower block and the top surface of the upper block are electrically grounded ($\phi = 0$). After reaching the prescribed displacement, the upper block is gradually retracted until the minimum gap between the two surfaces reaches $20\,\ell_q$ (i.e., 4.8~nm), at which point the tunneling channel is fully closed and no further charge transfer occurs.

\begin{figure}[!htbp]
	\centering
	\begin{subfigure}{0.32\textwidth}
		\includegraphics[width=\textwidth]{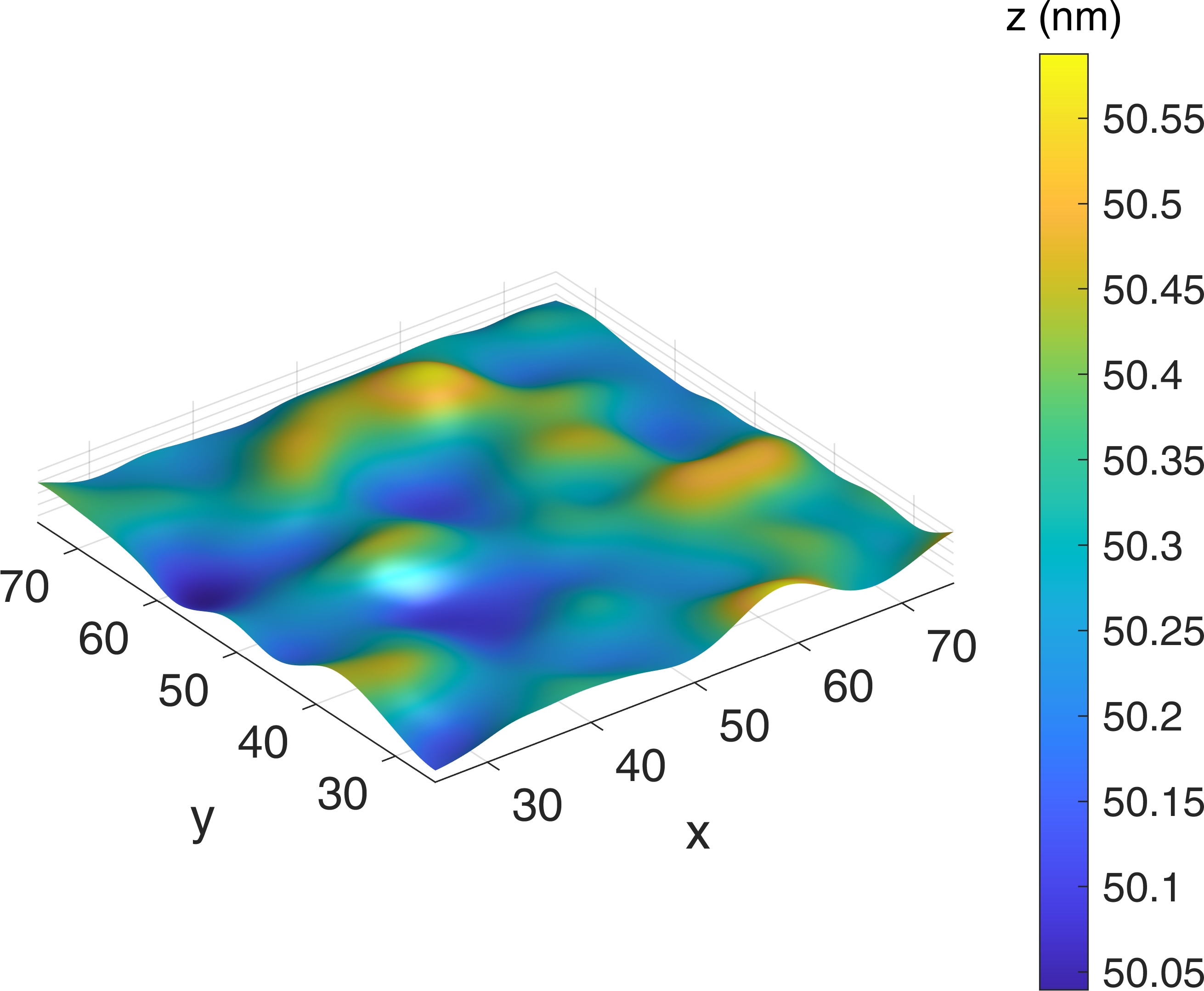}
		\caption{}
	\end{subfigure}
	\hfill
	\begin{subfigure}{0.32\textwidth}
		\includegraphics[width=\textwidth]{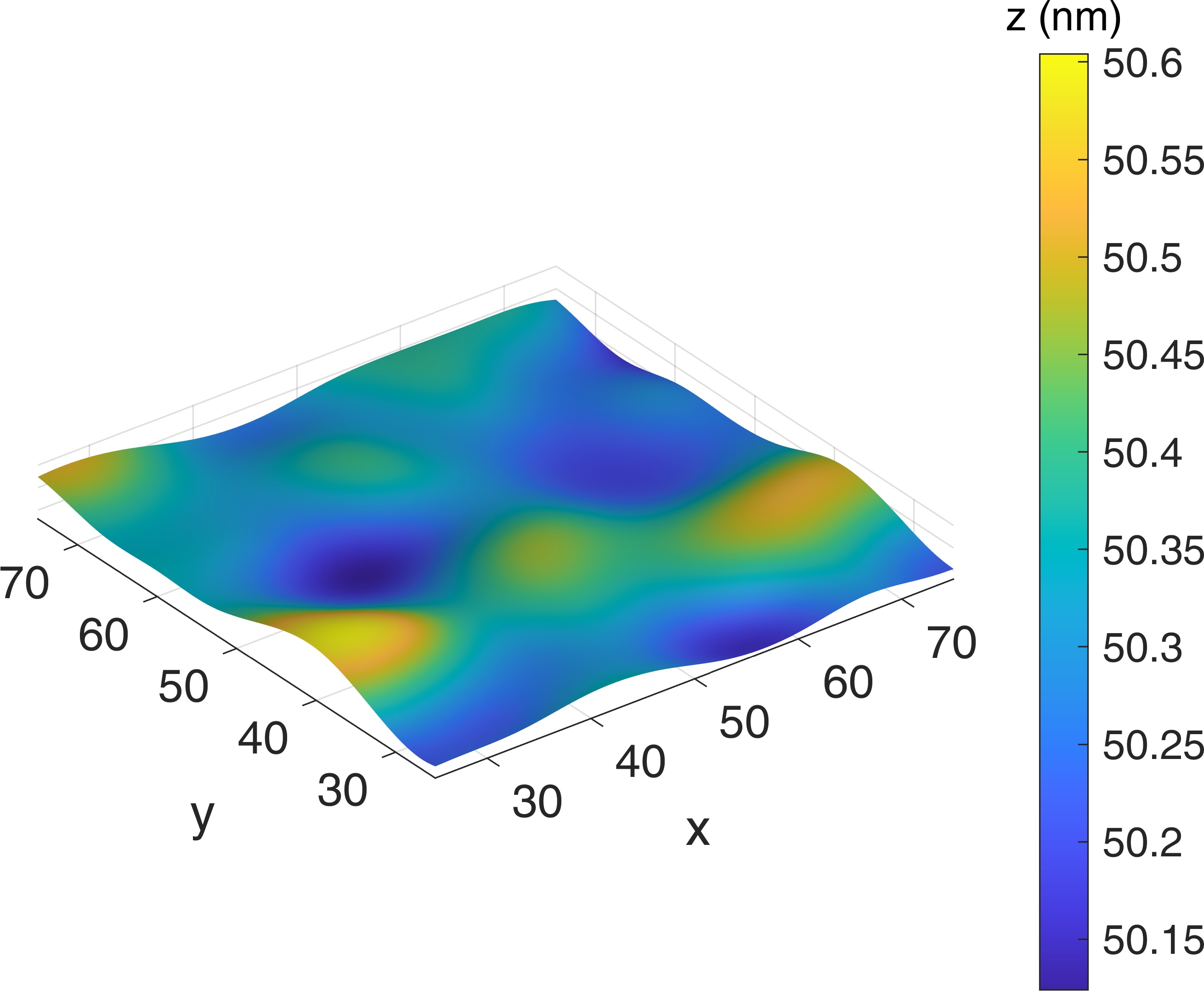}
		\caption{}
	\end{subfigure}
	\hfill
	\begin{subfigure}{0.32\textwidth}
		\includegraphics[width=\textwidth]{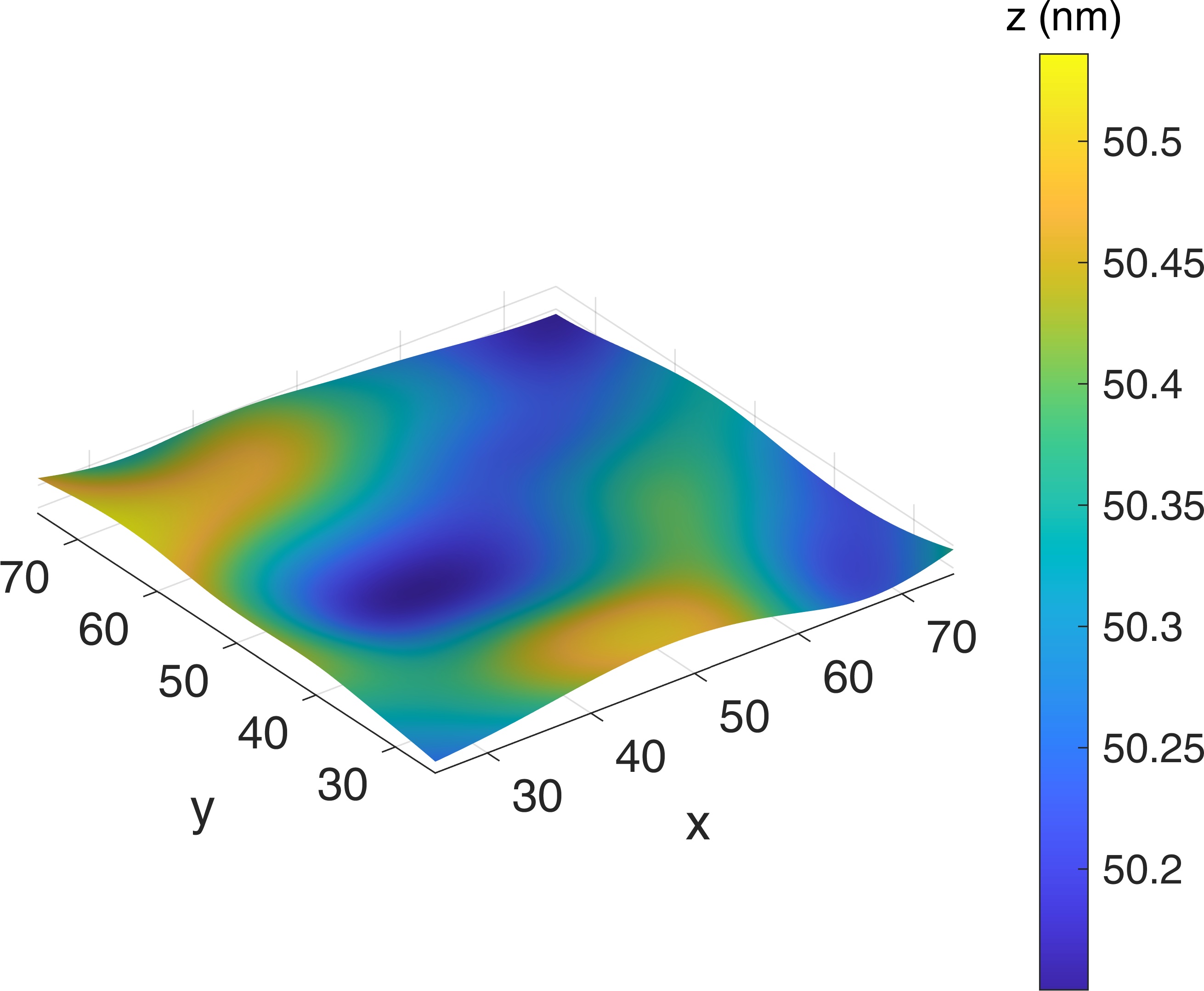}
		\caption{}
	\end{subfigure}
	\caption{Random rough surface profiles of the upper body for three correlation lengths: (a)~$l_\mathrm{corr} = 6$~nm; (b)~$l_\mathrm{corr} = 10$~nm; (c)~$l_\mathrm{corr} = 14$~nm. Shorter $l_\mathrm{corr}$ corresponds to finer, more irregular roughness.}
	\label{fig:rough_surfaces}
\end{figure}

To study the effect of surface randomness, we vary the correlation length $l_\mathrm{corr}$ across three values: 6, 10, and 14~nm, while keeping the RMS amplitude fixed. A shorter $l_\mathrm{corr}$ produces a rougher surface with more frequent, sharper asperities; a longer $l_\mathrm{corr}$ yields a smoother, more gently undulating profile. The resulting surface topographies are shown in Fig.~\ref{fig:rough_surfaces}. At $l_\mathrm{corr} = 6$~nm, the surface exhibits fine-scale oscillations with peak-to-valley heights spanning roughly 0.5~nm. At $l_\mathrm{corr} = 14$~nm, the surface is gentler, with broad hills and shallow valleys.

\begin{figure}[!htbp]
	\centering
	\begin{subfigure}{0.32\textwidth}
		\includegraphics[width=\textwidth]{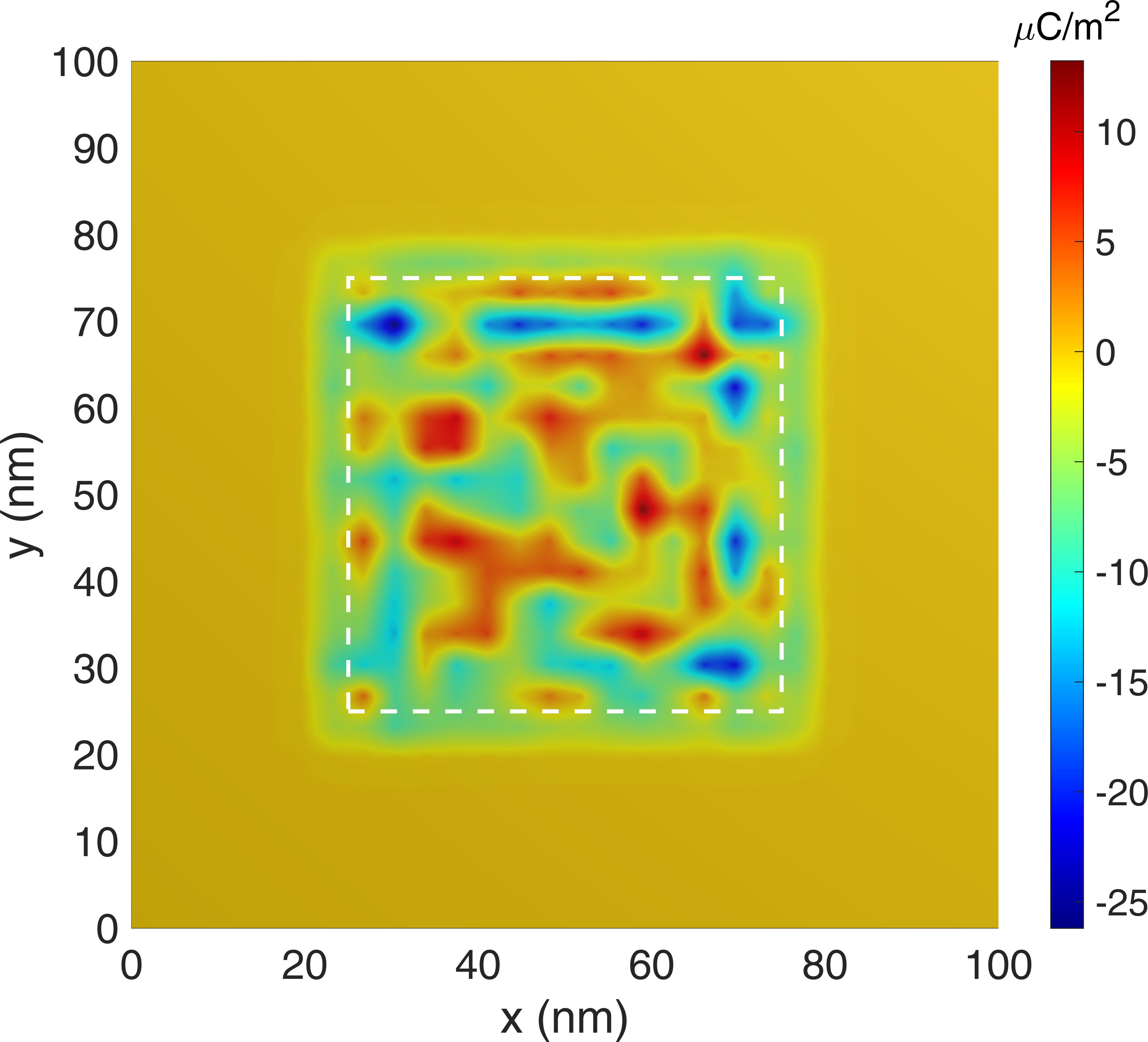}
		\caption{}
	\end{subfigure}
	\hfill
	\begin{subfigure}{0.32\textwidth}
		\includegraphics[width=\textwidth]{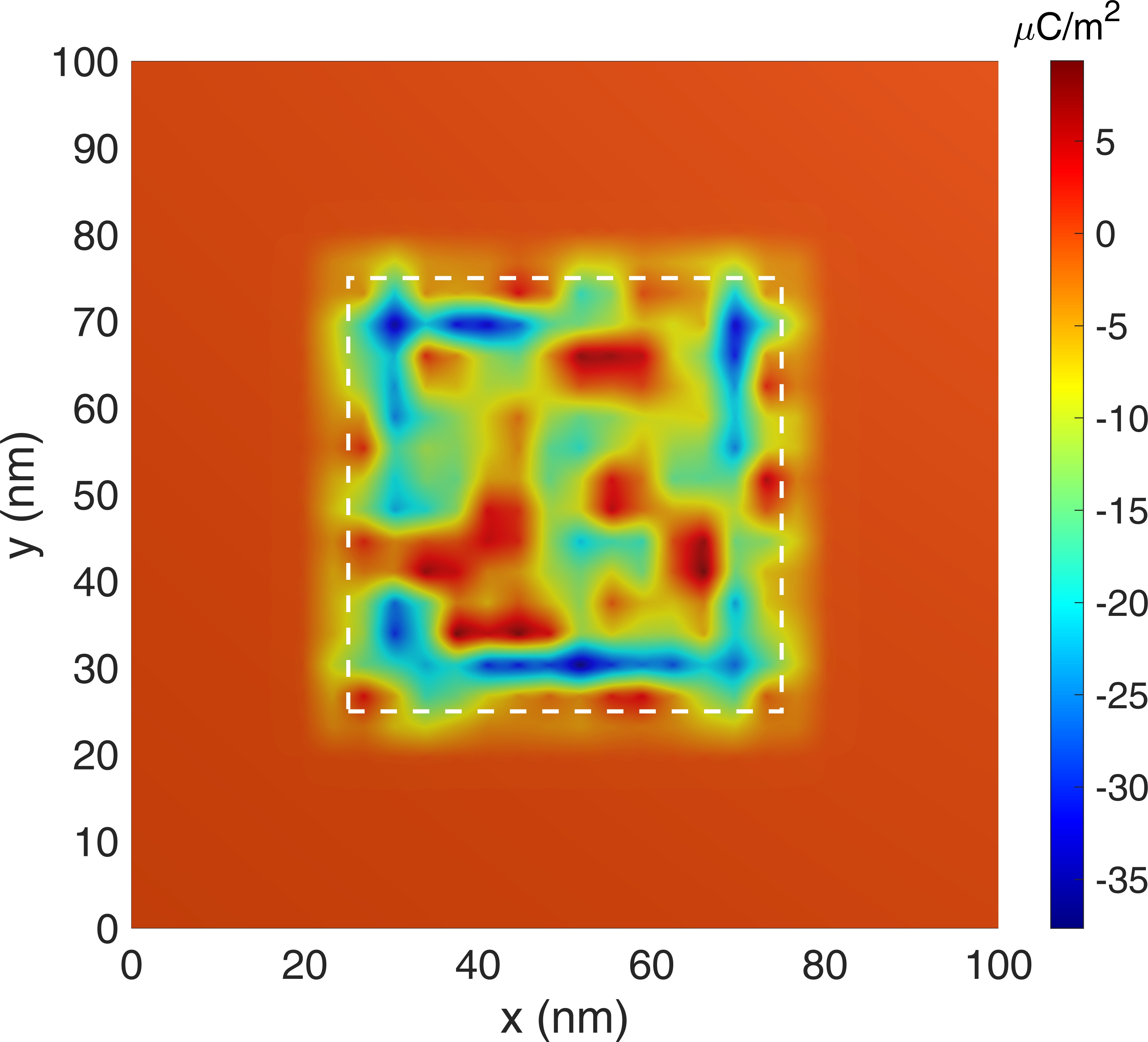}
		\caption{}
	\end{subfigure}
	\hfill
	\begin{subfigure}{0.32\textwidth}
		\includegraphics[width=\textwidth]{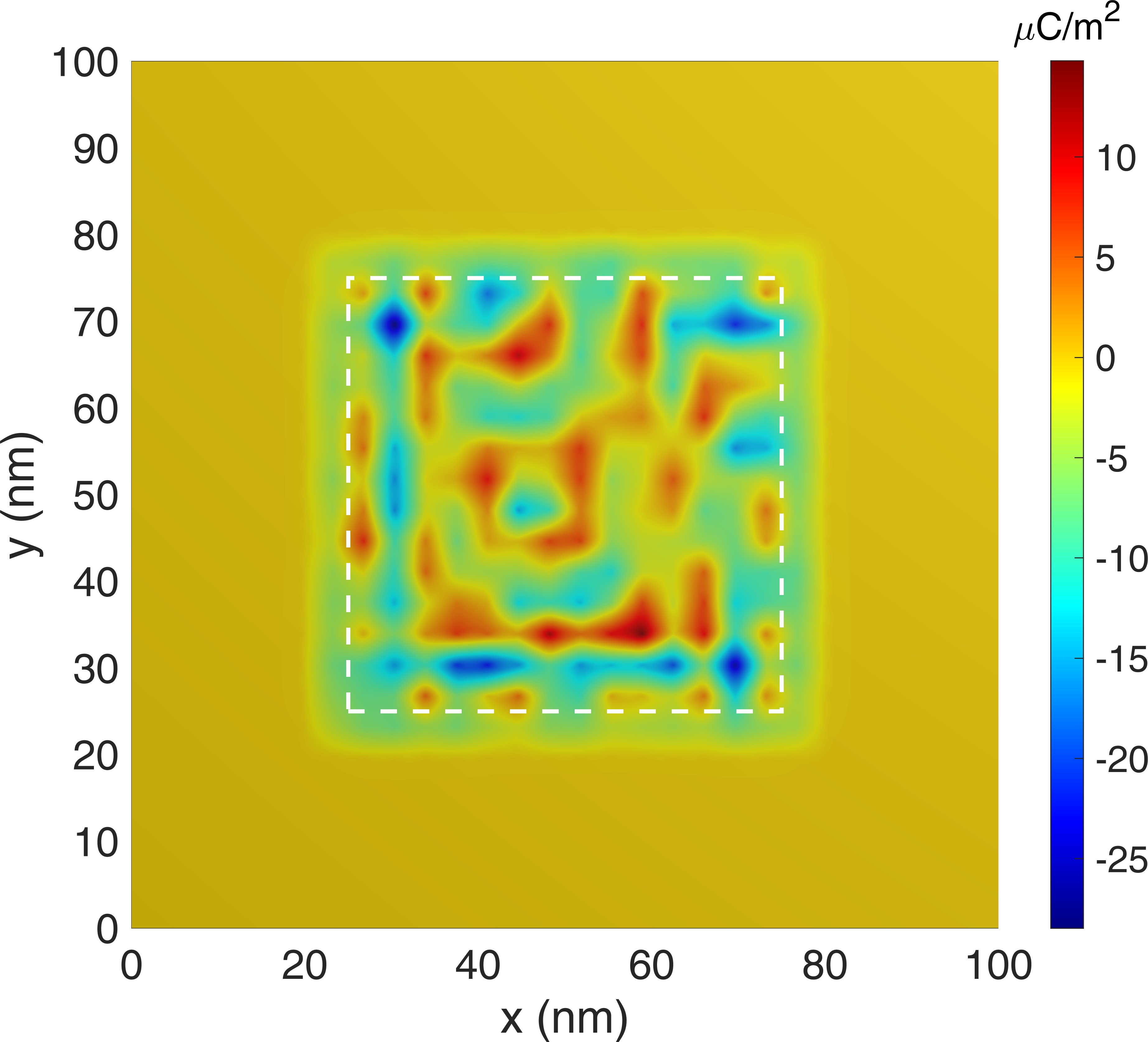}
		\caption{}
	\end{subfigure}
	\\[6pt]
	\begin{subfigure}{0.32\textwidth}
		\includegraphics[width=\textwidth]{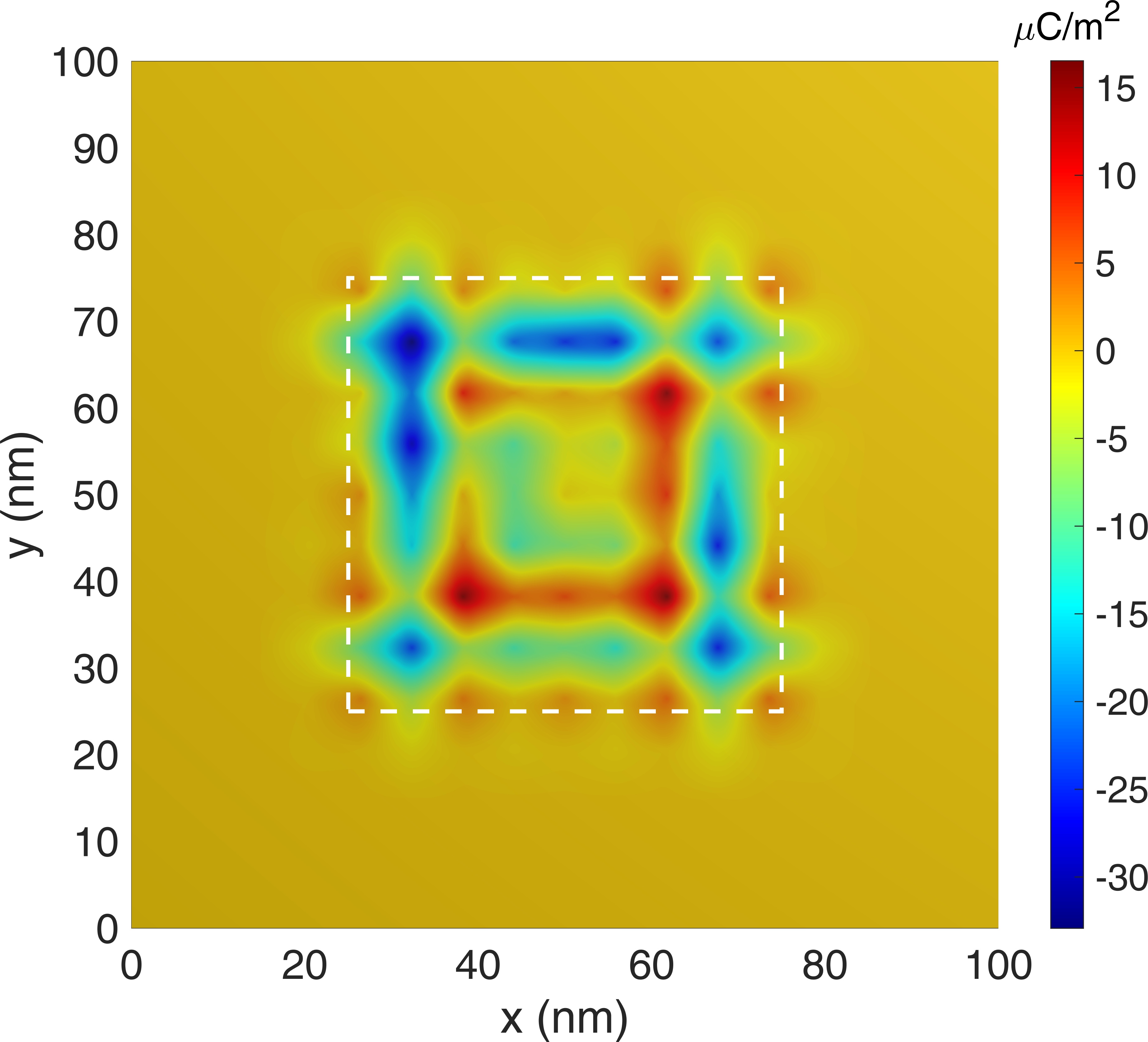}
		\caption{}
	\end{subfigure}
	\hfill
	\begin{subfigure}{0.32\textwidth}
		\includegraphics[width=\textwidth]{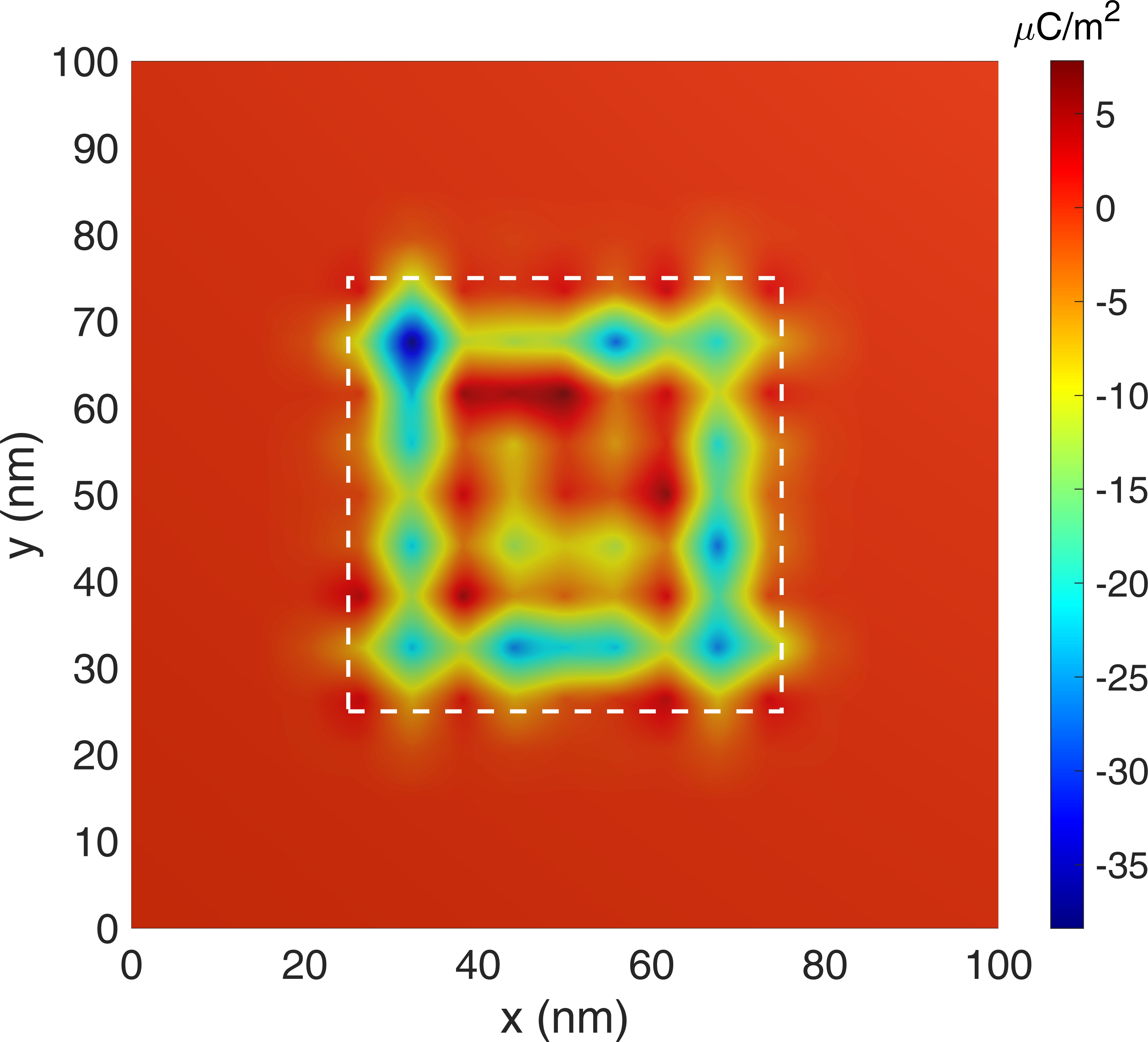}
		\caption{}
	\end{subfigure}
	\hfill
	\begin{subfigure}{0.32\textwidth}
		\includegraphics[width=\textwidth]{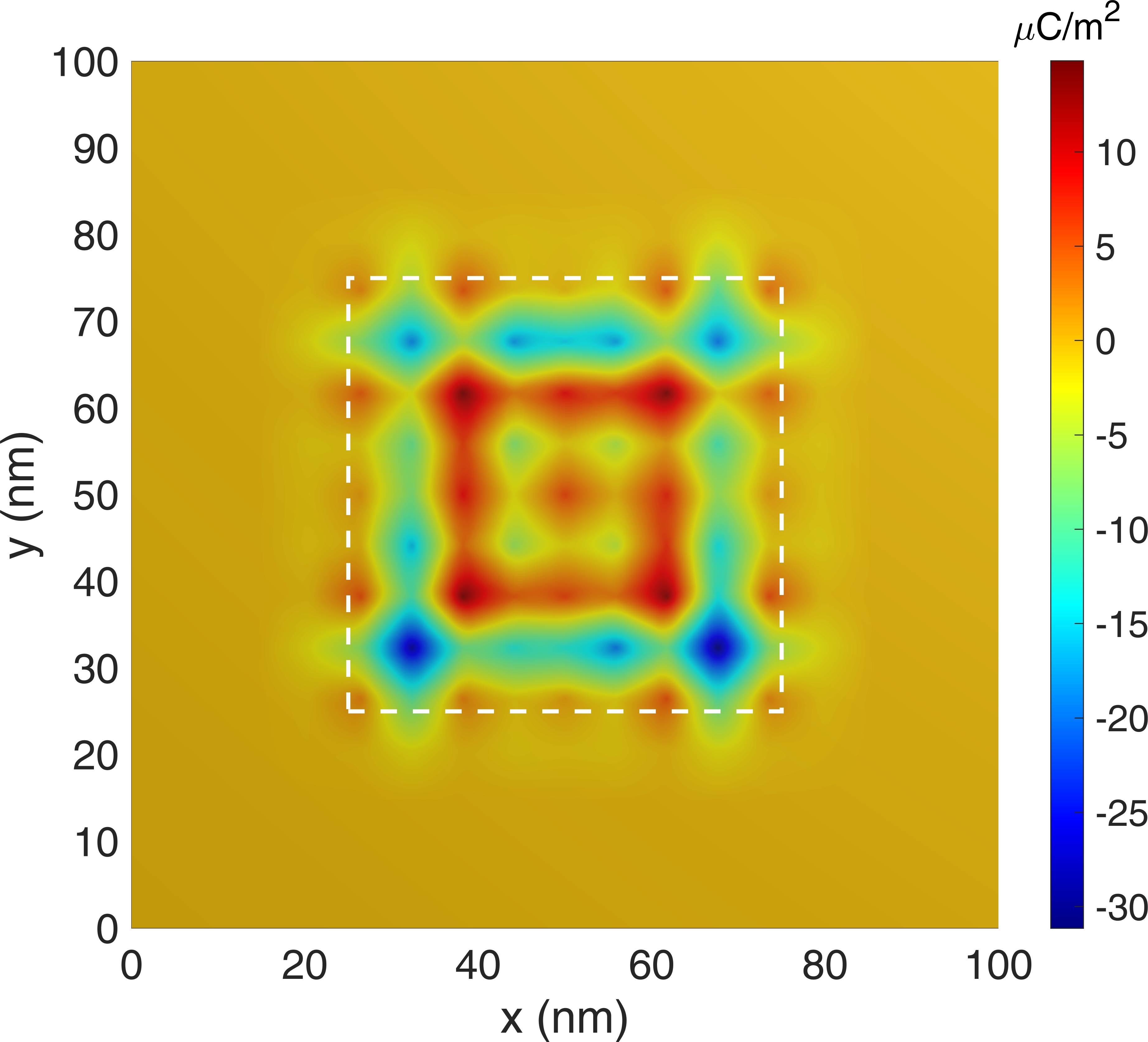}
		\caption{}
	\end{subfigure}
	\\[6pt]
	\begin{subfigure}{0.32\textwidth}
		\includegraphics[width=\textwidth]{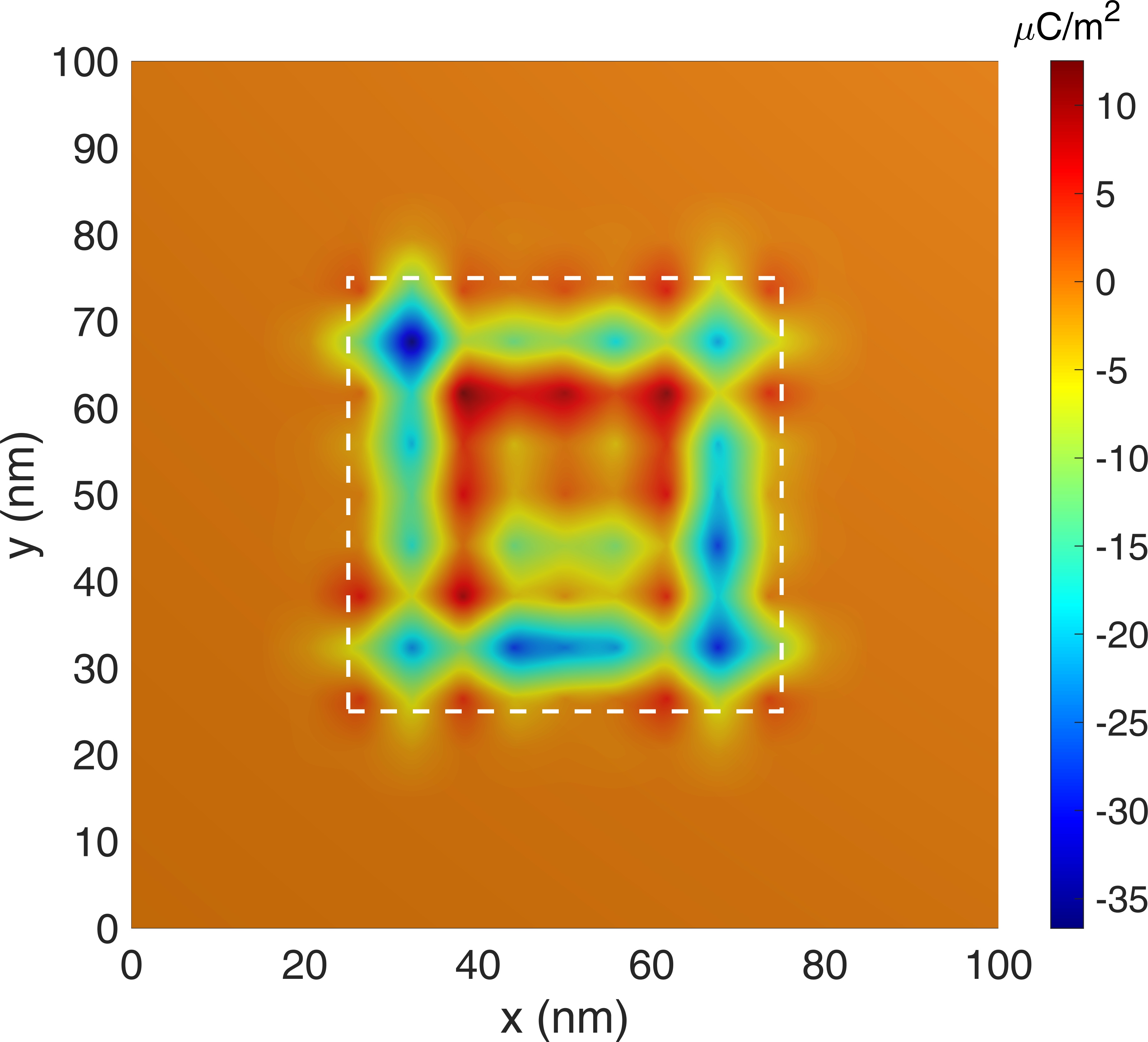}
		\caption{}
	\end{subfigure}
	\hfill
	\begin{subfigure}{0.32\textwidth}
		\includegraphics[width=\textwidth]{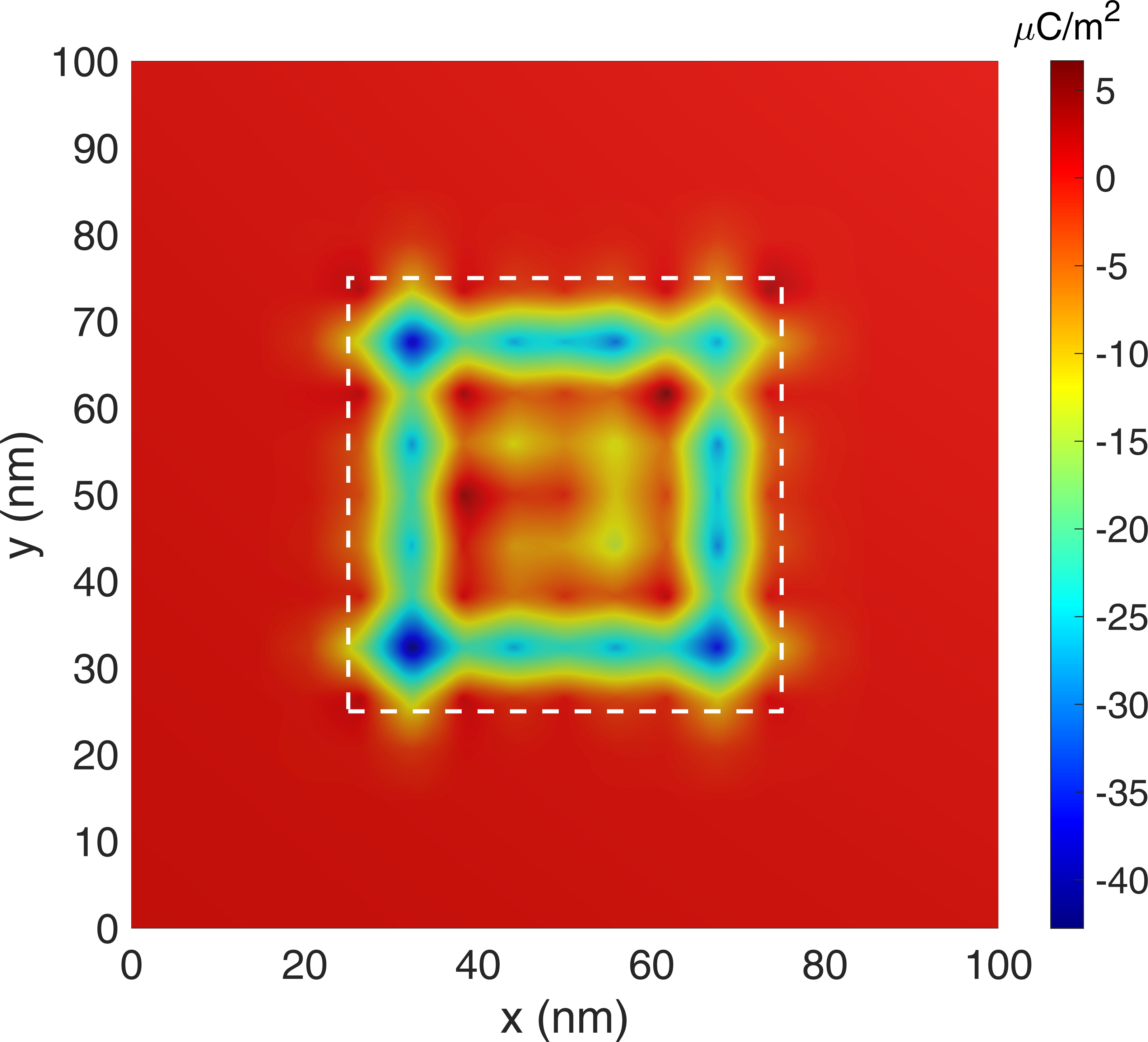}
		\caption{}
	\end{subfigure}
	\hfill
	\begin{subfigure}{0.32\textwidth}
		\includegraphics[width=\textwidth]{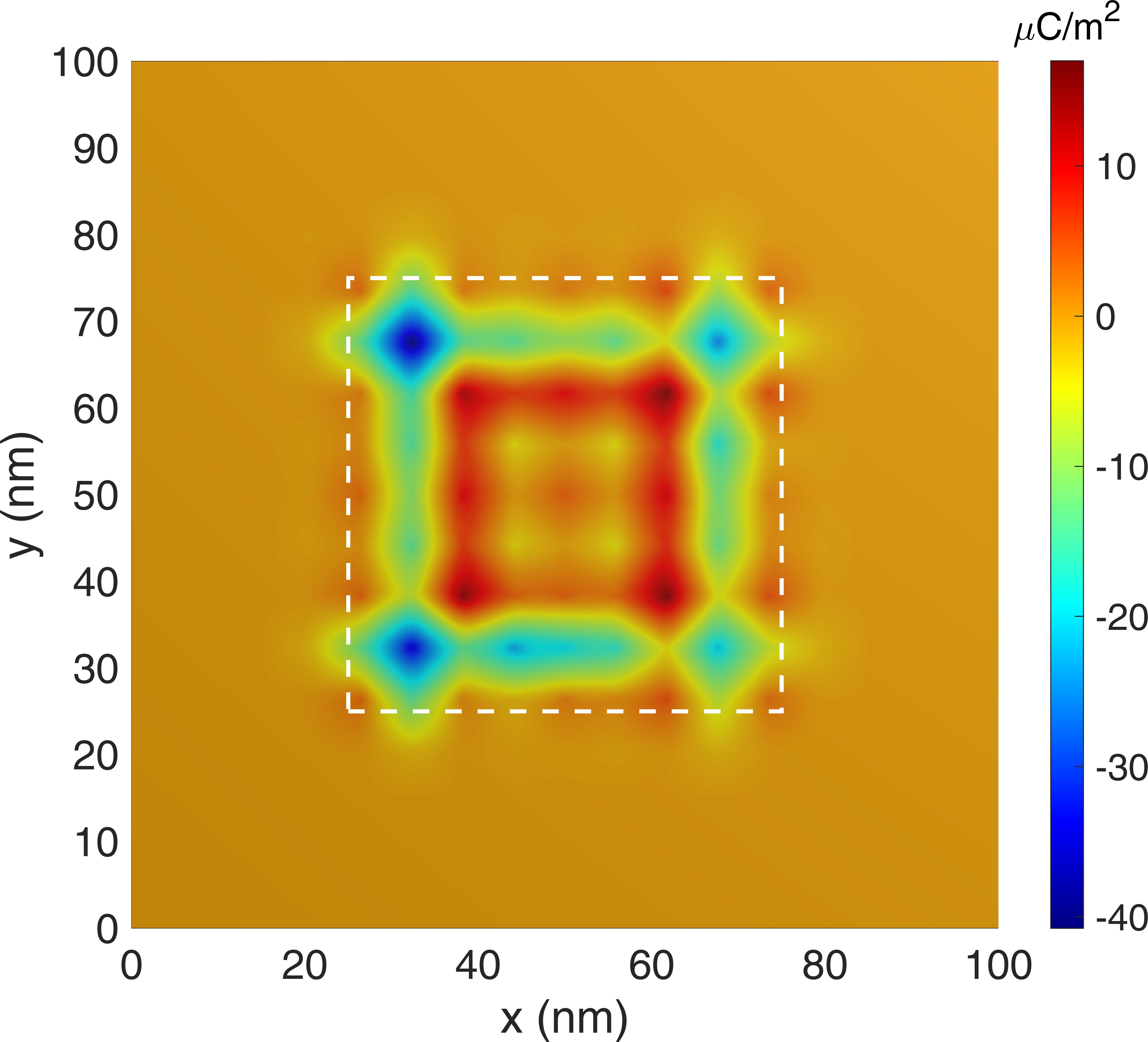}
		\caption{}
	\end{subfigure}
	\caption{Residual charge distribution on the master surface after separation for three correlation lengths and three independent surface realizations. Rows from top to bottom: $l_\mathrm{corr} = 6$, 10, and 14~nm. Columns correspond to different random seeds. The dashed rectangle marks the footprint of the slave block. Shorter $l_\mathrm{corr}$ produces a more pronounced ``mosaic'' of alternating positive and negative charge, and this trend is consistent across all realizations.}
	\label{fig:mosaic}
\end{figure}

The residual charge distributions on the master surface after complete separation are shown in Fig.~\ref{fig:mosaic} for three independent surface realizations at each correlation length. At $l_\mathrm{corr} = 6$~nm, every realization produces a ``mosaic'' of positive and negative patches scattered across the contact region without obvious spatial order, although the specific patch arrangement differs from one realization to another. When $l_\mathrm{corr}$ increases to 10~nm, the patches become broader and fewer, and the charge tends to concentrate near the edges of the contact region. At $l_\mathrm{corr} = 14$~nm, the mosaic character disappears in all three realizations: intense negative bands run along the contact edges while the interior carries only a weak, nearly uniform positive charge. The consistency of these trends across independent realizations indicates that they are governed by the correlation length of the roughness rather than by the details of any single random surface.

These results can be explained through the mechanism of flexoelectric polarization competition established in Section~\ref{sec:wavy_contact}. For contact between identical dielectrics, the local electrification polarity depends on which surface undergoes more intensive local deformation and consequently develops a stronger flexoelectric response. On a highly irregular surface with a short correlation length, the dominant flexoelectric response shifts rapidly between the upper and lower bodies across the contact interface. This spatial alternation in polarization dominance yields an irregular ``mosaic'' of positive and negative charge patches, with dimensions that scale with the roughness correlation length. As $l_\mathrm{corr}$ increases, the surface profile becomes smoother and the contact-induced local strain gradients become more uniform. Consequently, the charge distribution aggregates into distinct, macroscopic regions rather than a scattered mosaic.

Notably, these results are consistent with the ``mosaic charge'' pattern reported by Baytekin et al.\ \cite{Baytekin2011mosaic}, who observed that nominally identical polymer surfaces develop spatially heterogeneous charge distributions upon contact. Our simulations, which resolve the charge transfer between the two deformable bodies through loading and separation, show that flexoelectricity and nanoscale surface roughness together are sufficient to reproduce this effect in the residual charge, extending the analysis of the flexoelectric surface potential in rough contacts~\cite{persson2020role}. The random topography ensures that local deformation is distributed unequally between the two contacting bodies. This asymmetry in localized strain drives an unbalanced flexoelectric response across the interface, resulting in a spatially heterogeneous charge transfer without the need for material inhomogeneity or chemical asymmetry.

\begin{figure}[!htbp]
	\centering
	\begin{subfigure}{0.48\textwidth}
		\includegraphics[width=\textwidth]{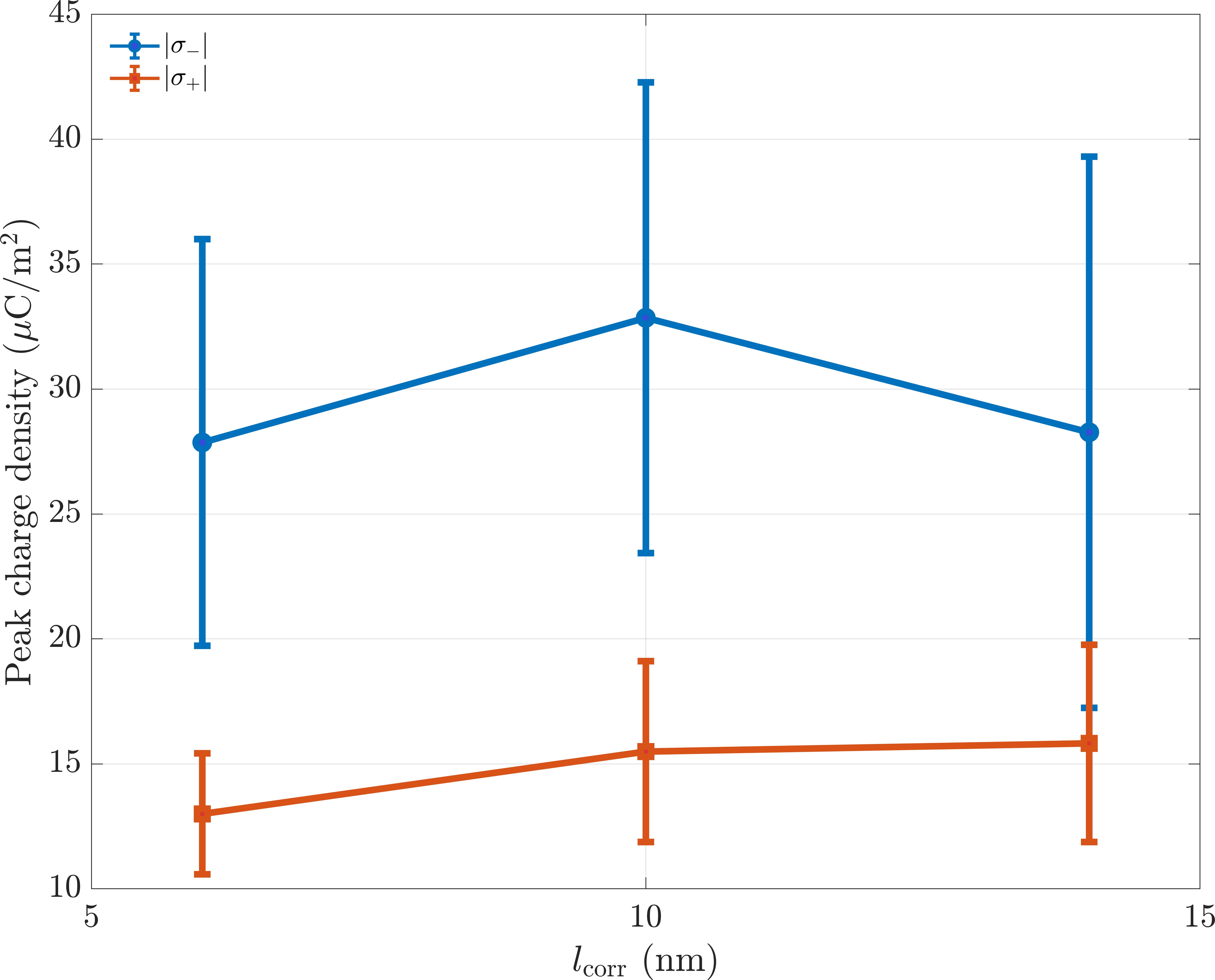}
		\caption{}
	\end{subfigure}
	\hfill
	\begin{subfigure}{0.48\textwidth}
		\includegraphics[width=\textwidth]{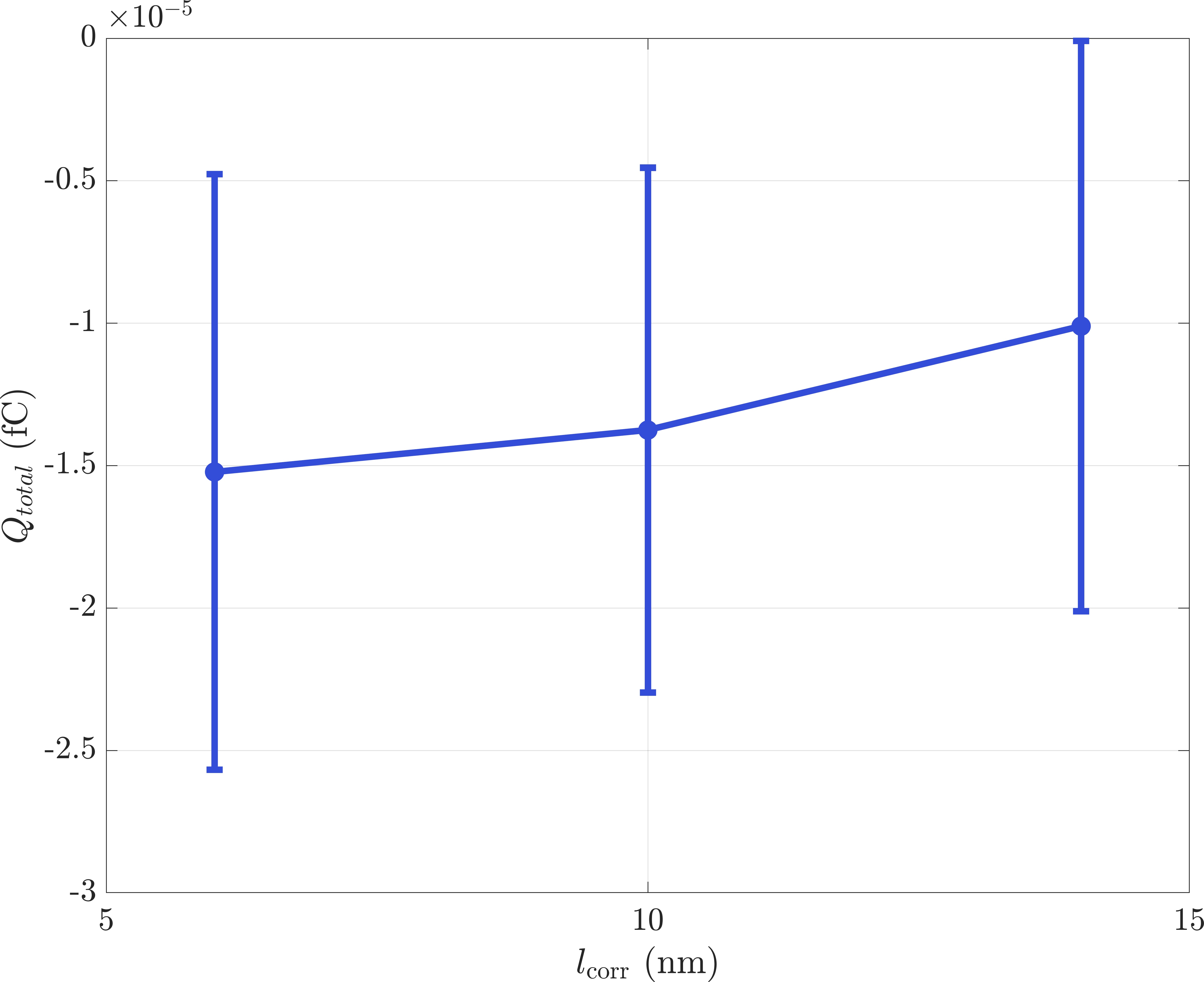}
		\caption{}
	\end{subfigure}
	\caption{Statistics of residual charge across ten random surface realizations for each $l_\mathrm{corr}$. (a)~Peak positive ($|\sigma_+|$) and peak negative ($|\sigma_-|$) charge density vs.\ $l_\mathrm{corr}$. (b)~Total residual charge $Q_\mathrm{total}$ on the master surface vs.\ $l_\mathrm{corr}$.}
	\label{fig:3D_statistics}
\end{figure}

To assess how robust these observations are, we repeat the simulation with ten independent random surface realizations for each correlation length and collect per-realization charge statistics. Figure~\ref{fig:3D_statistics}(a) shows the peak positive and peak negative charge densities on the master surface, averaged over the ten realizations, as a function of $l_\mathrm{corr}$. The peak negative charge density $|\sigma_-|$ stays between 28 and 33~$\mu$C/m$^2$ across the three correlation lengths, while $|\sigma_+|$ is roughly half that, around 13--16~$\mu$C/m$^2$. This asymmetry is attributed to the geometric asymmetry between the slave and master blocks. Neither quantity shows a statistically significant dependence on $l_\mathrm{corr}$, which is reasonable because the RMS amplitude is fixed and hence the overall contact force distribution remains similar across correlation lengths. What changes with $l_\mathrm{corr}$ is the spatial arrangement of charge, not its magnitude. The total residual charge $Q_\mathrm{total}$ on the master surface, shown in Fig.~\ref{fig:3D_statistics}(b), provides complementary information. Because the positive and negative charge patches in each realization largely cancel, $Q_\mathrm{total}$ corresponds to an average net charge density of only about 4--6~$\mu$C/m$^2$ over the contact region, well below the peak densities. The magnitude of its mean value shows a weak decreasing trend from roughly $1.5 \times 10^{-5}$~fC at $l_\mathrm{corr} = 6$~nm to about $1.0 \times 10^{-5}$~fC at 14~nm, though the large inter-realization variance renders this trend statistically insignificant. In other words, the peak charge density is a sample-robust quantity, whereas the net charge on each surface is sensitive to the specific random geometry.

\section{Discussion}
\label{sec:discussion}

\subsection{Limitations and assumptions}
\label{sec:limitations}

The present model assumes quasi-static loading and unloading throughout, so dynamic and rate-dependent effects that may arise during high-speed contacts are not captured. All simulations are performed at room temperature, and the temperature dependence of tunneling probability, surface state occupation, and material stiffness is neglected. A further simplification is that only electron transfer driven by flexoelectric polarization is considered. The metal tip is likewise treated as an ideal conductor at a prescribed potential, which neglects the strain-induced change of its mean inner potential. Ion transfer and material transfer, both of which may contribute under certain experimental conditions, lie outside the scope of this framework.

The flexoelectric coefficients used for the polymer substrates deserve particular attention. These coefficients are not well characterized experimentally for most polymers, and the values adopted here are assumed within a physically reasonable range rather than taken from independent measurements. 
As demonstrated in Section~\ref{sec:unbiased_results}, the predicted charge density is linear in the flexoelectric coefficients, so the discrepancy with experiment, which ranges from a factor of 1.7 for PDAP to about one order of magnitude for PMMA, may be attributed mainly to the uncertainty in the assumed values of these material parameters.
The qualitative trends, spatial patterns, and force dependence predicted by the model are independent of the coefficient magnitude and are therefore more robust conclusions than the absolute charge densities. Direct experimental measurement of the full flexoelectric tensor for the polymers of interest would substantially reduce this uncertainty and enable quantitative predictions.

Finally, the dielectric-dielectric contact model assumes a spatially uniform surface state density. Real polymer surfaces exhibit defects, contamination, and processing-induced heterogeneity that could modulate the local tunneling rate and charge capacity. Incorporating a spatially varying surface state distribution is a natural extension for future work.

\subsection{Comparison with existing models}
\label{sec:comparison}
Two recent models have addressed the role of flexoelectricity in contact electrification, each from a different angle. Silavnieks et al.\ \cite{Silavnieks2025flexo} derived an analytical expression for flexoelectric charge transfer at a Hertzian contact, including the unloading phase and the resulting residual charge. Their closed-form scaling laws offer physical transparency but are limited to smooth geometries. Olson and Marks \cite{Olson2025quantitative} modeled metal-semiconductor junctions with depletion layers, focusing on Schottky barrier modulation by flexoelectric fields. The numerical model developed in this work complements these approaches by treating arbitrary surface topographies, including the 3D random rough surfaces, and by targeting polymer dielectrics with surface states across the full loading-unloading cycle.

Beyond these two works, the present computational model offers two further features. First, by incorporating a tunneling transparency function that is grounded in the tunneling kinetics of interfacial charge exchange and regulates charge transfer based on the local interfacial gap, the model captures both the evolution of charge accumulation during loading and its irreversible freezing during separation. Second, the model provides a unified theoretical treatment for diverse contact scenarios. The same multiphysics formulation simulates unbiased metal-dielectric contact, biased metal-dielectric contact, and contact between identical dielectrics, differing only in the target charge, the applied boundary conditions and the material parameters. This unified approach makes it possible to systematically compare charging phenomena across widely different experimental configurations within a single, consistent framework.

\section{Conclusions}
\label{sec:conclusions}
This work presents a computational model for contact electrification that fully couples finite-deformation flexoelectric theory, contact mechanics with cohesive-zone adhesion, and physically motivated charge transfer rules. A central element of the model is the tunneling transparency function, which regulates the charge transfer channel based on the local interfacial gap throughout the loading-unloading cycle. During separation, the gap opens earliest at the contact periphery, causing the tunneling channel to close progressively inward. This naturally captures the irreversible trapping of surface charge and produces residual charge patterns after separation.

Three contact scenarios have been investigated within the same formulation. In unbiased metal-dielectric contact, the spatial variation of strain gradients beneath the indenter produces a characteristic bipolar charge pattern on PMMA and PDAP substrates, with one polarity at the contact center and the opposite at the edge. After separation, residual charge concentrates at the original contact periphery, and its magnitude increases with contact force, consistent with the AFM measurements of Lin et al.\ \cite{Lin2022flexo}. Parametric studies further show that peak charge density scales approximately linearly with inverse tip radius, and that a shorter tunneling length traps more charge by closing the tunneling channel earlier. When a voltage bias is applied to the tip, the projection mechanism restricts charge transfer to carriers of a single sign. This polarity restriction suppresses edge charging and produces residual charge that is nearly independent of the applied force.

The same model has been applied to contact between identical dielectric bodies.
In two-dimensional simulations with a wavy surface, the net charge polarity on each body reverses at a critical wavenumber. This reversal is driven by a shift in which contacting surface undergoes the more intensive local deformation, thus determining which body develops the dominant flexoelectric response. This polarity reversal suggests a geometric origin for the pressure-dependent charge reversal observed in experiments \cite{Xu2019curvature, Sow2012reversal}. 
Three-dimensional simulations on random rough surfaces demonstrate that the spatial scale of the resulting charge mosaic tracks the correlation length of the roughness: short correlation lengths produce fine, randomly interspersed positive and negative patches, consistent with the mosaic charge patterns first reported by Baytekin et al.\ \cite{Baytekin2011mosaic}. Statistical analysis over multiple random realizations shows that the peak charge density is robust across surface realizations, while the net charge on each surface is sensitive to the specific geometry, reflecting the cancellation between the positive and negative contributions. 
These results show that flexoelectric polarization and nanoscale geometric heterogeneity together are sufficient to produce spatially non-uniform charge distributions between chemically identical surfaces, without requiring any material inhomogeneity.

Several questions remain open for future investigation. Because the predicted charge density scales linearly with the flexoelectric coefficients, the quantitative accuracy of the model hinges on independent measurements of these coefficients for the polymers of interest. Dedicated experiments to determine the full flexoelectric tensor would therefore be a prerequisite for moving from the qualitative trends demonstrated here to quantitative predictions. Further experiments could test the predictions of the model. For instance, Kelvin probe force microscopy after single-point indentations at different forces would reveal whether the residual charge forms a ring with a common spatial profile. Contact between identical polymers with controlled surface waviness would allow the predicted polarity reversal to be examined. Extensions to dynamic loading conditions, temperature-dependent tunneling, and competing transfer mechanisms such as ion and material transfer would broaden the range of applicability.

\appendix
\counterwithin*{figure}{section}
\counterwithin*{table}{section}
\renewcommand{\thefigure}{\Alph{section}.\arabic{figure}}
\renewcommand{\thetable}{\Alph{section}.\arabic{table}}
\setcounter{figure}{0}
\setcounter{table}{0}
\section{Detailed representation of higher-order boundary tractions}
\label{Apdx1}
The terms $\widetilde{\mathbf{T}}^{t}$ and $\widetilde{\mathbf{T}}^{n}$ in the higher-order mechanical boundary conditions originate from the surface boundary integral during integration by parts. The surface gradient operator $D^{(t)}_J$ is defined by splitting the variation $\delta\chi$ into tangential and normal gradient components as follows:
\begin{equation}
	\begin{aligned}
		\delta\chi_{k,J} &= D^{(t)}_J\delta\chi_k + N_J D^{(n)}\delta \chi_k,\\
		D^{(t)}_J & \equiv (\delta_{JL} - N_J N_L)\partial_L,\\
		D^{(n)} & \equiv N_L\partial_L.
	\end{aligned}
\end{equation}
Consequently, the final formulation of the traction surface projections $\widetilde{\mathbf{T}}^{t}$ and $\widetilde{\mathbf{T}}^{n}$ in index notation can be obtained as
\begin{equation}
	\begin{aligned}
		\widetilde{T}^{(t)}_k &= \frac{1}{2}\big[\Sigma_{IJK}N_K(N_J\,\chi_{k,I} + N_I\,\chi_{k,J})D^{(t)}_LN_L - N_K(\chi_{k,I}D^{(t)}_J\Sigma_{IJK} + \chi_{k,J}D^{(t)}_I\Sigma_{IJK})\\
		&\qquad - \Sigma_{IJK}(\chi_{k,I}D^{(t)}_JN_K + \chi_{k,J}D^{(t)}_IN_K)\big],\\
		\widetilde{T}^{(n)}_k &= \frac{1}{2}\Sigma_{IJK}N_K\big(\chi_{k,I}N_J + \chi_{k,J}N_I \big).
	\end{aligned}
\end{equation}

\section{Contact Mechanics Implementation}
\label{app:contact}

This appendix presents the detailed contact formulation in a general three-dimensional setting, including surface parametrization, mortar treatment, and consistent linearization. Greek indices $\alpha, \beta \in \{1, 2\}$ denote surface parametric directions, with summation implied over repeated indices.

\subsection{Surface parametrization}

The master contact surface is parametrized by convective coordinates $\boldsymbol{\xi}^\alpha = (\xi^1, \xi^2)$. The covariant tangent vectors and unit normal are defined as
\begin{equation}
	\boldsymbol{\tau}_\alpha = \frac{\partial \bar{\mathbf{x}}^m}{\partial \xi^\alpha}, \qquad
	\mathbf{n} = \frac{\boldsymbol{\tau}_1 \times \boldsymbol{\tau}_2}{\|\boldsymbol{\tau}_1 \times \boldsymbol{\tau}_2\|},
\end{equation}
where $\bar{\mathbf{x}}^m = \bar{\mathbf{x}}^m(\boldsymbol{\xi}^\alpha)$ denotes the position on the master surface. The metric tensor $m_{\alpha\beta}$ (with its contravariant inverse denoted as $m^{\alpha\beta}$) and curvature tensor $k_{\alpha\beta}$ are defined as
\begin{equation}
	m_{\alpha\beta} = \boldsymbol{\tau}_\alpha \cdot \boldsymbol{\tau}_\beta, \qquad
	k_{\alpha\beta} = \bar{\mathbf{x}}^m_{,\alpha\beta} \cdot \mathbf{n}.
\end{equation}

For each slave point $\mathbf{x}^s$, the closest point projection onto the master surface determines the projection parameter $\bar{\boldsymbol{\xi}}^\alpha$ through
\begin{equation}
	(\mathbf{x}^s - \bar{\mathbf{x}}^m) \cdot \boldsymbol{\tau}_\alpha\big|_{\bar{\boldsymbol{\xi}}^\alpha} = 0.
\end{equation}
The normal gap is $g_N = (\mathbf{x}^s - \bar{\mathbf{x}}^m) \cdot \mathbf{n}$, and the potential gap is $g_\phi = \phi_s - \bar{\phi}_m$, where $\bar{\phi}_m = \phi^m(\bar{\boldsymbol{\xi}}^\alpha)$ is the master potential at the projection point.

\subsection{Variations and consistent linearization}

The variation of the normal gap is
\begin{equation}
	\delta g_N = (\delta\mathbf{u}^s - \delta\bar{\mathbf{u}}^m) \cdot \mathbf{n}.
\end{equation}

The variation of the projection parameter is obtained from the stationarity condition:
\begin{equation}
	\label{eq:param_variation}
	\delta\bar{\xi}^\alpha = (m_{\alpha\beta} - g_N k_{\alpha\beta})^{-1} \big[ (\delta\mathbf{u}^s - \delta\bar{\mathbf{u}}^m) \cdot \boldsymbol{\tau}_\beta + g_N\, \mathbf{n} \cdot \delta\bar{\mathbf{u}}^m_{,\beta} \big].
\end{equation}

For the potential gap:
\begin{equation}
	\delta g_\phi = \delta\phi_s - \delta\bar{\phi}_m - \bar{\phi}^m_{,\alpha}\, \delta\bar{\xi}^\alpha,
\end{equation}
where $\bar{\phi}^m_{,\alpha} = \partial\phi^m / \partial\xi^\alpha$ evaluated at the projection point.

For Newton--Raphson iteration, consistent linearization requires the second variation. From standard contact mechanics~\cite{wriggers2006computational}:
\begin{equation}
	\label{eq:gap_linearization}
	\begin{aligned}
		\Delta\delta g_N = &-(\delta\bar{\mathbf{u}}^m_{,\alpha}\, \Delta\bar{\xi}^\alpha + \Delta\bar{\mathbf{u}}^m_{,\alpha}\, \delta\bar{\xi}^\alpha + \bar{\mathbf{x}}^m_{,\alpha\beta}\, \delta\bar{\xi}^\alpha \Delta\bar{\xi}^\beta) \cdot \mathbf{n} \\
		&+ g_N m^{\alpha\beta} (\delta\bar{\mathbf{u}}^m_{,\alpha} + \bar{\mathbf{x}}^m_{,\alpha\gamma}\, \delta\bar{\xi}^\gamma)\, \mathbf{n} \otimes \mathbf{n}\, (\Delta\bar{\mathbf{u}}^m_{,\beta} + \bar{\mathbf{x}}^m_{,\beta\delta}\, \Delta\bar{\xi}^\delta).
	\end{aligned}
\end{equation}

\subsection{Mortar treatment}

The mortar method projects local contact quantities onto nodal values via weighted averaging with NURBS basis functions $R_A$. For each control point $A$, the nodal gap, transparency, and potential gap are
\begin{equation}
	\label{eq:mortar_average}
	g_{NA} = \frac{\int_{\partial\mathcal{B}^C_t} R_A\, g_N\, \mathrm{d}\gamma}{A_A}, \qquad
	\Transparency_A = \frac{\int_{\partial\mathcal{B}^C_t} R_A\, \Transparency\, \mathrm{d}\gamma}{A_A}, \qquad
	g_{\phi A} = \frac{\int_{\partial\mathcal{B}^C_t} R_A\, g_\phi\, \mathrm{d}\gamma}{A_A},
\end{equation}
where $A_A = \int_{\partial\mathcal{B}^C_t} R_A\, \mathrm{d}\gamma$ is the tributary area. The nodal contact traction and surface charge are:
\begin{equation}
	\label{eq:nodal_traction}
	t_{NA} = \begin{cases}
		\varepsilon_N g_{NA} & \text{if } g_{NA} \leq 0, \\[3pt]
		\displaystyle\frac{\phi_N}{(g_N^\mathrm{max})^2}\,g_{NA} \exp\left(-\frac{g_{NA}}{g_N^\mathrm{max}}\right) & \text{if } g_{NA} > 0,
	\end{cases}
	\qquad
	q_{CA} = \varepsilon_\phi \Transparency_A g_{\phi A}.
\end{equation}
The first case of $t_{NA}$ approximates impenetrability via penalty, while the second applies the adhesive cohesive traction~\eqref{eq:adhesion}.

The contact virtual work~\eqref{eq:contact_virtual_work} can then be expressed in mortar form as
\begin{equation}
	\delta\bar{\Pi}_C = \sum_A \big( t_{NA}\, \delta g_{NA} + q_{CA}\, \delta g_{\phi A} \big) A_A.
\end{equation}

\subsection{NURBS discretization}

Let the element contain $n_e^s$ slave and $n_e^m$ master control points. We define the following discretization vectors by assembling slave and master contributions:
\begin{equation}
	\mathbf{N}^e = 
	\begin{bmatrix}
		R_1^s\, \mathbf{n} \\ \vdots \\ R_{n_e^s}^s\, \mathbf{n} \\
		-R_1^m\, \mathbf{n} \\ \vdots \\ -R_{n_e^m}^m\, \mathbf{n}
	\end{bmatrix}, \qquad
	\mathbf{T}_\alpha^e = 
	\begin{bmatrix}
		R_1^s\, \boldsymbol{\tau}_\alpha \\ \vdots \\ R_{n_e^s}^s\, \boldsymbol{\tau}_\alpha \\
		-R_1^m\, \boldsymbol{\tau}_\alpha \\ \vdots \\ -R_{n_e^m}^m\, \boldsymbol{\tau}_\alpha
	\end{bmatrix}, \qquad
	\mathbf{P}^e = 
	\begin{bmatrix}
		R_1^s \\ \vdots \\ R_{n_e^s}^s \\
		-R_1^m \\ \vdots \\ -R_{n_e^m}^m
	\end{bmatrix},
\end{equation}
where shape functions are evaluated at the slave point and its master projection, respectively. The parameter gradient vectors involve only master shape function derivatives:
\begin{equation}
	\mathbf{N}_\alpha^e = 
	\begin{bmatrix}
		\mathbf{0} \\ \vdots \\ \mathbf{0} \\
		-R_{1,\alpha}^m\, \mathbf{n} \\ \vdots \\ -R_{n_e^m,\alpha}^m\, \mathbf{n}
	\end{bmatrix}.
\end{equation}

The parametric response vector, derived from~\eqref{eq:param_variation}, captures the change in projection parameters due to displacement variations:
\begin{equation}
	\mathbf{D}_\alpha^e = (m_{\alpha\beta} - g_N k_{\alpha\beta})^{-1} \big( \mathbf{T}_\beta^e - g_N \mathbf{N}_\beta^e \big).
\end{equation}
The geometric coupling vector accounts for curvature effects in the normal direction:
\begin{equation}
	\bar{\mathbf{N}}_\alpha^e = \mathbf{N}_\alpha^e - k_{\alpha\beta}\, \mathbf{D}_\beta^e.
\end{equation}

\subsection{Tangent stiffness matrices}

The contact contribution to the tangent stiffness matrix is obtained by substituting the discretized variations into the linearized weak form. The mechanical block comprises both material and geometric contributions:
\begin{equation}
	\begin{aligned}
		\mathbf{K}_C^{uu} = \sum_A \Bigg[ &\frac{\varepsilon_N^A}{A_A} \int_{\partial\mathcal{B}^C_t} R_A\, \mathbf{N}^e\, \mathrm{d}\gamma \int_{\partial\mathcal{B}^C_t} R_A\, \mathbf{N}^{e\mathrm{T}}\, \mathrm{d}\gamma \\
		&+ \int_{\partial\mathcal{B}^C_t} t_{NA} R_A \Big( g_N m^{\alpha\beta} \bar{\mathbf{N}}_\alpha^e \bar{\mathbf{N}}_\beta^{e\mathrm{T}} + \mathbf{D}_\alpha^e \mathbf{N}_\alpha^{e\mathrm{T}} + \mathbf{N}_\alpha^e \mathbf{D}_\alpha^{e\mathrm{T}} - k_{\alpha\beta}\, \mathbf{D}_\alpha^e \mathbf{D}_\beta^{e\mathrm{T}} \Big)\, \mathrm{d}\gamma \Bigg],
	\end{aligned}
\end{equation}
where the effective stiffness $\varepsilon_N^A$ switches between penalty and adhesive contributions:
\begin{equation}
	\varepsilon_N^A = \begin{cases}
		\varepsilon_N & \text{if } g_{NA} \leq 0, \\[3pt]
		\displaystyle\frac{\phi_N}{(g_N^\mathrm{max})^2}\left(1 - \frac{g_{NA}}{g_N^\mathrm{max}}\right)\exp\left(-\frac{g_{NA}}{g_N^\mathrm{max}}\right) & \text{if } g_{NA} > 0.
	\end{cases}
\end{equation}

The electrical block is
\begin{equation}
	\mathbf{K}_C^{\phi\phi} = \sum_A \frac{\varepsilon_\phi \Transparency_A}{A_A} \int_{\partial\mathcal{B}^C_t} R_A\, \mathbf{P}^e\, \mathrm{d}\gamma \int_{\partial\mathcal{B}^C_t} R_A\, \mathbf{P}^{e\mathrm{T}}\, \mathrm{d}\gamma.
\end{equation}

The internal force vectors are
\begin{equation}
	\mathbf{F}_C^u = \sum_A t_{NA} \int_{\partial\mathcal{B}^C_t} R_A\, \mathbf{N}^e\, \mathrm{d}\gamma, \qquad
	\mathbf{F}_C^\phi = \sum_A q_{CA} \int_{\partial\mathcal{B}^C_t} R_A\, \mathbf{P}^e\, \mathrm{d}\gamma.
\end{equation}

\begin{remark}
	For two-dimensional problems, the surface reduces to a curve with a single parameter $\xi^1 = \xi$. The metric tensor becomes the scalar $m_{11} = \|\boldsymbol{\tau}_1\|^2$ and the curvature reduces to $k_{11} = \bar{\mathbf{x}}^m_{,\xi\xi} \cdot \mathbf{n}$. All tensorial expressions collapse accordingly.
\end{remark}

\section{Mesh and Load-Step Convergence}
\label{app:convergence}

\begin{figure}[!htbp]
	\centering
	\includegraphics[width=0.7\textwidth]{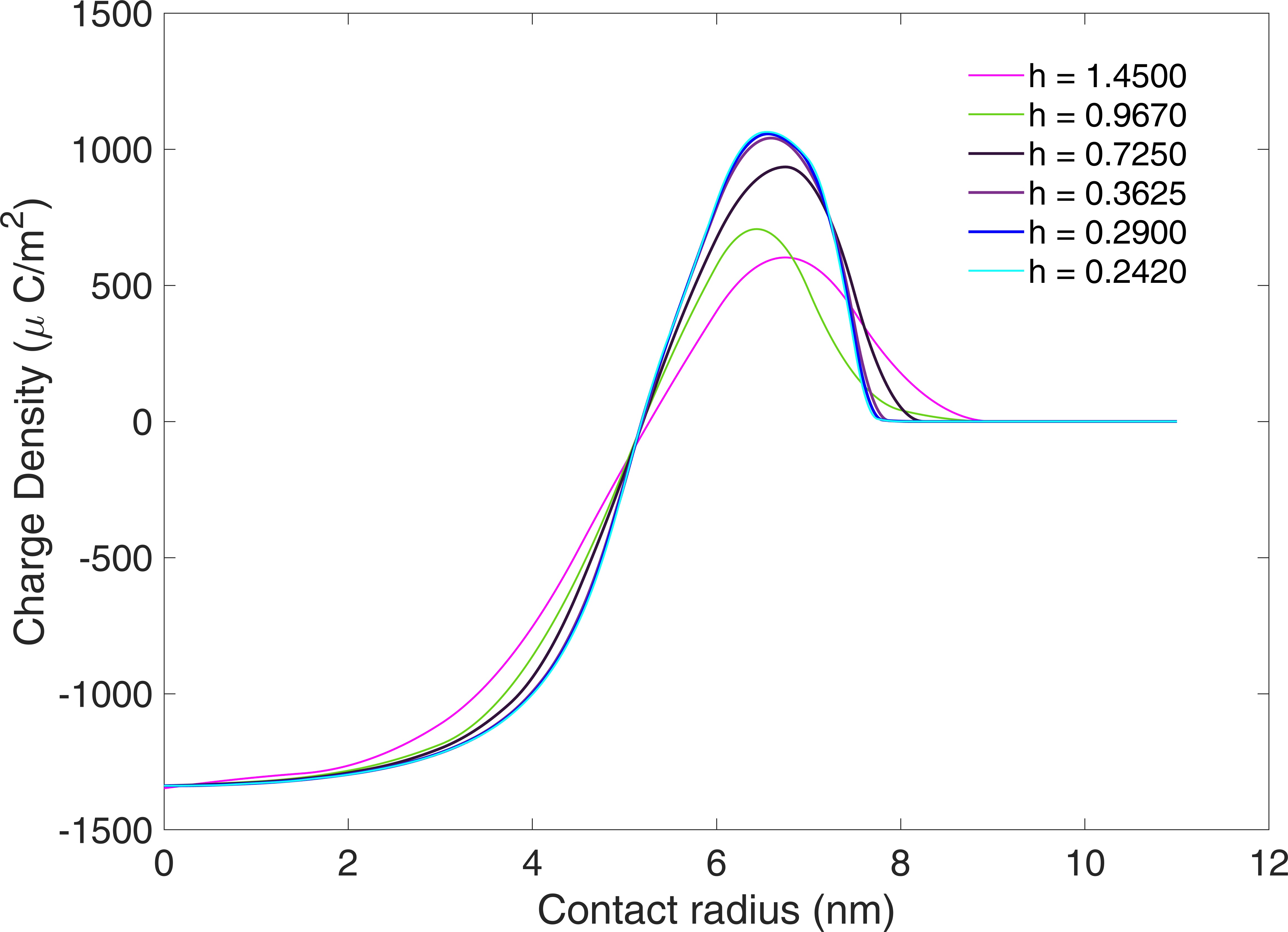}
	\caption{Mesh convergence of the surface charge density at the end of loading for unbiased PMMA contact at 50~nN. The element size $h$ (in nm) refers to the contact-zone mesh.}
	\label{fig:mesh_convergence}
\end{figure}

\begin{figure}[!htbp]
	\centering
	\includegraphics[width=0.7\textwidth]{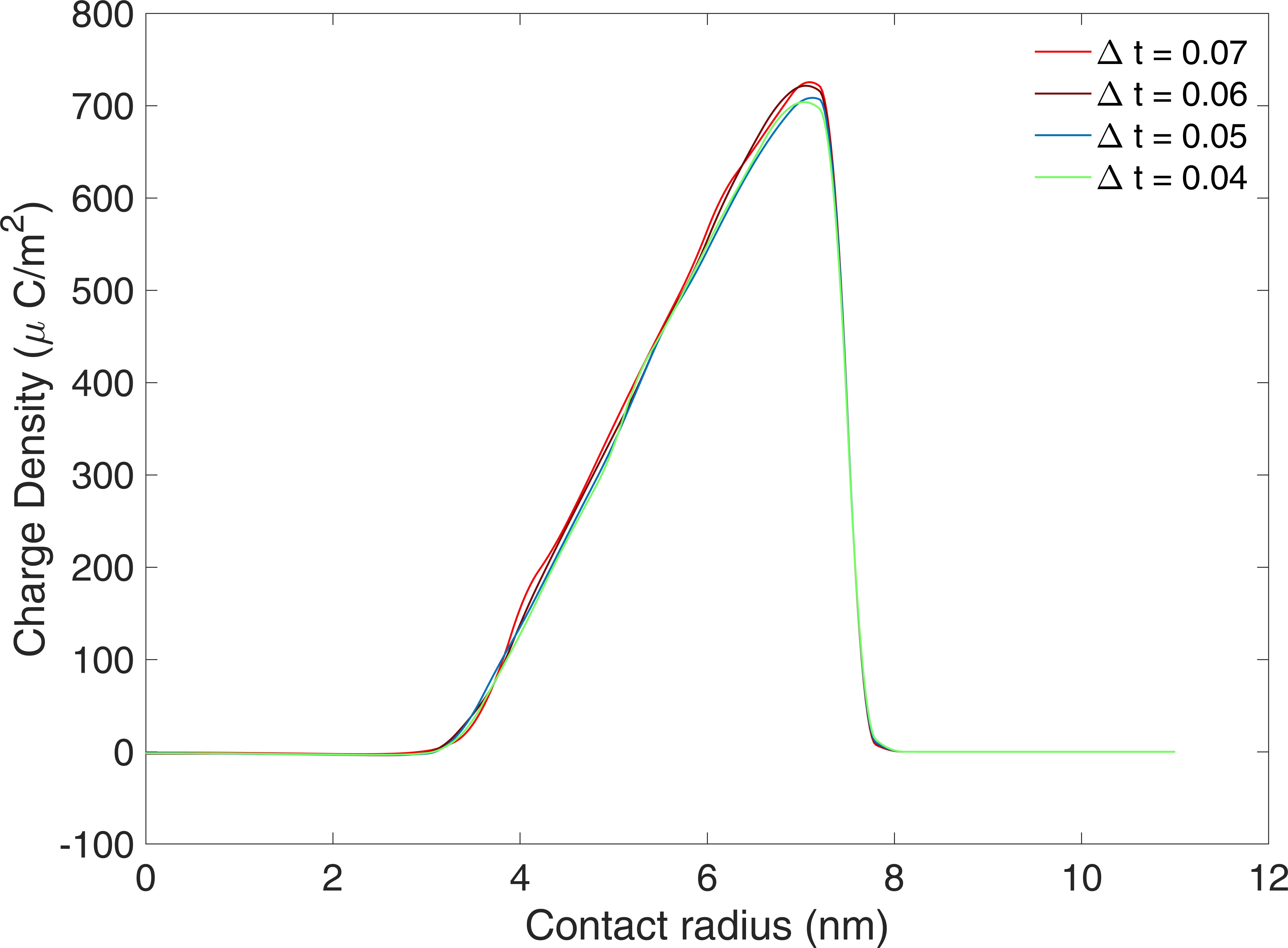}
	\caption{Load-step convergence of the residual charge density for unbiased PMMA contact at 50~nN. $\Delta t$ is the pseudo-time increment.}
	\label{fig:timestep_convergence}
\end{figure}

To verify that the numerical results are independent of the spatial and pseudo-time step discretization, we perform a convergence study for the representative case of unbiased PMMA contact at a loading force of 50~nN. Figure~\ref{fig:mesh_convergence} shows the surface charge density at the end of loading for six successively refined meshes. The profiles for $h = 0.29$~nm and $h = 0.24$~nm are virtually indistinguishable, confirming that the adopted mesh captures the charge distribution with sufficient accuracy. Figure~\ref{fig:timestep_convergence} presents the residual charge density after complete separation for four pseudo-time increments. The curves for $\Delta t = 0.05$ and $\Delta t = 0.04$ largely overlap, indicating that the chosen step size introduces no substantial discretization error.

\FloatBarrier

\section*{Data availability}
The data supporting the findings of this study are contained within the article and its appendices. Additional simulation outputs and post-processing files are available from the corresponding author upon reasonable request.

\section*{CRediT authorship contribution statement}
Han Hu: Conceptualization, Methodology, Software, Formal analysis, Visualization, Writing - original draft. Xiaoying Zhuang: Conceptualization, Supervision, Funding acquisition, Writing - review \& editing. Timon Rabczuk: Supervision, Writing - review \& editing.

\section*{Declaration of generative AI and AI-assisted technologies in the manuscript preparation process}
During the preparation of this manuscript, the authors used ChatGPT to improve language and readability. After using this tool/service, the authors reviewed and edited the content as needed and assume full responsibility for the content of the publication.

\section*{Declaration of competing interest}
The authors declare that they have no known competing financial interests or personal relationships that could have appeared to influence the work reported in this paper.

\section*{Acknowledgments}
The authors acknowledge financial support from the Deutsche Forschungsgemeinschaft (DFG, German Research Foundation) under Germany's Excellence Strategy within the Cluster of Excellence PhoenixD (EXC 2122, Project ID 390833453).

\bibliography{references}

@article{Wang2019,
  author  = {Wang, Zhong Lin and Wang, Aurelia Chi},
  title   = {On the origin of contact-electrification},
  journal = {Materials Today},
  volume  = {30},
  pages   = {34--51},
  year    = {2019},
  doi     = {10.1016/j.mattod.2019.05.016}
}

@article{Pan2019,
  author  = {Pan, S. and Zhang, Z.},
  title   = {Fundamental theories and basic principles of triboelectric effect: A review},
  journal = {Friction},
  volume  = {7},
  number  = {1},
  pages   = {2--17},
  year    = {2019},
  doi     = {10.1007/s40544-018-0217-7}
}

@article{Lowell1980,
  author  = {Lowell, J. and Rose-Innes, A. C.},
  title   = {Contact electrification},
  journal = {Advances in Physics},
  volume  = {29},
  number  = {6},
  pages   = {947--1023},
  year    = {1980},
  doi     = {10.1080/00018738000101466}
}

@book{Harper1967,
  author    = {Harper, William R.},
  title     = {Contact and Frictional Electrification},
  publisher = {Clarendon Press},
  address   = {Oxford},
  year      = {1967}
}

@article{Wong2015,
  author  = {Wong, J. and Kwok, P. C. L. and Chan, H.-K.},
  title   = {Electrostatics in pharmaceutical solids},
  journal = {Chemical Engineering Science},
  volume  = {125},
  pages   = {225--237},
  year    = {2015},
  doi     = {10.1016/j.ces.2014.05.037}
}

@book{Schein2013,
  author    = {Schein, Lawrence B.},
  title     = {Electrophotography and Development Physics},
  publisher = {Springer},
  year      = {1992},
  edition   = {2nd},
  doi       = {10.1007/978-3-642-77744-8}
}

@article{Saunders2008,
  author  = {Saunders, C.},
  title   = {Charge separation mechanisms in clouds},
  journal = {Space Science Reviews},
  volume  = {137},
  number  = {1},
  pages   = {335--353},
  year    = {2008},
  doi     = {10.1007/s11214-008-9345-0}
}

@article{Lacks2011,
  title     = {Contact electrification of insulating materials},
  author    = {Lacks, Daniel J and Sankaran, R Mohan},
  journal   = {Journal of Physics D: Applied Physics},
  volume    = {44},
  number    = {45},
  pages     = {453001},
  year      = {2011},
  publisher = {IOP Publishing},
  doi       = {10.1088/0022-3727/44/45/453001}
}

@book{Greason1987,
  title     = {Electrostatic Damage in Electronics: Devices and Systems},
  author    = {Greason, William D.},
  publisher = {Research Studies Press},
  address   = {Letchworth},
  year      = {1987},
  isbn      = {9780863800535}
}

@article{Glor1985,
  title     = {Hazards due to electrostatic charging of powders},
  author    = {Glor, Martin},
  journal   = {Journal of Electrostatics},
  volume    = {16},
  number    = {2-3},
  pages     = {175--191},
  year      = {1985},
  publisher = {Elsevier},
  doi       = {10.1016/0304-3886(85)90041-5}
}

@article{Matsusaka2010,
  title     = {Triboelectric charging of powders: A review},
  author    = {Matsusaka, Shuji and Maruyama, Hiroyuki and Matsuyama, Tatsushi and Ghadiri, Mojtaba},
  journal   = {Chemical Engineering Science},
  volume    = {65},
  number    = {22},
  pages     = {5781--5807},
  year      = {2010},
  publisher = {Elsevier},
  doi       = {10.1016/j.ces.2010.07.005}
}

@article{Fan2012,
  title     = {Flexible triboelectric generator},
  author    = {Fan, Feng-Ru and Tian, Zhong-Qun and Wang, Zhong Lin},
  journal   = {Nano Energy},
  volume    = {1},
  number    = {2},
  pages     = {328--334},
  year      = {2012},
  publisher = {Elsevier},
  doi       = {10.1016/j.nanoen.2012.01.004}
}

@article{Wang2020,
  title     = {Triboelectric nanogenerator ({TENG})---sparking an energy and sensor revolution},
  author    = {Wang, Zhong Lin},
  journal   = {Advanced Energy Materials},
  volume    = {10},
  number    = {17},
  pages     = {2000137},
  year      = {2020},
  publisher = {Wiley Online Library},
  doi       = {10.1002/aenm.202000137}
}

@article{Zi2015,
  title     = {Standards and figure-of-merits for quantifying the performance of triboelectric nanogenerators},
  author    = {Zi, Yunlong and Niu, Simiao and Wang, Jie and Wen, Zhen and Tang, Wei and Wang, Zhong Lin},
  journal   = {Nature Communications},
  volume    = {6},
  number    = {1},
  pages     = {8376},
  year      = {2015},
  publisher = {Nature Publishing Group UK London},
  doi       = {10.1038/ncomms9376}
}

@article{chen2018scavenging,
  title     = {Scavenging wind energy by triboelectric nanogenerators},
  author    = {Chen, Bo and Yang, Ya and Wang, Zhong Lin},
  journal   = {Advanced Energy Materials},
  volume    = {8},
  number    = {10},
  pages     = {1702649},
  year      = {2018},
  publisher = {Wiley Online Library},
  doi       = {10.1002/aenm.201702649}
}

@article{wang2016sustainably,
  title     = {Sustainably powering wearable electronics solely by biomechanical energy},
  author    = {Wang, Jie and Li, Shengming and Yi, Fang and Zi, Yunlong and Lin, Jun and Wang, Xiaofeng and Xu, Youlong and Wang, Zhong Lin},
  journal   = {Nature Communications},
  volume    = {7},
  number    = {1},
  pages     = {12744},
  year      = {2016},
  publisher = {Nature Publishing Group UK London},
  doi       = {10.1038/ncomms12744}
}

@article{wang2015progress,
  title     = {Progress in triboelectric nanogenerators as a new energy technology and self-powered sensors},
  author    = {Wang, Zhong Lin and Chen, Jun and Lin, Long},
  journal   = {Energy \& Environmental Science},
  volume    = {8},
  number    = {8},
  pages     = {2250--2282},
  year      = {2015},
  publisher = {Royal Society of Chemistry},
  doi       = {10.1039/C5EE01532D}
}

@article{cheng2024influence,
  title     = {The Influence of Discontinuity-Induced Fringing Effect on the Output Performance of Contact-Separation Mode Triboelectric Nanogenerators: Experiment and Modeling Studies},
  author    = {Cheng, Teresa and Hu, Han and Valizadeh, Navid and Qiong, Liu and Bittner, Florian and Yang, Ling and Rabczuk, Timon and Jiang, Xiaoning and Zhuang, Xiaoying},
  journal   = {Advanced Energy and Sustainability Research},
  volume    = {5},
  number    = {10},
  pages     = {2400002},
  year      = {2024},
  publisher = {Wiley Online Library},
  doi       = {10.1002/aesr.202400002}
}

@article{liu2021promoting,
  title     = {Promoting smart cities into the 5{G} era with multi-field Internet of Things ({IoT}) applications powered with advanced mechanical energy harvesters},
  author    = {Liu, Long and Guo, Xinge and Lee, Chengkuo},
  journal   = {Nano Energy},
  volume    = {88},
  pages     = {106304},
  year      = {2021},
  publisher = {Elsevier},
  doi       = {10.1016/j.nanoen.2021.106304}
}

@article{yi2019recent,
  title     = {Recent advances in triboelectric nanogenerator-based health monitoring},
  author    = {Yi, Fang and Zhang, Zheng and Kang, Zhuo and Liao, Qingliang and Zhang, Yue},
  journal   = {Advanced Functional Materials},
  volume    = {29},
  number    = {41},
  pages     = {1808849},
  year      = {2019},
  publisher = {Wiley Online Library},
  doi       = {10.1002/adfm.201808849}
}

@article{dharmasena2017triboelectric,
  title     = {Triboelectric nanogenerators: providing a fundamental framework},
  author    = {Dharmasena, Randunu Devage Ishara Gihan and Jayawardena, KDGI and Mills, CA and Deane, JHB and Anguita, JV and Dorey, RA and Silva, SRP},
  journal   = {Energy \& Environmental Science},
  volume    = {10},
  number    = {8},
  pages     = {1801--1811},
  year      = {2017},
  publisher = {Royal Society of Chemistry},
  doi       = {10.1039/C7EE01139C}
}

@article{tian2021self,
  title     = {Self-powered room-temperature ethanol sensor based on brush-shaped triboelectric nanogenerator},
  author    = {Tian, Jingwen and Wang, Fan and Ding, Yafei and Lei, Rui and Shi, Yuxiang and Tao, Xinglin and Li, Shuyao and Yang, Ya and Chen, Xiangyu},
  journal   = {Research},
  volume    = {2021},
  pages     = {8564780},
  year      = {2021},
  publisher = {AAAS},
  doi       = {10.34133/2021/8564780}
}

@article{xu2018coupled,
  title     = {Coupled triboelectric nanogenerator networks for efficient water wave energy harvesting},
  author    = {Xu, Liang and Jiang, Tao and Lin, Pei and Shao, Jia Jia and He, Chuan and Zhong, Wei and Chen, Xiang Yu and Wang, Zhong Lin},
  journal   = {ACS Nano},
  volume    = {12},
  number    = {2},
  pages     = {1849--1858},
  year      = {2018},
  publisher = {ACS Publications},
  doi       = {10.1021/acsnano.7b08674}
}

@article{zi2016effective,
  title     = {Effective energy storage from a triboelectric nanogenerator},
  author    = {Zi, Yunlong and Wang, Jie and Wang, Sihong and Li, Shengming and Wen, Zhen and Guo, Hengyu and Wang, Zhong Lin},
  journal   = {Nature Communications},
  volume    = {7},
  number    = {1},
  pages     = {10987},
  year      = {2016},
  publisher = {Nature Publishing Group UK London},
  doi       = {10.1038/ncomms10987}
}

@article{cheng2023triboelectric,
  title     = {Triboelectric nanogenerators},
  author    = {Cheng, Tinghai and Shao, Jiajia and Wang, Zhong Lin},
  journal   = {Nature Reviews Methods Primers},
  volume    = {3},
  number    = {1},
  pages     = {39},
  year      = {2023},
  publisher = {Nature Publishing Group UK London},
  doi       = {10.1038/s43586-023-00220-3}
}

@article{cao2024progress,
  title     = {Progress in techniques for improving the output performance of triboelectric nanogenerators},
  author    = {Cao, Chen and Li, Zhongjie and Shen, Fan and Zhang, Qin and Gong, Ying and Guo, Hengyu and Peng, Yan and Wang, Zhong Lin},
  journal   = {Energy \& Environmental Science},
  volume    = {17},
  number    = {3},
  pages     = {885--924},
  year      = {2024},
  publisher = {Royal Society of Chemistry},
  doi       = {10.1039/D3EE03520D}
}

@article{zou2019quantifying,
  title     = {Quantifying the triboelectric series},
  author    = {Zou, Haiyang and Zhang, Ying and Guo, Litong and Wang, Peihong and He, Xu and Dai, Guozhang and Zheng, Haiwu and Chen, Chaoyu and Wang, Aurelia Chi and Xu, Cheng and others},
  journal   = {Nature Communications},
  volume    = {10},
  number    = {1},
  pages     = {1427},
  year      = {2019},
  publisher = {Nature Publishing Group UK London},
  doi       = {10.1038/s41467-019-09461-x}
}

@article{chen2024quantifying,
  title     = {Quantifying triboelectric series of polymers based on the measurement of triboelectrification with {NaCl} solution},
  author    = {Chen, Qin and Shang, Hongfei and Cheng, Bingxue and Lu, Chaoze and Wang, Yihan and Zhang, Yang and Shao, Tianmin},
  journal   = {Chemical Engineering Journal},
  volume    = {488},
  pages     = {150871},
  year      = {2024},
  publisher = {Elsevier},
  doi       = {10.1016/j.cej.2024.150871}
}

@article{Lin2022flexo,
  author  = {Lin, Shiquan and Zheng, Mingli and Xu, Liang and Zhu, Laipan and Wang, Zhong Lin},
  title   = {Electron transfer driven by tip-induced flexoelectricity in contact electrification},
  journal = {Journal of Physics D: Applied Physics},
  volume  = {55},
  number  = {31},
  pages   = {315502},
  year    = {2022},
  doi     = {10.1088/1361-6463/ac6f2e}
}

@article{Qiao2021mixed,
  author  = {Qiao, Huimin and Zhao, Pin and Kwon, Owoong and Sohn, Ahrum and Zhuo, Fangping and Lee, Dong-Min and Sun, Changhyo and Seol, Daehee and Lee, Daesu and Kim, Sang-Woo and Kim, Yunseok},
  title   = {Mixed triboelectric and flexoelectric charge transfer at the nanoscale},
  journal = {Advanced Science},
  volume  = {8},
  number  = {20},
  pages   = {2101793},
  year    = {2021},
  doi     = {10.1002/advs.202101793}
}

@article{Mizzi2019does,
  author  = {Mizzi, Christopher A. and Lin, Audrey Y. W. and Marks, Laurence D.},
  title   = {Does flexoelectricity drive triboelectricity?},
  journal = {Physical Review Letters},
  volume  = {123},
  number  = {11},
  pages   = {116103},
  year    = {2019},
  doi     = {10.1103/PhysRevLett.123.116103}
}

@article{persson2020role,
  title     = {On the role of flexoelectricity in triboelectricity for randomly rough surfaces},
  author    = {Persson, BNJ},
  journal   = {Europhysics Letters},
  volume    = {129},
  number    = {1},
  pages     = {10006},
  year      = {2020},
  publisher = {IOP Publishing},
  doi       = {10.1209/0295-5075/129/10006}
}

@article{Olson2025complex,
  author  = {Olson, Karl P. and Marks, Laurence D.},
  title   = {Is triboelectricity confusing, confused, or complex?},
  journal = {Reports on Progress in Physics},
  volume  = {88},
  number  = {10},
  pages   = {104501},
  year    = {2025},
  doi     = {10.1088/1361-6633/ae08cb}
}

@article{xia2020material,
  title     = {On the material-dependent charge transfer mechanism of the contact electrification},
  author    = {Xia, Xin and Wang, Haoyu and Guo, Hengyu and Xu, Cheng and Zi, Yunlong},
  journal   = {Nano Energy},
  volume    = {78},
  pages     = {105343},
  year      = {2020},
  publisher = {Elsevier},
  doi       = {10.1016/j.nanoen.2020.105343}
}

@article{Lowell1979,
  title     = {Tunnelling between metals and insulators and its role in contact electrification},
  author    = {Lowell, J},
  journal   = {Journal of Physics D: Applied Physics},
  volume    = {12},
  number    = {9},
  pages     = {1541--1554},
  year      = {1979},
  publisher = {IOP Publishing},
  doi       = {10.1088/0022-3727/12/9/016}
}

@article{Harper1960,
  title     = {Contact electrification of semiconductors},
  author    = {Harper, W. R.},
  journal   = {British Journal of Applied Physics},
  volume    = {11},
  number    = {8},
  pages     = {324--327},
  year      = {1960},
  publisher = {IOP Publishing},
  doi       = {10.1088/0508-3443/11/8/304}
}

@article{Xu2018electron,
  title     = {On the electron-transfer mechanism in the contact-electrification effect},
  author    = {Xu, Cheng and Zi, Yunlong and Wang, Aurelia Chi and Zou, Haiyang and Dai, Yejing and He, Xu and Wang, Peihong and Wang, Yi-Cheng and Feng, Peizhong and Li, Dawei and others},
  journal   = {Advanced Materials},
  volume    = {30},
  number    = {15},
  pages     = {1706790},
  year      = {2018},
  publisher = {Wiley Online Library},
  doi       = {10.1002/adma.201706790}
}

@article{Lin2020overlap,
  title     = {The overlapped electron-cloud model for electron transfer in contact electrification},
  author    = {Lin, Shiquan and Xu, Cheng and Xu, Liang and Wang, Zhong Lin},
  journal   = {Advanced Functional Materials},
  volume    = {30},
  number    = {11},
  pages     = {1909724},
  year      = {2020},
  publisher = {Wiley Online Library},
  doi       = {10.1002/adfm.201909724}
}

@article{Vick1953,
  author  = {Vick, F. A.},
  title   = {Theory of contact electrification},
  journal = {British Journal of Applied Physics},
  volume  = {4},
  number  = {Supplement 2},
  pages   = {S1--S5},
  year    = {1953},
  doi     = {10.1088/0508-3443/4/S2/301}
}

@article{Cowley1965,
  author  = {Cowley, A. M. and Sze, S. M.},
  title   = {Surface states and barrier height of metal-semiconductor systems},
  journal = {Journal of Applied Physics},
  volume  = {36},
  number  = {10},
  pages   = {3212--3220},
  year    = {1965},
  doi     = {10.1063/1.1702952}
}

@article{Diaz1993,
  title     = {An ion transfer model for contact charging},
  author    = {Diaz, A. F. and Fenzel-Alexander, D.},
  journal   = {Langmuir},
  volume    = {9},
  number    = {4},
  pages     = {1009--1015},
  year      = {1993},
  publisher = {ACS Publications},
  doi       = {10.1021/la00028a021}
}

@article{Liu2008,
  title     = {Electrostatic electrochemistry at insulators},
  author    = {Liu, Chongyang and Bard, Allen J.},
  journal   = {Nature Materials},
  volume    = {7},
  number    = {6},
  pages     = {505--509},
  year      = {2008},
  publisher = {Nature Publishing Group UK London},
  doi       = {10.1038/nmat2160}
}

@article{ma2025quantifying,
  title     = {Quantifying Electron and Ion Transfers in Contact Electrification with Ionomers},
  author    = {Ma, Xiaoting and Zhou, Jiaming and Kim, Eunjong and Gao, Jingyi and Hu, Wenyi and Shin, Dong-Myeong},
  journal   = {Advanced Functional Materials},
  volume    = {36},
  number    = {5},
  pages     = {e06471},
  year      = {2026},
  publisher = {Wiley Online Library},
  doi       = {10.1002/adfm.202506471}
}

@article{Pence1994,
  title     = {Effect of surface moisture on contact charge of polymers containing ions},
  author    = {Pence, S. and Novotny, V. J. and Diaz, A. F.},
  journal   = {Langmuir},
  volume    = {10},
  number    = {2},
  pages     = {592--596},
  year      = {1994},
  publisher = {ACS Publications},
  doi       = {10.1021/la00014a042}
}

@article{McCarty2008,
  author  = {McCarty, L. S. and Whitesides, G. M.},
  title   = {Electrostatic charging due to separation of ions at interfaces: Contact electrification of ionic electrets},
  journal = {Angewandte Chemie International Edition},
  volume  = {47},
  number  = {12},
  pages   = {2188--2207},
  year    = {2008},
  doi     = {10.1002/anie.200701812}
}

@article{Salaneck1976,
  title     = {Double mass transfer during polymer-polymer contacts},
  author    = {Salaneck, W. R. and Paton, A. and Clark, D. T.},
  journal   = {Journal of Applied Physics},
  volume    = {47},
  number    = {1},
  pages     = {144--147},
  year      = {1976},
  publisher = {American Institute of Physics},
  doi       = {10.1063/1.322306}
}

@article{Baytekin2012,
  title   = {Material transfer and polarity reversal in contact charging},
  author  = {Baytekin, H. Tarik and Baytekin, Bilge and Incorvati, Jared T. and Grzybowski, Bartosz A.},
  journal = {Angewandte Chemie International Edition},
  volume  = {51},
  number  = {20},
  pages   = {4843--4847},
  year    = {2012},
  doi     = {10.1002/anie.201200057}
}

@article{Willatzen2019,
  title     = {Contact electrification by quantum-mechanical tunneling},
  author    = {Willatzen, Morten and Wang, Zhong Lin},
  journal   = {Research},
  volume    = {2019},
  pages     = {6528689},
  year      = {2019},
  publisher = {AAAS},
  doi       = {10.34133/2019/6528689}
}

@article{Pandey2018,
  title     = {Correlating material transfer and charge transfer in contact electrification},
  author    = {Pandey, Rakesh K and Kakehashi, Hiroto and Nakanishi, Hideyuki and Soh, Siowling},
  journal   = {The Journal of Physical Chemistry C},
  volume    = {122},
  number    = {28},
  pages     = {16154--16160},
  year      = {2018},
  publisher = {ACS Publications},
  doi       = {10.1021/acs.jpcc.8b04357}
}

@article{Olson2025quantitative,
  author  = {Olson, Karl P. and Marks, Laurence D.},
  title   = {A quantitative model of triboelectric charge transfer},
  journal = {Friction},
  volume  = {13},
  number  = {2},
  pages   = {9440937},
  year    = {2025},
  doi     = {10.26599/FRICT.2025.9440937}
}

@article{Tan2021electron,
  author  = {Tan, Dan and Zhang, Yi and Liu, Yang and Chen, Lei and Wang, Ning and Cai, Zengxia and Tang, Wei and Liu, Haibo and Wang, Zhong Lin},
  title   = {Electron transfer in the contact-electrification between corrugated 2{D} materials: A first-principles study},
  journal = {Nano Energy},
  volume  = {79},
  pages   = {105386},
  year    = {2021},
  doi     = {10.1016/j.nanoen.2020.105386}
}

@article{Xu2019curvature,
  author  = {Xu, Cheng and Zhang, Bojing and Wang, Aurelia Chi and Zou, Haiyang and Liu, Guoxu and Ding, Wenbo and Wu, Changsheng and Ma, Ming and Feng, Puguang and Lin, Zhaoling and Wang, Zhong Lin},
  title   = {Contact-electrification between two identical materials: Curvature effect},
  journal = {ACS Nano},
  volume  = {13},
  number  = {2},
  pages   = {2034--2041},
  year    = {2019},
  doi     = {10.1021/acsnano.8b08533}
}

@article{Sow2012reversal,
  author  = {Sow, Mark and Widenor, Rachel and Kumar, Ashish and Lee, Seung Whan and Lacks, Daniel J. and Sankaran, R. Mohan},
  title   = {Strain-induced reversal of charge transfer in contact electrification},
  journal = {Angewandte Chemie International Edition},
  volume  = {51},
  number  = {11},
  pages   = {2695--2698},
  year    = {2012},
  doi     = {10.1002/anie.201107256}
}

@article{Wang2019progress,
  author  = {Wang, Biao and Gu, Yuantai and Zhang, Shengping and Chen, Long-Qing},
  title   = {Flexoelectricity in solids: Progress, challenges, and perspectives},
  journal = {Progress in Materials Science},
  volume  = {106},
  pages   = {100570},
  year    = {2019},
  doi     = {10.1016/j.pmatsci.2019.05.003}
}

@article{zubko2013flexoelectric,
  title     = {Flexoelectric effect in solids},
  author    = {Zubko, Pavlo and Catalan, Gustau and Tagantsev, Alexander K.},
  journal   = {Annual Review of Materials Research},
  volume    = {43},
  number    = {1},
  pages     = {387--421},
  year      = {2013},
  publisher = {Annual Reviews},
  doi       = {10.1146/annurev-matsci-071312-121634}
}

@article{nguyen2013nanoscale,
  title     = {Nanoscale flexoelectricity},
  author    = {Nguyen, Thanh D. and Mao, Sheng and Yeh, Yao-Wen and Purohit, Prashant K. and McAlpine, Michael C.},
  journal   = {Advanced Materials},
  volume    = {25},
  number    = {7},
  pages     = {946--974},
  year      = {2013},
  publisher = {Wiley Online Library},
  doi       = {10.1002/adma.201203852}
}

@article{ma2001large,
  title     = {Large flexoelectric polarization in ceramic lead magnesium niobate},
  author    = {Ma, Wenhui and Cross, L. Eric},
  journal   = {Applied Physics Letters},
  volume    = {79},
  number    = {26},
  pages     = {4420--4422},
  year      = {2001},
  publisher = {American Institute of Physics},
  doi       = {10.1063/1.1426690}
}

@article{shu2011symmetry,
  title     = {Symmetry of flexoelectric coefficients in crystalline medium},
  author    = {Shu, Longlong and Wei, Xiaoyong and Pang, Ting and Yao, Xi and Wang, Chunlei},
  journal   = {Journal of Applied Physics},
  volume    = {110},
  number    = {10},
  pages     = {104106},
  year      = {2011},
  publisher = {AIP Publishing},
  doi       = {10.1063/1.3662196}
}

@article{Olson2022band,
  author  = {Olson, Karl P. and Mizzi, Christopher A. and Marks, Laurence D.},
  title   = {Band bending and ratcheting explain triboelectricity in a flexoelectric contact diode},
  journal = {Nano Letters},
  volume  = {22},
  number  = {10},
  pages   = {3914--3921},
  year    = {2022},
  doi     = {10.1021/acs.nanolett.2c00107}
}

@article{Forward2009,
  title     = {Charge segregation depends on particle size in triboelectrically charged granular materials},
  author    = {Forward, Keith M. and Lacks, Daniel J. and Sankaran, R. Mohan},
  journal   = {Physical Review Letters},
  volume    = {102},
  number    = {2},
  pages     = {028001},
  year      = {2009},
  publisher = {APS},
  doi       = {10.1103/PhysRevLett.102.028001}
}

@article{Bilici2014,
  title     = {Particle size effects in particle-particle triboelectric charging studied with an integrated fluidized bed and electrostatic separator system},
  author    = {Bilici, Mihai A. and Toth, Joseph R. and Sankaran, R. Mohan and Lacks, Daniel J.},
  journal   = {Review of Scientific Instruments},
  volume    = {85},
  number    = {10},
  pages     = {103903},
  year      = {2014},
  publisher = {AIP Publishing},
  doi       = {10.1063/1.4897182}
}

@article{Marks2025flexo,
  author  = {Marks, Laurence D. and Olson, Karl P.},
  title   = {Flexoelectricity, triboelectricity, and free interfacial charges},
  journal = {Small},
  volume  = {21},
  number  = {28},
  pages   = {2310546},
  year    = {2025},
  doi     = {10.1002/smll.202310546}
}

@article{mao2014insights,
  title     = {Insights into flexoelectric solids from strain-gradient elasticity},
  author    = {Mao, Sheng and Purohit, Prashant K.},
  journal   = {Journal of Applied Mechanics},
  volume    = {81},
  number    = {8},
  pages     = {081004},
  year      = {2014},
  publisher = {American Society of Mechanical Engineers},
  doi       = {10.1115/1.4027451}
}

@article{Mao2016,
  title     = {Mixed finite-element formulations in piezoelectricity and flexoelectricity},
  author    = {Mao, Sheng and Purohit, Prashant K. and Aravas, Nikolaos},
  journal   = {Proceedings of the Royal Society A: Mathematical, Physical and Engineering Sciences},
  volume    = {472},
  number    = {2190},
  pages     = {20150879},
  year      = {2016},
  publisher = {The Royal Society Publishing},
  doi       = {10.1098/rspa.2015.0879}
}

@article{deng2017mixed,
  title     = {Mixed finite elements for flexoelectric solids},
  author    = {Deng, Feng and Deng, Qian and Yu, Wenshan and Shen, Shengping},
  journal   = {Journal of Applied Mechanics},
  volume    = {84},
  number    = {8},
  pages     = {081004},
  year      = {2017},
  publisher = {American Society of Mechanical Engineers},
  doi       = {10.1115/1.4036939}
}

@article{Abdollahi2014,
  title     = {Computational evaluation of the flexoelectric effect in dielectric solids},
  author    = {Abdollahi, Amir and Peco, Christian and Millan, Daniel and Arroyo, Marino and Arias, Irene},
  journal   = {Journal of Applied Physics},
  volume    = {116},
  number    = {9},
  pages     = {093502},
  year      = {2014},
  publisher = {AIP Publishing},
  doi       = {10.1063/1.4893974}
}

@article{Abdollahi2015,
  title     = {Revisiting pyramid compression to quantify flexoelectricity: A three-dimensional simulation study},
  author    = {Abdollahi, Amir and Mill{\'a}n, Daniel and Peco, Christian and Arroyo, Marino and Arias, Irene},
  journal   = {Physical Review B},
  volume    = {91},
  number    = {10},
  pages     = {104103},
  year      = {2015},
  publisher = {APS},
  doi       = {10.1103/PhysRevB.91.104103}
}

@article{Hughes2005,
  title     = {Isogeometric analysis: {CAD}, finite elements, {NURBS}, exact geometry and mesh refinement},
  author    = {Hughes, Thomas J. R. and Cottrell, John A. and Bazilevs, Yuri},
  journal   = {Computer Methods in Applied Mechanics and Engineering},
  volume    = {194},
  number    = {39-41},
  pages     = {4135--4195},
  year      = {2005},
  publisher = {Elsevier},
  doi       = {10.1016/j.cma.2004.10.008}
}

@article{nguyen2015isogeometric,
  title     = {Isogeometric analysis: an overview and computer implementation aspects},
  author    = {Nguyen, Vinh Phu and Anitescu, Cosmin and Bordas, St{\'e}phane PA and Rabczuk, Timon},
  journal   = {Mathematics and Computers in Simulation},
  volume    = {117},
  pages     = {89--116},
  year      = {2015},
  publisher = {Elsevier},
  doi       = {10.1016/j.matcom.2015.05.008}
}

@article{Ghasemi2017,
  title     = {A level-set based {IGA} formulation for topology optimization of flexoelectric materials},
  author    = {Ghasemi, Hamid and Park, Harold S. and Rabczuk, Timon},
  journal   = {Computer Methods in Applied Mechanics and Engineering},
  volume    = {313},
  pages     = {239--258},
  year      = {2017},
  publisher = {Elsevier},
  doi       = {10.1016/j.cma.2016.09.029}
}

@article{nanthakumar2017topology,
  title     = {Topology optimization of flexoelectric structures},
  author    = {Nanthakumar, S. S. and Zhuang, Xiaoying and Park, Harold S. and Rabczuk, Timon},
  journal   = {Journal of the Mechanics and Physics of Solids},
  volume    = {105},
  pages     = {217--234},
  year      = {2017},
  publisher = {Elsevier},
  doi       = {10.1016/j.jmps.2017.05.010}
}

@article{Yvonnet2017,
  title     = {A numerical framework for modeling flexoelectricity and {Maxwell} stress in soft dielectrics at finite strains},
  author    = {Yvonnet, Julien and Liu, L. P.},
  journal   = {Computer Methods in Applied Mechanics and Engineering},
  volume    = {313},
  pages     = {450--482},
  year      = {2017},
  publisher = {Elsevier},
  doi       = {10.1016/j.cma.2016.09.007}
}

@article{Nguyen2019,
  author  = {Nguyen, B. H. and Zhuang, X. and Rabczuk, T.},
  title   = {{NURBS-based} formulation for nonlinear electro-gradient elasticity in semiconductors},
  journal = {Computer Methods in Applied Mechanics and Engineering},
  volume  = {346},
  pages   = {1074--1095},
  year    = {2019},
  doi     = {10.1016/j.cma.2018.08.026}
}

@article{zhuang2025variationally,
  title     = {Variationally consistent {Maxwell} stress in flexoelectric structures under finite deformation and immersed in free space},
  author    = {Zhuang, Xiaoying and Hu, Han and Nanthakumar, S. S. and Tran, Quoc-Thai and Gong, Yanpeng and Rabczuk, Timon},
  journal   = {Applied Mathematical Modelling},
  volume    = {150},
  pages     = {116327},
  year      = {2026},
  publisher = {Elsevier},
  doi       = {10.1016/j.apm.2025.116327}
}

@article{hu2025computational,
  title     = {Computational flexoelectronics: A framework for analyzing {PN} junctions in flexoelectric semiconductors},
  author    = {Hu, Han and Liu, Zhaowei and Liu, Qiong and Jiang, Xiaoning and Zhuang, Xiaoying and Rabczuk, Timon},
  journal   = {International Journal of Mechanical Sciences},
  volume    = {305},
  pages     = {110740},
  year      = {2025},
  publisher = {Elsevier},
  doi       = {10.1016/j.ijmecsci.2025.110740}
}

@article{codony2021modeling,
  title     = {Modeling flexoelectricity in soft dielectrics at finite deformation},
  author    = {Codony, David and Gupta, Prakhar and Marco, Onofre and Arias, Irene},
  journal   = {Journal of the Mechanics and Physics of Solids},
  volume    = {146},
  pages     = {104182},
  year      = {2021},
  publisher = {Elsevier},
  doi       = {10.1016/j.jmps.2020.104182}
}

@article{xu1993void,
  title     = {Void nucleation by inclusion debonding in a crystal matrix},
  author    = {Xu, X.-P. and Needleman, Alan},
  journal   = {Modelling and Simulation in Materials Science and Engineering},
  volume    = {1},
  number    = {2},
  pages     = {111--132},
  year      = {1993},
  publisher = {IOP Publishing},
  doi       = {10.1088/0965-0393/1/2/001}
}

@article{van2006improved,
  title     = {An improved description of the exponential {Xu} and {Needleman} cohesive zone law for mixed-mode decohesion},
  author    = {Van den Bosch, M. J. and Schreurs, P. J. G. and Geers, M. G. D.},
  journal   = {Engineering Fracture Mechanics},
  volume    = {73},
  number    = {9},
  pages     = {1220--1234},
  year      = {2006},
  publisher = {Elsevier},
  doi       = {10.1016/j.engfracmech.2005.12.006}
}

@article{dimitri2015coupled,
  title     = {Coupled cohesive zone models for mixed-mode fracture: A comparative study},
  author    = {Dimitri, Rossana and Trullo, Marco and De Lorenzis, Laura and Zavarise, Giorgio},
  journal   = {Engineering Fracture Mechanics},
  volume    = {148},
  pages     = {145--179},
  year      = {2015},
  publisher = {Elsevier},
  doi       = {10.1016/j.engfracmech.2015.09.029}
}

@article{simmons1963generalized,
  title     = {Generalized formula for the electric tunnel effect between similar electrodes separated by a thin insulating film},
  author    = {Simmons, John G.},
  journal   = {Journal of Applied Physics},
  volume    = {34},
  number    = {6},
  pages     = {1793--1803},
  year      = {1963},
  publisher = {American Institute of Physics},
  doi       = {10.1063/1.1702682}
}

@article{tang2022mechanoelectric,
  title     = {Mechanoelectric coupling model of polymethyl methacrylate under impact load},
  author    = {Tang, Enling and Leng, Bingyu and Han, Yafei and Xu, Mingyang and Chen, Chuang and Chang, Mengzhou and Guo, Kai and He, Liping},
  journal   = {Waves in Random and Complex Media},
  volume    = {35},
  number    = {7},
  pages     = {13185--13207},
  year      = {2022},
  publisher = {Taylor \& Francis},
  doi       = {10.1080/17455030.2022.2139427}
}

@article{ishiyama2002effects,
  title     = {Effects of humidity on {Young's} modulus in poly(methyl methacrylate)},
  author    = {Ishiyama, C. and Higo, Y.},
  journal   = {Journal of Polymer Science Part B: Polymer Physics},
  volume    = {40},
  number    = {5},
  pages     = {460--465},
  year      = {2002},
  publisher = {Wiley Online Library},
  doi       = {10.1002/polb.10107}
}

@book{harper2000modern,
  title     = {Modern plastics handbook},
  author    = {Harper, Charles A},
  year      = {2000},
  publisher = {McGraw Hill Professional}
}

@book{guo2017thermosets,
  title     = {Thermosets: structure, properties, and applications},
  author    = {Guo, Qipeng},
  year      = {2017},
  publisher = {Woodhead Publishing},
  doi       = {10.1016/C2015-0-06205-0}
}

@article{jamieson1910electrification,
  title     = {The electrification of insulating materials},
  author    = {Jamieson, Walter},
  journal   = {Nature},
  volume    = {83},
  number    = {2113},
  pages     = {189},
  year      = {1910},
  publisher = {Nature Publishing Group UK London},
  doi       = {10.1038/083189a0}
}

@article{Lowell1986,
  title     = {Triboelectrification of identical insulators. {I}. {An} experimental investigation},
  author    = {Lowell, J. and Truscott, W. S.},
  journal   = {Journal of Physics D: Applied Physics},
  volume    = {19},
  number    = {7},
  pages     = {1273--1280},
  year      = {1986},
  publisher = {IOP Publishing},
  doi       = {10.1088/0022-3727/19/7/017}
}

@article{hu2022isogeometric,
  title     = {An isogeometric analysis based method for frictional elastic contact problems with randomly rough surfaces},
  author    = {Hu, Han and Batou, Anas and Ouyang, Huajiang},
  journal   = {Computer Methods in Applied Mechanics and Engineering},
  volume    = {394},
  pages     = {114865},
  year      = {2022},
  publisher = {Elsevier},
  doi       = {10.1016/j.cma.2022.114865}
}

@article{apodaca2010contact,
  title   = {Contact electrification between identical materials},
  author  = {Apodaca, Mario M. and Wesson, Paul J. and Bishop, Kyle J. M. and Ratner, Mark A. and Grzybowski, Bartosz A.},
  journal = {Angewandte Chemie International Edition},
  volume  = {49},
  number  = {5},
  pages   = {946--949},
  year    = {2010},
  doi     = {10.1002/anie.200905281}
}

@article{lu2019temperature,
  title     = {Temperature dependence of flexoelectric coefficient for bulk polymer polyvinylidene fluoride},
  author    = {Lu, Jianfeng and Liang, Xu and Yu, Wenshan and Hu, Shuling and Shen, Shengping},
  journal   = {Journal of Physics D: Applied Physics},
  volume    = {52},
  number    = {7},
  pages     = {075302},
  year      = {2019},
  publisher = {IOP Publishing},
  doi       = {10.1088/1361-6463/aaf543}
}

@article{deng2014flexoelectricity,
  title     = {Flexoelectricity in soft materials and biological membranes},
  author    = {Deng, Qian and Liu, Liping and Sharma, Pradeep},
  journal   = {Journal of the Mechanics and Physics of Solids},
  volume    = {62},
  pages     = {209--227},
  year      = {2014},
  publisher = {Elsevier},
  doi       = {10.1016/j.jmps.2013.09.021}
}

@article{Baytekin2011mosaic,
  author  = {Baytekin, H. Tarik and Patashinski, Alexander Z. and Branicki, Michał and Baytekin, Bilge and Soh, Siowling and Grzybowski, Bartosz A.},
  title   = {The mosaic of surface charge in contact electrification},
  journal = {Science},
  volume  = {333},
  number  = {6040},
  pages   = {308--312},
  year    = {2011},
  doi     = {10.1126/science.1201512}
}

@article{Yudin2013fundamentals,
  author  = {Yudin, P. V. and Tagantsev, A. K.},
  title   = {Fundamentals of flexoelectricity in solids},
  journal = {Nanotechnology},
  volume  = {24},
  number  = {43},
  pages   = {432001},
  year    = {2013},
  doi     = {10.1088/0957-4484/24/43/432001}
}

@article{Codony2021mathematical,
  author  = {Codony, D. and Mocci, A. and Barcel{\'o}-Mercader, J. and Arias, I.},
  title   = {Mathematical and computational modeling of flexoelectricity},
  journal = {Journal of Applied Physics},
  volume  = {130},
  number  = {23},
  pages   = {231102},
  year    = {2021},
  doi     = {10.1063/5.0067852}
}

@article{Zhuang2020computational,
  author  = {Zhuang, Xiaoying and Nguyen, Binh Huy and Nanthakumar, Subbiah Srivilliputtur and Tran, Thai Quoc and Alajlan, Naif and Rabczuk, Timon},
  title   = {Computational modeling of flexoelectricity---{A} review},
  journal = {Energies},
  volume  = {13},
  number  = {6},
  pages   = {1326},
  year    = {2020},
  doi     = {10.3390/en13061326}
}

@article{Silavnieks2025flexo,
  author  = {Silavnieks, Ulvis and Jing, Qingshen and Gadegaard, Nikolaj and Mulvihill, Daniel M. and Xu, Yang},
  title   = {Flexoelectricity driven elastic contact-separation model for triboelectrification},
  journal = {Friction},
  volume  = {14},
  number  = {3},
  pages   = {9441115},
  year    = {2026},
  doi     = {10.26599/FRICT.2025.9441115}
}

@book{wriggers2006computational,
  author    = {Wriggers, Peter},
  title     = {Computational Contact Mechanics},
  publisher = {Springer},
  address   = {Berlin, Heidelberg},
  year      = {2006},
  edition   = {2nd},
  doi       = {10.1007/978-3-540-32609-0}
}
\end{document}